\documentclass[12pt]{article}

\usepackage{axodraw}
\usepackage{color}
\usepackage{pstricks}

\usepackage{amsmath}

\input{epsf}
\def\slash#1{\ooalign{$\hfil/\hfil$\crcr$#1$}} 

\def\ltap{\raisebox{-.6ex}{\rlap{$\,\sim\,$}} \raisebox{.4ex}{$\,<\,$}}

\def\naive{na\"{\i}ve}

\newcommand\as{\alpha_{\mathrm{S}}}
\newcommand\g{g_{\mathrm{S}}}
\newcommand\gep{\mu^\epsilon \,g_{\mathrm{S}}}

\newcommand\f[2]{\frac{#1}{#2}}
\def\ep{\epsilon}
\def\ktil{\widetilde k}

\def\wp{\widetilde P}

\def\beq{\begin{equation}}
\def\eeq{\end{equation}}
\def\beeq{\begin{eqnarray}}
\def\eeeq{\end{eqnarray}}
\def\cm{{\cal M}}
\def\bom#1{{\mbox{\boldmath $#1$}}}
\def\to{\rightarrow}

\newcommand{\la}{\langle}
\newcommand{\ra}{\rangle}

\def\nn{\nonumber}

\def\ID{1 \kern -.45 em 1}

\def\sp{{\bom {Sp}}}

\def\ket#1{|{#1}\ra}
\def\bra#1{\la{#1}|}

\def\ubar{{\overline u}}
\def\vbar{{\overline v}}
\def\vep{{\varepsilon}}

\def\msbar{{\overline {\rm MS}}}

\def\cmbar{{\overline {\cal M}}}
\def\Mbar{{\overline M}}

\def\imc{{\bom I}^{(1)}_{m\,C}}
\def\isc{{\bom I}^{(1)}_{2\,C}}
\def\dmc{{\bom \Delta}^{(1)}_{m\,C}}
\def\imctwo{{\bom I}^{(2)}_{m\,C}}
\def\imp{{\bom I}^{(1)}_{m\,P}}
\def\itwoc{{\bom I}^{(1)}_{2\,C}}
\def\itwopart{{\bom I}^{(2)}_{2\,C}}
\def\cbet0{b_0}
\def\bone{b_1}

\def\itc{{\widetilde{\bom I}}_C}
\def\itp{{\widetilde  I}_P}

\begin{document}

\begin{titlepage}
\renewcommand{\thefootnote}{\fnsymbol{footnote}}
\begin{flushright}
     LPN11-94\\ IFIC/11-72\\ ZU-TH 27/11
     \end{flushright}
\par \vspace{10mm}

\begin{center}
{\Large \bf
Space-like (vs. time-like) collinear limits in QCD:\\[1ex] 
is factorization violated?
}
\end{center}

\par \vspace{2mm}
\begin{center}
{\bf Stefano Catani}~$^{(a)}$, 
{\bf Daniel de Florian}~$^{(b) (c)}$ and 
{\bf Germ\'an Rodrigo}~$^{(d)}$

\vspace{5mm}

${}^{(a)}$INFN, Sezione di Firenze and
Dipartimento di Fisica e Astronomia,\\ 
Universit\`a
di Firenze,
I-50019 Sesto Fiorentino, 
Florence, Italy \\
\vspace*{2mm}
${}^{(b)}$Departamento de F\'\i sica and IFIBA, FCEYN, Universidad de Buenos Aires, \\
(1428) Pabell\'on 1 Ciudad Universitaria, 
Capital Federal, Argentina \\
\vspace*{2mm}
${}^{(c)}$Institut f\"ur Theoretische Physik, Universit\"at Z\"urich, 
CH-8057 Z\"urich,
Switzerland \\
\vspace*{2mm}
${}^{(d)}$Instituto de F\'{\i}sica Corpuscular, 
UVEG - Consejo Superior de Investigaciones Cient\'{\i}ficas,
Parc Cient\`{\i}fic, E-46980 Paterna (Valencia), Spain \\

\vspace{5mm}

\end{center}

\par \vspace{2mm}
\begin{center} {\large \bf Abstract} \end{center}
\begin{quote}
\pretolerance 10000

We consider the singular behaviour of QCD scattering amplitudes
in kinematical configurations where two or more momenta 
of the external partons become collinear.
At the tree level, this behaviour is known to be controlled by factorization
formulae in which the singular collinear factor is universal (process
independent).
We show that this strict (process-independent) factorization is not valid
at one-loop and higher-loop orders in the case of the collinear limit in
space-like regions 
(e.g., collinear radiation from initial-state partons).
We introduce a generalized version of all-order collinear factorization,
in which the space-like singular factors retain some dependence on the 
momentum and colour charge of the non-collinear partons. We present explicit
results on one-loop and two-loop amplitudes for both the two-parton
and multiparton collinear limits. 
At the level of squared amplitudes and, more generally, cross sections
in hadron--hadron collisions, the violation of strict collinear factorization
has implications 
on the non-abelian
structure of logarithmically-enhanced terms in perturbative calculations 
(starting from the next-to-next-to-leading order)
and on various factorization issues of mass singularities 
(starting from the next-to-next-to-next-to-leading order).

\end{quote}

\vspace*{\fill}
\begin{flushleft}
     LPN11-94\\ December 2011 

\end{flushleft}
\end{titlepage}

\renewcommand{\thefootnote}{\fnsymbol{footnote}}

\section{Introduction}
\label{sec:in}

A relevant topic in QCD and, more generally, gauge field theories
is the structure of the perturbative scattering amplitudes in various
infrared (soft and collinear) regions.
{\em Virtual} partonic fluctuations (i.e. partons circulating in loops)
in the infrared (IR) region lead to divergent contributions to scattering
amplitudes in four space-time dimensions. {\em Real} radiation of soft 
and collinear partons produces kinematical singularities, which also lead to
IR divergent contributions after integration over the phase space of the
emitted partons.

In the context of dimensional regularization, the (virtual)
IR divergences of QCD amplitudes have been studied
\cite{Catani:1998bh, Sterman:2002qn, Aybat:2006wq, 
Dixon:2008gr, Becher:2009cu, Gardi:2009qi, Becher:2009qa, Dixon:2009ur}
at one-loop, two-loop and higher-loop orders.
The singularity structure related to (single and multiple) soft-parton 
radiation has been explicitly worked out 
\cite{Bassetto:1984ik, Berends:1988zn, Catani:1999ss, Bern:1998sc, 
Catani:2000pi, Badger:2004uk}
in the cases of tree-level, one-loop and two-loop scattering amplitudes.

These studies and the ensuing understanding of (virtual) IR divergences
and (real) soft-parton singularities rely on general factorization 
properties of QCD. The structure of the IR divergences and of the soft-parton
singularities is described by corresponding {\em factorization formulae}.
The divergent or singular behaviour is captured by factors that have a high
degree of universality or, equivalently, a minimal process dependence
(i.e. a minimal dependence on the specific scattering amplitude).
To be precise, the divergent or singular factors depend on the momenta and
quantum numbers (flavour, colour) of the {\em external} QCD partons in the
scattering amplitude,
while the detailed internal structure of the scattering amplitude plays no 
active role. Of course, in the case of soft-parton radiation, the singular
factors also depend on the momenta and quantum numbers of the emitted soft
partons.

In this paper we deal with collinear-parton singularities. The singular
behaviour of QCD amplitudes, in kinematical configurations where two or more
external-parton momenta become collinear, is also described by 
factorization formulae.

Considering the case of two collinear partons at the tree level,
the collinear-factorization formula for QCD {\em squared amplitudes}
was derived in a celebrated paper \cite{Altarelli:1977zs}.
The corresponding factorization for QCD {\em amplitudes} (rather than 
squared amplitudes) was introduced in 
Refs.~\cite{Berends:1987me, Mangano:1990by}.
At the tree level, the multiple collinear limit of three, four or more partons 
has been studied 
\cite{Campbell:1997hg, Catani:1998nv, Catani:1999ss, DelDuca:1999ha,
Birthwright:2005ak, Birthwright:2005vi}
for both amplitudes and squared amplitudes.
In the case of {\em one-loop} QCD amplitudes, collinear factorization was
introduced in Refs.~\cite{Bern:1993qk, Bern:1995ix, Bern:1998sc, 
Kosower:1999rx},
by explicitly treating the collinear limit of two partons.
Explicit, though partial, results for the triple collinear limit of one-loop
amplitudes were presented in Ref.~\cite{Catani:2003vu}.
The two-parton collinear limit of {\em two-loop} amplitudes
was explicitly computed in 
Refs.~\cite{Bern:2004cz, Badger:2004uk}.
The structure of collinear factorization of higher-loop amplitudes 
was discussed in Ref.~\cite{Kosower:1999xi}.

The collinear-factorization formulae that we have just recalled are similar to
the factorization formulae that apply to virtual IR divergences and soft-parton
singularities. However, the collinear-factorization formulae are `more
universal'. Indeed, the collinear singular factors {\em only} depend on the
momenta and quantum numbers (flavour, colour, spin) of the collinear partons.
In other words, the collinear singular factors have no dependence on the
external {\em non-collinear} partons of the QCD amplitudes. 
Throughout this paper,
this feature of collinear-parton factorization is denoted 
`strict' collinear factorization.

Despite so many established results, in this paper we show that strict 
collinear factorization of QCD amplitudes is actually {\em not} valid 
{\em beyond} the tree level.

We are not going to show that some of the known results at one-loop, two-loop
or higher-loop levels are not correct. We simply start from the observation
that these results refer (either explicitly 
\cite{Catani:2003vu, Bern:2004cz, Badger:2004uk} or implicitly) to the collinear
limit in a specific kinematical configuration. This is the configuration
where {\em all} partons with collinear momenta are produced in the {\em final
state} of the physical process that is described by the QCD amplitude. 
We refer to this
configuration as the time-like (TL) collinear limit.

In the TL collinear limit, strict collinear factorization is valid. In all the
other kinematical configurations, generically denoted as 
space-like (SL) collinear limits, we find that strict collinear 
factorization is not valid (modulo some exceptional cases)
beyond the tree level. 
We also show that,
in the SL collinear limits, QCD amplitudes fulfill
{\em generalized} factorization formulae, in which the collinear singular factors
retain some dependence on the momenta and quantum numbers of the external
non-collinear partons of the scattering amplitude.

The violation of strict collinear factorization is due to long-range (gauge)
loop interactions between the collinear and non-collinear partons.
These virtual radiative corrections produce absorptive contributions that,
due to causality, distinguish initial-state from final-state interactions.
In the TL collinear region, all the collinear partons are produced in the final
state and strict factorization is recovered because of QCD colour coherence
(i.e., the coherent action of the system of collinear partons). The SL collinear
region involves collinear partons in both the initial and final states and,
therefore, causality limits the factorization power of colour coherence.

Owing to their absorptive ('imaginary') origin, strict-factorization 
breaking effects partly cancel at the level of {\em squared amplitudes}
and, hence, in order-by-order perturbative calculations of physical observables.
Indeed, we find that such a cancellation is complete up to the next-to-leading
order (NLO). Nonetheless, strict factorization is violated at higher orders.
For instance, the simplest subprocess in which
strict collinear factorization is definitely violated at the
squared amplitude level is $2 \to 3$ parton scattering, in the kinematical
configurations where one of the three final-state partons is collinear
or almost collinear to one of the two initial-state partons.
In this subprocess we find non-abelian factorization breaking effects that
first occur at the two-loop level. Therefore, these effects
contribute to hard-scattering processes in hadron--hadron collisions:
they produce next-to-next-to-leading order (NNLO) logarithmic contributions to three jet
production with one low-$p_T$ jet (the low-$p_T$ jet is originated by the
final-state parton that is almost collinear to one of initial-state partons),
and next-to-next-to-next-to-leading order (N$^3$LO)
contributions to one-jet and di-jet inclusive production.

The strict factorization breaking effects uncovered
in the simple example of $2 \to 3$ parton scattering
have more general implications in the context of perturbative QCD
computations of jet and hadron production in hadron--hadron collisions.
Starting from the N$^3$LO in perturbation theory, these effects severely
complicate the mechanism of cancellation of IR divergences that leads
to the factorization theorem of mass (collinear) singularities
\cite{Collins:1989gx}.
These complications challenge the universal (process-independent) validity
of mass-singularity factorization,
and they are related to issues that arise in the context of
factorization of transverse-momentum dependent distributions
\cite{Bomhof:2004aw, Bacchetta:2005rm, Rogers:2010dm}.
The perturbative resummation of large logarithmic terms
produced by collinear parton evolution is also affected by the violation
of strict collinear factorization: parton evolution gets tangled with
the colour and kinematical structure of the hard-scattering subprocess,
and this leads to the appearance of `entangled logarithms'.
An example of entangled logarithms is represented by the class
of `super-leading'  non-global logarithms discovered \cite{Forshaw:2006fk}
in the N$^4$LO computation of the dijet cross section
with a large rapidity gap between the two jets.
Indeed, the physical mechanism that produces those super-leading
logarithms \cite{Seymour:2007vw} is directly related to the
mechanism that generates the violation of strict collinear factorization.


The outline of the paper is as follows. Sections~\ref{sec:tree}--\ref{sec:genlim}
are devoted to the two-parton collinear limit. In Sect.~\ref{sec:tree}, we consider
tree-level amplitudes; we introduce our notation and, in particular, 
the colour space formulation based on the collinear splitting matrix.
In Sect.~\ref{sec:tl}, we review the known results on the TL collinear limit
of one-loop amplitudes.
The SL collinear limit at one-loop level is considered in Sect.~\ref{sec:genlim}.
Here, we illustrate the violation of strict factorization, we introduce our
generalized form of collinear factorization, and we present the result of the
one-loop splitting matrix to all-orders in the dimensional regularization parameter
$\ep$.
In Sect.~\ref{sec:multiall}, the study of the collinear behaviour of QCD amplitudes
is extended to the multiparton collinear limit and beyond the one-loop level.
In particular, we illustrate the violation of strict collinear factorization
by deriving the explicit expression of the IR divergences (i.e., the $\ep$ poles)
of the one-loop multiparton splitting matrix. 
In Sect.~\ref{sec:bey2}, we consider the all-order IR structure of the collinear
splitting matrix, we present the explicit IR divergent terms at the two-loop level,
and we discuss the ensuing new features of strict collinear factorization.
In Sect.~\ref{sec:square}, we use our results on the collinear splitting matrix
to compute the singular collinear behaviour of squared amplitudes.
We explicitly show that strict collinear factorization is violated also at 
the squared amplitude level, and we comment on the implications 
for QCD calculations of
hard-scattering cross sections in hadron--hadron collisions.
In Sect.~\ref{sec:fin}, we briefly summarize the main results.
Additional technical 
details are presented in the Appendices. 
In Appendix~\ref{sec:appa}, we illustrate the violation of strict collinear
factorization within the (colour-stripped) formulation in terms of colour
subamplitudes and splitting amplitudes.
In Appendix~\ref{sec:appb}, we explicitly compute the IR divergences of 
the two-loop splitting matrix. 
In Appendix~\ref{sec:tlircon}, we discuss how strict collinear factorization is
recovered in the TL collinear region.


\section{Collinear limit and tree-level amplitudes}
\label{sec:tree}

We consider a generic scattering process that involves external QCD partons 
(gluons and massless\footnote{The case of external massive quarks
and antiquarks is not considered in this paper.} quarks and antiquarks)
and, possibly, additional non-QCD particles (e.g. partons with no colour
such as leptons, photons, electroweak vector bosons, Higgs bosons and so forth).
The corresponding $S$-matrix element (i.e., the on-shell scattering amplitude)
is denoted by $\cm(p_1,p_2,\dots,p_n)$,  where $p_i$ $(i=1,\dots,n)$
is the momentum of the QCD parton $A_i$
($A_i= g, q$ or ${\bar q}$ ),
while the dependence on the momenta 
of additional colourless particles is always understood. 

The external QCD
partons are
on-shell $(p_i^2=0)$ and with physical spin polarizations (thus, $\cm$
includes the corresponding spin wave functions). Note, however, that we always
define the external momenta $p_i$'s as {\em outgoing} momenta.
In particular, the time-component (i.e. the `energy') $p_i^0$ of 
the momentum vector
$p_i^\mu$ ($\mu=0, 1, \dots, d-1$) in $d$ space-time dimensions is {\em not}
positive definite. Different types of physical processes
with $n$ external partons are described by applying crossing symmetry
to the same matrix element $\cm(p_1,p_2,\dots,p_n)$. 
According to our definition of the momenta,
if $p_i$ has positive energy, $\cm(p_1,p_2,\dots,p_n)$ describes a physical
process that produces the {\em parton} $A_i$ in the final state;
if $p_i$ has negative energy, $\cm(p_1,p_2,\dots,p_n)$ describes a physical
process produced by the collision of the {\em antiparton} 
${\overline A}_i$
in the initial state.

The matrix element $\cm(p_1,p_2,\dots)$ can be evaluated in QCD perturbation
theory as a power series expansion (i.e., loop expansion) in the QCD 
coupling $\g$
(or, equivalently, in the strong coupling $\as=\g^2/(4\pi)$).
We write
\beq
\label{loopexnog}
\cm = \cm^{(0)} 
+ \cm^{(1)} + \cm^{(2)}  + \dots \;\;,
\eeq
where $\cm^{(0)}$ is the tree-level\footnote{Precisely
speaking, $\cm^{(0)}$ is not necessarily a tree amplitude, but rather the
lowest-order amplitude for that given process. Thus, $\cm^{(1)}$
is the corresponding one-loop correction. For instance, in the
case of the process $\gamma \gamma \to gg$, $\cm^{(0)}$ involves a quark loop.} 
scattering amplitude,
$\cm^{(1)}$ is the one-loop scattering amplitude, 
$\cm^{(2)}$ is the two-loop scattering amplitude, and so forth.
Note that in Eq.~(\ref{loopexnog}) we have not written down any power
of $\g$. Thus, $\cm^{(0)}$ includes an integer power of $\g$ as overall factor,
and $\cm^{(1)}$ includes an extra factor of $\g^2$
(i.e., $\cm^{(1)}/\cm^{(0)} \propto \g^2$). 
Throughout the first part of the paper 
(Sects.~\ref{sec:tree}--\ref{sec:allorder}),
we always consider unrenormalized matrix elements,
and $\g$ denotes the bare (unrenormalized) coupling constant. 

Physical processes take place in four-dimensional space time.
In the four-dimensional evaluation of the one-loop amplitude $\cm^{(1)}$ 
one encounters ultraviolet 
and IR
divergences
that have to be properly regularized.
The most efficient method to simultaneously regularize both 
kind of divergences
in gauge theories is 
dimensional regularization 
in $d \neq 4$ space-time dimensions.
We work in $d=4 - 2\ep$ space-time dimensions,
and the dimensional-regularization scale is denoted by $\mu$. 
Unless otherwise stated,
throughout the paper we formally consider expressions for arbitrary values of
$d=4 - 2\ep$
(equivalently, in terms of $\ep$-expansions, the expressions are valid to
all orders in~$\ep$).

We are interested in studying the behaviour of  
$\cm(p_1,p_2,\dots,p_n)$ in the kinematical configuration where 
two of the external parton momenta become (almost) collinear. Without loss of
generality, we assume that these momenta are $p_1$ and $p_2$. We parametrize
these momenta as follows:
\beq
\label{kin2}
p_i^\mu = x_i \,p^\mu +k_{\perp i}^\mu - \frac{k_{\perp i}^2}{x_i} 
\frac{n^\mu}{2p \cdot n} \;, \quad \quad i=1,2 \;,
\eeq
where the light-like ($p^2=0$) vector $p^\mu$ denotes 
the collinear direction, while $n^\mu$
is an auxiliary light-like ($n^2=0$) vector, which 
is necessary to specify the transverse components $k_{\perp i}$ 
($k_{\perp i}\cdot p = k_{\perp i}\cdot n = 0$, with
$k_{\perp i}^2<0$) 
or, equivalently, to specify
how the collinear direction is approached.
No other
constraints are imposed on the longitudinal and transverse variables 
$x_i$ and $k_{\perp i}$
(in particular, we have $x_1 + x_2 \neq 1$ and 
$k_{\perp 1} + k_{\perp 2} \neq 0$).
Thus, we can 
consider any (asymmetric) collinear limits 
at once.
Note, however, that the collinear limit is invariant 
under longitudinal boosts along the direction of the total momentum
$p_{12}^\mu = p_1^\mu + p_2^\mu$. 
Thus, the relevant (independent) kinematical
variables are the following boost-invariant quantities: a single
transverse-momentum variable ${\ktil}^\mu$ 
(${\ktil}^\mu= z_2 k_{\perp 1}^\mu - z_1 k_{\perp 2}^\mu$,
${\ktil}^2 < 0$) 
and a single 
longitudinal-momentum fraction, which can be either $z_1$ or $z_2$
(or the ratio between $z_1$ and $z_2$). The longitudinal-momentum 
fractions $z_1$ and $z_2$ are
\beq
\label{zvar12}
z_i = \frac{x_i}{x_1+x_2} \;\;, \quad \quad
z_1 + z_2 = 1 \;\;.
\eeq
In terms of these boost-invariant variables,
the invariant mass squared $s_{12}=(p_1+p_2)^2$ of the system 
of the two `collinear' partons is written as
\beq
\label{s12var}
s_{12}=2 p_1 \cdot p_2= - \f{{\ktil}^2}{z_1 \,z_2} \;\;.
\eeq
We also define 
the following light-like $({\widetilde P}^2=0)$
momentum ${\widetilde P}^\mu$:
\beq
\label{ptil}
{\widetilde P}^\mu = (p_1 + p_2)^\mu - 
\frac{s_{12} \; n^\mu}{2(p_1 + p_2) \cdot n} \;\;. 
\eeq

In the kinematical configuration where the parton momenta $p_1$ and
$p_2$ become collinear, their invariant mass $s_{12}$ vanishes, and
the matrix element $\cm(p_1,p_2,\dots,p_n)$ becomes singular.
To precisely 
define the collinear limit,
we rescale 
the transverse momenta $k_{\perp i}$ in Eq.~(\ref{kin2})
by an overall factor $\lambda$ 
(namely, $k_{\perp i} \to \lambda \; k_{\perp i}$ with $i=1,2$),
and then we perform the limit $\lambda \to 0$. 
In this limit, the behaviour of the matrix element 
$\cm(p_1,p_2,\dots,p_n)$ is proportional to $1/\lambda$. We are interested
in explicitly evaluating the matrix element contribution that 
controls this singular behaviour order by order in the perturbative expansion. 
More precisely, in $d=4 - 2\ep$ dimensions,
the four-dimensional scaling behaviour
in the collinear limit is modified by powers of $(\lambda^2)^{-\ep}$.
Since we work with fixed $\ep$, we treat the powers of $(\lambda^2)^{-\ep}$
as contributions of order unity in the collinear limit.

In summary, considering the limit $s_{12} \to 0$, we are interested in the
singular behaviour:
\beq
\label{mscale12}
\cm(p_1,p_2,\dots,p_n) \sim \f{1}{\sqrt {s_{12}}} \;{\rm mod}\,(\ln^k s_{12}) 
\;\Bigl[ \,1 + {\cal O}({\sqrt {s_{12}}}\;) \,\Bigr] \;\;,
\eeq
where the logarithmic contributions $\ln^k s_{12}$ ($k=0,1,2,\dots)$ ultimately arise from the power series expansion in $\ep$
of terms such as $(s_{12})^{-\ep}$. These logarithmic contributions
are taken into account in our calculation, while the corrections of relative
order ${\cal O}({\sqrt {s_{12}}}\;)$ are systematically neglected.

As is well known 
\cite{Berends:1987me, Mangano:1990by}, 
the singular behaviour of tree-level scattering
amplitudes in the collinear limit is universal (process independent) and
factorized. The factorization structure is usually presented at the level of 
colour subamplitudes \cite{Mangano:1990by}, in a colour-stripped form. 
In Ref.~\cite{Catani:2003vu},
we proposed a formulation of collinear factorization that is valid directly in
colour space. Here, we follow this colour space formulation, which turns out to be
particularly suitable to the main purpose of the present
paper, namely, the general study of the SL collinear limit at one-loop and 
higher-loop orders. 

To directly work in colour space, we use the notation of Ref.~\cite{csdip}
(see also Ref.~\cite{Catani:1998bh}). The scattering amplitude $\cm$ depends on
the colour indices $\{c_1,c_2,\dots\}$ and on the 
spin (e.g. helicity) indices $\{s_1,s_2,\dots\}$
of the external QCD partons; we write
\beq
\label{mel}
\cm^{c_1,c_2,\dots,c_n;s_1,s_2,\dots,s_n}(p_1,p_2,\dots,p_n) \;\;.
\eeq
We formally treat the colour and spin structures by
introducing an orthonormal basis
$\{ \ket{c_1,c_2,\dots,c_n} \otimes \ket{s_1,s_2,\dots,s_n} \}$
in colour + spin space. The scattering amplitude in Eq.~(\ref{mel})
can 
be written as 
\beeq
\label{cmvdef}
\cm^{c_1,c_2,\dots;s_1,s_2,\dots}(p_1,p_2,\dots)
\equiv
\Bigl( \bra{c_1,c_2,\dots} \otimes \bra{s_1,s_2,\dots} \Bigr) \;
\ket{\cm(p_1,p_2,\dots)} \;\;.
\eeeq
Thus $\ket{\cm(p_1,p_2,\dots,p_n)}$
is a vector in colour + spin (helicity) space.

As stated at the beginning of this section, we 
define the external momenta $p_i$'s as {\em outgoing} momenta.
The colour indices $\{c_1,c_2,\dots\,c_n\}$ 
are consistently treated as {\em outgoing} colour indices:
$c_i$ is the colour index of the parton $A_i$ with outgoing momentum  
$p_i$
(if $p_i$ has negative energy, $c_i$ is the colour index of the physical 
parton ${\overline A}_i$ that collides in the initial state).
An analogous comment applies to spin indices.

Having introduced our notation, we can write down the colour-space factorization
formula \cite{Catani:2003vu} for the collinear limit of 
the tree-level amplitude $\cm^{(0)}$. We have
\beq
\label{fact12t}
\ket{\cm^{(0)}(p_1,p_2,\dots,p_n)}
\simeq \sp^{(0)}(p_1,p_2;{\widetilde P}) 
\;\;\ket{\cm^{(0)}({\widetilde P},\dots,p_n)}
\;\;,
\eeq
which is valid in any number $d=4-2\ep$ of dimensions.
The only approximation
      (which is denoted\footnote{The symbol `$\simeq$'
      is used throughout the paper with
      the same meaning (namely, neglecting terms that are
      non-singular or vanishing in the collinear
      limit) as in Eq.~(\ref{fact12t}).} by the symbol `$\simeq$')
      involved
on the right-hand side amounts to neglecting terms that are less singular
in the collinear limit (i.e. the contributions denoted by the term 
${\cal O}({\sqrt {s_{12}}}\;)$ in Eq.~(\ref{mscale12})).

The tree-level factorization formula (\ref{fact12t}) relates the original
matrix element (on the left-hand side)
with $n$ partons to a corresponding matrix element (on the
right-hand side) with $n-1$ partons. 
The latter is obtained from the former
by replacing the two collinear partons $A_1$ and $A_2$
(with momentum $p_1$ and $p_2$, respectively)
with a single parent parton $A$, whose momentum is ${\widetilde P}$ 
(see Eq.~(\ref{ptil})) and whose 
flavour is determined by flavour conservation of the QCD interactions.
More precisely, $A$ is a quark (an antiquark) if the two collinear partons
are a quark (an antiquark) and a gluon, and $A$ is a gluon otherwise. 

The process dependence of Eq.~(\ref{fact12t})
is entirely embodied
in the matrix elements on both sides.
The tree-level factor $\sp^{(0)}(p_1,p_2;{\widetilde P})$, which
encodes the singular behaviour in the collinear limit,
is universal (process independent) and it does not depend on the non-collinear
partons with momenta $p_3, \dots, p_n$. 
It depends on the momenta and quantum
numbers (flavour, spin, colour) of the 
partons that 
are involved in
the collinear
splitting $A \to A_1 A_2$. 
According to our notation,  
$\sp^{(0)}(p_1,p_2;{\widetilde P})$ is a matrix in colour+spin space, named
the {\em splitting matrix} \cite{Catani:2003vu}.

The splitting matrix acts between the colour space of the $n-1$ partons of
$\ket{\cm^{(0)}({\widetilde P},..,p_n)}$
and the colour space of the $n$
partons of the original amplitude 
$\ket{\cm^{(0)}(p_1,p_2,..,p_n)}$.
Because $\sp^{(0)}(p_1,p_2;{\widetilde P})$ does not depend on the non-collinear
partons, their colour is left unchanged in Eq.~(\ref{fact12t})
(precisely speaking, $\sp^{(0)}$ is proportional to the unit matrix in the colour
subspace of the non-collinear partons). The only non-trivial dependence on the
colour (and spin) indices is due to the partons that
undergo 
the collinear splitting $A \to A_1 A_2$. Making this dependence explicit,
we have
\beq
\label{colsp02}
Sp^{(0)\;(c_1,c_2;\,c)}(p_1,p_2;{\widetilde P}) \equiv
\bra{c_1,c_2} \;\sp^{(0)}(p_1,p_2;{\widetilde P})
\; \ket{c} \;\;,
\eeq
where $c_1, c_2$ and $c$ are the colour indices of the partons $A_1, A_2$
and the parent parton $A$. The colour indices of gluons, quarks and antiquarks
are actually different; we use the notation $c = \{a\}=1,\dots,N_c^2-1$ for
gluons and $c=\{\alpha\}=1,\dots,N_c$ for quarks and antiquarks, where $N_c$
is the number of colours. A colour matrix of the fundamental representation of
the gauge group is denoted by $t^a_{\alpha_1 \alpha_2}\,$, and the structure
constants are $f_{abc}\,$; we use the following normalization:
\beq
\label{tnor}
[ t^a, t^b ] = i f_{abc} t^c \;\;, \quad 
{\rm Tr} \left( t^a t^b \right) = \frac{1}{2} \,\delta^{a b} \;\;.
\eeq
The matrices $t^a$ are hermitian ($(t^a)^\dagger = (t^a)$)
	    and the structure constants $f_{abc}$
	    are real ($f_{abc}^* = f_{abc}$).
There are four different flavour-conserving configurations $A \to A_1 A_2$. The
corresponding explicit form of the tree-level splitting matrix is:

$q \to q_1 g_2$
\beq
\label{qqg0}
Sp_{q_1 g_2}^{(0)\;(\alpha_1,a_2;\,\alpha)}(p_1,p_2;{\widetilde P}) = 
\gep \;t^{a_2}_{\alpha_1 \alpha} \; \frac{1}{s_{12}} \; \ubar(p_1)\, 
{\slash \vep}(p_2) \,u({\widetilde P}) \;\;,
\eeq

${\bar q} \to {\bar q}_1 g_2$
\beq
Sp_{{\bar q}_1 g_2}^{(0)\;(\alpha_1,a_2;\,\alpha)}(p_1,p_2;{\widetilde P}) = 
\gep \;\left( - t^{a_2}_{\alpha \,\alpha_1} \right) \; \frac{1}{s_{12}} \; 
\vbar({\widetilde P})\, 
{\slash \vep}(p_2) \,v(p_1) \;\;,
\eeq

$g \to q_1 {\bar q}_2$
\beq
\label{gqqbar0}
Sp_{q_1 {\bar q}_2}^{(0)\;(\alpha_1,\alpha_2;\,a)}(p_1,p_2;{\widetilde P}) = 
\gep \;t^{a}_{\alpha_1 \alpha_2} \; \frac{1}{s_{12}} \; \ubar(p_1)\, 
{\slash \vep}^*({\widetilde P}) \,v(p_2) \;\;,
\eeq

$g \to g_1 g_2$
\beeq
\label{ggg0}
Sp_{g_1 g_2}^{(0)\;(a_1,a_2;\,a)}(p_1,p_2;{\widetilde P})&& \!\!\!\!\!\!\!\!\!
= \gep \;i \;f_{a_1 a_2 \,a} 
\;\frac{2}{s_{12}}  \\
&&\!\!\!\!\!\! \!\!\!\!\!\!\!\!\!\!\!\!\!\!\!\!\!\!\!\! 
\!\!\!\!\!\!\!\!\!\!\!\! \!\!\!\!\!\times
\left[ 
\vep(p_1) \cdot \vep(p_2) \;p_1 \cdot \vep^*({\widetilde P})
+ \vep(p_2) \cdot \vep^*({\widetilde P}) \;p_2 \cdot \vep(p_1)
- \vep(p_1) \cdot \vep^*({\widetilde P}) \;p_1 \cdot \vep(p_2)
\right] \;\;, \nn
\eeeq
where $u(p)$ and $v(p)$ are the customary Dirac spinors and
 $\vep^\mu(p)$ is the physical polarization
	      vector of the gluon ($\vep_\mu^*$ is the complex
	      conjugate of $\vep_\mu$).
Spin indices play no relevant active role in the context of the main discussion
of the present paper. They are embodied in the parton wave functions 
$u, v, \vep$   and are not explicitly denoted throughout the paper. The explicit
expressions
of Eqs.~(\ref{qqg0})--(\ref{ggg0})
in a definite helicity basis can be found in the 
literature (see, for instance, the Appendix~A of the second
paper in Ref.~\cite{Bern:1998sc}).

We briefly comment on the relation between Eq.~(\ref{fact12t}) and the customary
collinear-factorization formulae for colour subamplitudes
(see also the Appendix~\ref{sec:appa}).
The colour-space factorization formula (\ref{fact12t}) is valid for
a generic 
matrix element $\ket{\cm^{(0)}(p_1,p_2,..,p_n)}$; in particular, the
factorization formula does not require
any specifications about the colour structure of the matrix element.
Collinear factorization of QCD scattering amplitudes is usually discussed
\cite{Berends:1987me}
upon colour decomposition of the matrix element.
The colour decomposition, whose actual form depends on the specific partonic
content of the matrix element (e.g., on the number of gluons and 
quark-antiquark pairs), factorizes the QCD colour from colourless 
kinematical coefficients,
which are called colour subamplitudes (see, e.g., Ref.~\cite{Mangano:1990by}).
Colour subamplitudes fulfil several process-independent properties, including
collinear factorization. In the region where the two parton momenta
$p_1$ and $p_2$ become collinear, the collinear-factorization formula 
of the colour subamplitudes is
a colour-stripped analogue of Eq.~(\ref{fact12t}): the colour vectors 
$\ket{\cm^{(0)}}$ on both sides are replaced by corresponding colour 
subamplitudes, and the colour matrix 
$\sp^{(0)}(p_1,p_2;{\widetilde P})$ is replaced by a universal kinematical
function, which is called {\em splitting amplitude} and is usually denoted by
${\rm Split}^{(0)}(p_1,p_2;{\widetilde P})$ 
(see also the Appendix~\ref{sec:appa}).

In the case of the {\em tree-level} collinear splitting of {\em two} 
partons, the 
relation between the splitting matrix $\sp^{(0)}$ and the splitting amplitude
${\rm Split}^{(0)}$ is particularly straightforward. Indeed, having fixed the
flavour of the partons in the collinear splitting process
$A \to A_1  A_2$, the corresponding splitting matrix $\sp^{(0)}$ involves
a {\em single} (and unique) colour structure and, therefore, we have a direct
proportionality relation:
\beq
\label{spvssplit}
\sp^{(0)}(p_1,p_2;{\widetilde P}) 
\propto
\left( {\bf colour \;\;\;matrix} \right) \times
{\rm Split}^{(0)}(p_1,p_2;{\widetilde P}) \;\;,
\eeq
where the colour matrix on the right-hand side is obtained by simple inspection
of Eqs.~(\ref{qqg0})--(\ref{ggg0})
(see the colour factors $t^a_{\alpha \, \alpha'}$ in 
Eqs.~(\ref{qqg0})--(\ref{gqqbar0}) 
and the colour factor $i f_{a_1 a_2 a}$ in Eq.~(\ref{ggg0})).

As discussed at the beginning of this section, the outgoing momenta $p_i$'s
of $\cm(p_1,\dots,p_n)$, depending on the sign of their energy, actually 
describe
different physical processes, which take place in different kinematical regions.
Correspondingly, the collinear splitting 
$A \to A_1 A_2$ 
formally
describes different physical subprocesses, which take place in either the TL
(if $p_1^0 p_2^0 > 0$) or SL (if $p_1^0 p_2^0 < 0$) regions. In these regions,
the collinear variables $z_1$, $z_2$ and $s_{12}$ in Eqs.~(\ref{zvar12})
and (\ref{s12var}) are constrained as follows:
\beeq
\label{tlc}
{\rm TL}: \quad \quad \quad \quad \quad &&
s_{12} > 0 \;\;, \quad z_1 z_2 > 0 \;\;,\\
\label{slc} 
{\rm SL}: \quad \quad \quad \quad \quad &&
s_{12} < 0 \;\;, \quad z_1 z_2 < 0 \;\;.
\eeeq
The most relevant physical subprocesses are the customary subprocesses: 
\begin{itemize}
\item TL $\;\;\;(p_1^0 > 0 \;, \; p_2^0 > 0$):
\end{itemize}
\beq
\label{tls}
A^* \to A_1(z) \; A_2(1-z) \;\;, 
\quad {\rm with} \quad z=z_1=1-z_2 \;, \;\;\;\;\, 0 < z <1 \;\;,
\eeq
\begin{itemize}
\item SL $\;\;\;(p_1^0 < 0 \;, \; p_2^0 > 0 \;, \; \wp^0 < 0$):
\end{itemize}
\beq
\label{sls}
{\overline A}_1 \to {\overline A}^{\;*}(z) \; A_2(1-z) \;\;, 
\quad {\rm with} \quad z=\f{1}{z_1}=\f{1}{1-z_2} \;, \;\;\; 0 < z <1 \;\;.
\eeq
In the TL subprocess of Eq.~(\ref{tls}), the partons $A_1$, $A_2$ and the parent
parton $A$  are physically produced into the final state; the collinear decay
of the parent parton $A^*$, which is slightly 
off-shell\footnote{We introduce the star 
superscript in $A^*$ to explicitly remind the reader that the parent parton $A$ is off-shell
before approaching the collinear limit.} 
(with positive virtuality)
in the vicinity of the collinear limit, transfers the longitudinal-momentum
fractions $z$ and $1-z$  to $A_1$ and $A_2$, respectively.
In the SL subprocess\footnote{The corresponding
subprocess with $p_1 \leftrightarrow p_2$ is trivially related to Eq.~(\ref{sls})
by the exchange $1 \leftrightarrow 2$ of the parton indices.}
 of Eq.~(\ref{sls}), the physical parton ${\overline A}_1$,
which collides in the initial state, radiates the physical parton $A_2$, with 
longitudinal-momentum fraction $1-z$, in the
final state; the remaining fraction, $z$, of longitudinal momentum is carried
by the accompanying (`parent') parton ${\overline A}^{\;*}$
(which is slightly off-shell, with negative virtuality,
in the vicinity of the collinear limit) that replaces ${\overline A}_1$
as physically colliding parton in the initial state.

There are two other physical subprocesses that are kinematically
allowed: the TL subprocess
${\overline A}_1 {\overline A}_2 \to {\overline A}^{\;*} $
(parton--parton fusion into an initial-state parton)
is allowed if $p_1^0$ and $p_2^0$ are both negative, 
and the SL subprocess
${\overline A}_1  A^* \to A_2 $
(parton--parton fusion into a final-state parton)
is allowed if $p_1^0 < 0 \;, \; p_2^0 > 0 \;,$ and $\wp^0 > 0$.
The subprocess ${\overline A}_1 {\overline A}_2 \to {\overline A}^{\;*}$
(the initial-state parton ${\overline A}^{\;*}$ is produced by the fusion of
the two initial-state collinear partons 
${\overline A}_1$ and  ${\overline A}_2$)
occurs if $\cm(p_1,p_2,\dots)$ corresponds to a physical process with
at least three colliding particles in the initial state (the partons 
${\overline A}_1, {\overline A}_2$ and, at least, one additional particle).
The subprocess ${\overline A}_1  A^* \to A_2 $ 
(the final-state parton $A_2 $ is produced by the fusion of the initial-state
parton ${\overline A}_1$ and the final-state parton $A^*$) 
occurs if $\cm(p_1,p_2,\dots)$ corresponds to a physical process
in which the initial state contains the parton ${\overline A}_1$
and, in addition, either one massive particle or (at least) 
two particles.  Owing to these
kinematical features, these subprocesses are less relevant in the context of
QCD hard-scattering processes. 

The splitting matrix $\sp^{(0)}(p_1,p_2;{\widetilde P})$ in the factorization
formula (\ref{fact12t}) applies to any physical subprocesses, in both the TL and
SL regions. Strictly speaking, the explicit expressions in
Eqs.~(\ref{qqg0})--(\ref{ggg0}) refer to the TL region where the energies of
$p_1, p_2$ and ${\widetilde P}$ are positive. The corresponding
expressions in other kinematical regions are straightforwardly obtained by
applying crossing symmetry. If the energy of the momentum $P$ ($P=p_1, p_2$
or ${\widetilde P}$) is negative, the crossing relation simply amounts to
the usual
replacement of the corresponding wave function (i.e.,
$u(P) \leftrightarrow v(-P)$ and $\vep(P) \leftrightarrow \vep^*(-P)$).

\section{One-loop amplitudes: time-like collinear limit}
\label{sec:tl}

In this section we consider the collinear behaviour of the one-loop QCD
amplitudes $\cm^{(1)}$ in Eq.~(\ref{loopexnog}). We use the same general
notation as in Sect.~\ref{sec:tree}. However, we anticipate that the results 
are valid only in the case of the TL collinear splitting (i.e., $s_{12} > 0$).

The singular behaviour of $\cm^{(1)}(p_1,p_2,\dots,p_n)$ in the region where the
two momenta $p_1$ and $p_2$ become collinear is also described by a 
factorization formula. The extension of the tree-level colour-space
formula (\ref{fact12t})
to one-loop amplitudes is \cite{Catani:2003vu}
\beeq
\label{fact12one}
\ket{\cm^{(1)}(p_1,p_2,\dots,p_n)}
&\simeq& \sp^{(1)}(p_1,p_2;{\widetilde P}) 
\;\;\ket{\cm^{(0)}(\wp,\dots,p_n)} \nn \\
&+& \sp^{(0)}(p_1,p_2;{\widetilde P}) 
\;\;\ket{\cm^{(1)}(\wp,\dots,p_n)}
\;\;.
\eeeq
The `reduced' matrix elements on the right-hand side are obtained from 
$\cm(p_1,p_2,\dots,p_n)$
by replacing the two collinear partons $A_1$ and $A_2$
(with momentum $p_1$ and $p_2$, respectively)
with their parent parton $A$, with momentum ${\widetilde P}$. 
The two contributions on the right-hand side are proportional to the 
reduced matrix element at the tree-level and at the one-loop order,
respectively. The splitting matrix $\sp^{(0)}$ is exactly the tree-level
splitting matrix that enters Eq.~(\ref{fact12t}).
The {\em one-loop} splitting matrix $\sp^{(1)}(p_1,p_2;{\widetilde P})$
encodes new (one-loop) information on the collinear splitting process
$A \to A_1 A_2$. Analogously to $\sp^{(0)}$, the one-loop factor 
$\sp^{(1)}(p_1,p_2;{\widetilde P})$ is a universal (process-independent)
matrix in colour+spin space,
and it only depends on the momenta and quantum
numbers of the partons involved in the collinear splitting subprocess.

Within the colour subamplitude formulation,
the collinear limit of two partons at the one-loop level was first discussed
in Ref.~\cite{Bern:1993qk} by introducing one-loop splitting amplitudes
$\;{\rm Split}^{(1)}(p_1,p_2;{\widetilde P})$, which are the one-loop analogues 
of the tree-level splitting amplitudes mentioned in Sect.~\ref{sec:tree}.
A proof of collinear factorization of one-loop colour subamplitudes was
presented in Ref.~\cite{Bern:1995ix}. Explicit results for the 
splitting amplitudes $\,{\rm Split}^{(1)}\,$  in $d=4-2\ep$ dimensions
(or, equivalently, the results to all orders in the $\ep$ expansion)
were obtained in Refs.~\cite{Bern:1998sc, Kosower:1999rx}.  

The relation between the one-loop factorization formula (\ref{fact12one})
and its colour subamplitude version is exactly the same as the
relation at the tree level (see also the Appendix~\ref{sec:appa}). 
The main point is that the one-loop splitting 
matrix $\sp^{(1)}(p_1,p_2;{\widetilde P})$ involves a {\em single}
colour structure (more precisely, there is a single colour structure for each 
flavour configuration of the splitting processes $A \to A_1  A_2$), and this
colour structure is the {\em same} structure that occurs in the tree-level  
splitting matrix $\sp^{(0)}(p_1,p_2;{\widetilde P})$. In other words,
the proportionality relation in Eq.~(\ref{spvssplit}) is valid also at the
one-loop level: we can simply perform the replacements  
$\sp^{(0)}(p_1,p_2;{\widetilde P}) \to \sp^{(1)}(p_1,p_2;{\widetilde P})$
and ${\rm Split}^{(0)}(p_1,p_2;{\widetilde P})
\to {\rm Split}^{(1)}(p_1,p_2;{\widetilde P})$. Therefore,
from the known $\,{\rm Split}^{(1)}\,$ \cite{Bern:1998sc, Kosower:1999rx}
we directly obtain the corresponding $\sp^{(1)}$.

We now comment on the kinematical structure of 
$\sp^{(1)}(p_1,p_2;{\widetilde P})$, i.e. on the momentum dependence of
${\rm Split}^{(1)}(p_1,p_2;{\widetilde P})$. Apart from the
overall proportionality to the wave functions $u,v,\vep$ of the collinear 
partons (which is analogous to that in Eqs.~(\ref{qqg0})--(\ref{ggg0})),
the kinematical structure
\cite{Bern:1998sc, Kosower:1999rx} depends on two different classes of
contributions. One class contains all the contributions that have a 
{\em rational} dependence on the momenta; the other class contains 
transcendental functions (e.g., logarithms and polylogarithms) and, 
in particular,
transcendental functions of the momentum fractions $z_1$ and $z_2$.

Considering $d=4-2\ep$ space-time dimensions, the one-loop integrals introduce
the dimensional factor $\mu^{2\ep}$. Since the one-loop corrections to
$\sp^{(0)}(p_1,p_2;{\widetilde P})$ are dimensionless, 
$\sp^{(1)}(p_1,p_2;{\widetilde P})$
necessarily includes the overall factor
\beq
\label{s12fact}
\left( \frac{-s_{12} -i0}{\mu^2} \right)^{-\ep} \;\;,
\eeq
where the $i0$ prescription follows from usual analyticity properties
of the scattering amplitudes. Apart from this overall factor,
the trascendental dependence of the two-parton collinear limit at one-loop
order turns out to be {\em entirely} captured \cite{Bern:1998sc, Kosower:1999rx} 
by a single hypergeometric function, namely, the function 
${}_2F_1(1, -\ep; 1-\ep; x)$.
The integral representation of this hypergeometric function is  
\beq
{}_2F_1(1, -\ep; 1-\ep; x) = - \,\ep
\int_0^1 \;dt \;t^{-1-\ep} \; (1-xt)^{-1} \;\;. 
\eeq
We thus define the following function:
\beq
\label{fep}
f(\ep;1/x) \equiv \f{1}{\ep} \;\Bigl[ \; {}_2F_1(1, -\ep; 1-\ep; 1-x) - 1 
\,\Bigr]
\;\;.
\eeq
The expansion of this function in powers of $\ep$ is as follows:
\beq
\label{fepex}
f(\ep;1/x)=   \;\ln x - \ep \left[ {\rm Li}_2(1-x)
+ \sum_{k=1}^{+\infty} \ep^k \;{\rm Li}_{k+2}(1-x) \right] \;,
\eeq
where the dilogarithm function ${\rm Li}_2$ is
\beq
{\rm Li}_2(x) \equiv - \int_0^x \;\frac{dt}{t} \; \ln(1-t) \;\;,
\eeq
and the polylogarithms ${\rm Li}_{k+1}$ (with $k=2,3,\dots$) are defined 
\cite{trafun} by
\beq
{\rm Li}_{k+1}(x) \equiv \frac{\left(-1\right)^k}{(k-1)!}
\;\int_0^1 \;\frac{dt}{t} \;\left( \ln t \right)^{k-1} \; \ln(1-xt) \;\;.
\eeq

As recalled in Sect.~\ref{sec:tree}, there are four different flavour
configurations in the collinear splitting process $A \to A_1 A_2$.
Considering the corresponding explicit results of 
Refs.~\cite{Bern:1998sc, Kosower:1999rx}, the one-loop splitting matrix
$\sp^{(1)}(p_1,p_2;{\widetilde P})$ of Eq.~(\ref{fact12one})
can be written in the following general
(and compact) form:
\beq
\label{sp112gen}
\sp^{(1)}(p_1,p_2;{\widetilde P}) = 
\sp^{(1)}_{H}(p_1,p_2;{\widetilde P}) + 
I_C(p_1,p_2;{\widetilde P}) \;
\sp^{(0)}(p_1,p_2;{\widetilde P}) \;\;.
\eeq
The factor $I_C(p_1,p_2;{\widetilde P})$ is specified below.
Having specified this factor, the term $\sp^{(1)}_{H}$ on the right-hand side
of Eq.~(\ref{sp112gen}) can be extracted, in explicit form, 
from the results in 
Refs.~\cite{Bern:1998sc, Kosower:1999rx}. 
We do not report the explicit form of $\sp^{(1)}_{H}$, since it has no relevant
role in our discussion of the relation between the TL and SL collinear limits.
In this respect, the {\em only} relevant property of $\sp^{(1)}_{H}$
(which follows from our definition of $I_C$)
is that it contains only terms with rational dependence\footnote{Precisely
speaking, this statement is true modulo the dependence on the overall factor
in Eq.~(\ref{s12fact}).} 
on the momenta
$p_1, p_2$ and ${\widetilde P}$. Moreover, all the terms of $\sp^{(1)}$
that are 
IR or ultraviolet divergent in $d=4$ dimensions (i.e. all the $\ep$ poles)
are collected in the factor $I_C$ and, thus, 
removed from $\sp^{(1)}_{H}$. Therefore, $\sp^{(1)}_{H}$ is 
finite if we set $\ep=0$.

The term  $I_C \times \sp^{(0)}$ on the right-hand side of
Eq.~(\ref{sp112gen}) contains all the IR and ultraviolet divergences 
of $\sp^{(1)}$ and,
more importantly, it collects the entire
            dependence of the collinear behaviour at one-loop order
            on transcendental functions
	     (modulo the function in
            Eq.~(\ref{s12fact}), which also appears in $\sp^{(1)}_{H}$).
            The explicit expression
of the factor $I_C$ for a generic splitting process 
$A \to A_1 A_2$ is
\beeq
\label{ictl}
I_C(p_1,p_2;{\widetilde P}) &=& 
 \; \g^2 \; c_{\Gamma} \;
\left( \frac{-s_{12} -i0}{\mu^2} \right)^{-\ep} \nn \\
&\times& \left\{  \; \frac{1}{\ep^2} \; \Bigl( C_{12} - C_1 - C_2 \Bigr)
+ \frac{1}{\ep} \; \Bigl( \gamma_{12} - \gamma_1 - \gamma_2 +
\cbet0 \Bigr)
\right.  \\
&-&\left. \; \frac{1}{\ep} \, \left[ \Bigl(C_{12} + C_1 - C_2 \Bigr) \;f(\ep;z_1)
 +  \Bigl(C_{12} + C_2 - C_1 \Bigr) \;f(\ep;z_2)\;
\right] \right\} \;\;,\nn 
\eeeq
where $c_{\Gamma}$  is the typical volume factor of 
$d$-dimensional one-loop integrals:
\beq
\label{cgammaf}
c_{\Gamma} \equiv \frac{\Gamma(1+\epsilon) 
\Gamma^2(1-\epsilon)}{\left(4\pi\right)^{2-\epsilon}\Gamma(1-2\epsilon)} \;\;.
\eeq
The coefficients $C_1, C_2$ and  $C_{12}$ are the
Casimir coefficients of the partons $A_1, A_2$ and $A$; explicitly,  
$C=C_F=(N_c^2-1)/(2N_c)$ if the parton is a quark or antiquark, and 
$C=C_A=N_c$ if the parton is a gluon. Analogously, the coefficients
$\gamma_1, \gamma_2$ and  $\gamma_{12}$ refer to the flavour of the 
partons $A_1, A_2$ and $A$; explicitly, we have
\beq
\label{gacoef}
\gamma_q=\gamma_{\bar q}= \f{3}{2} \; C_F \;\;, \quad
\quad \gamma_g= \f{1}{6} \;(11 \,C_A -2 \,N_f) \;\;,
\eeq
where $N_f$ is the number of flavours of massless quarks.
The coefficient $\cbet0$ 
is the first perturbative coefficient of the QCD $\beta$ function,
\beq
\label{norbeta0}
\cbet0=\f{1}{6} \;(11 \,C_A - 2 \,N_f) \;\;.
\eeq
Note that, in to our normalization, we have $\cbet0= \gamma_g$.

The coefficients of the $\ep$ poles in Eq.~(\ref{ictl}) agree with those of the
general structure presented in Eq.~(11) of Ref.~\cite{Catani:2003vu}.
The single-pole term proportional to $\cbet0$ is of ultraviolet origin;
it can be removed by renormalizing the splitting matrix
$\sp(p_1,p_2;\wp)$ (we recall that we are considering unrenormalized matrix
elements and, correspondingly, unrenormalized splitting matrices).
The other pole terms are of IR origin.  
The double-pole terms (which are proportional to the 
Casimir factors $C_F$ and $C_A$)
originate from one-loop contributions where the loop
momentum is nearly on-shell, soft and parallel to the momentum of one of the
three partons involved in the collinear splitting subprocess.
The single-pole terms with $\gamma$ coefficients are produced by
contributions where the loop
momentum is not soft, though it is nearly on-shell
and parallel to the momentum of one of the collinear partons. According to
Eq.~(\ref{fepex}), the 
$\ep$ expansion of the transcendental function gives
$f(\ep;z)= - \ln z + {\cal O}(\ep)$; therefore, $f(\ep;z_1)$ and $f(\ep;z_2)$
contribute to Eq.~(\ref{ictl}) with single-pole terms. The coefficients of these
single-pole terms are controlled by the Casimir factors $C_F$ and $C_A$ and,
hence, they originate from one-loop configurations with soft momentum;
they are produced by contributions where the loop
momentum is nearly on-shell, soft and at large angle with respect to the
directions of the collinear partons.
The specific combination of Casimir factors in these single-pole terms
(namely, $C_{12} + C_1 - C_2$ and 
$C_{12} + C_2 - C_1$) originates from a colour coherence effect
(see Eq.~(\ref{d12gentl}) and the comments below it).

As anticipated at the beginning of this section, the one-loop
factorization formula
(\ref{fact12one}) and the explicit results in Eqs.~(\ref{sp112gen}) and 
(\ref{ictl})
(or, equivalently, the one-loop splitting amplitudes 
in Refs.~\cite{Bern:1993qk}--\cite{Kosower:1999rx})
are valid in the case of the TL collinear limit (see Eq.~(\ref{tlc})).
At the tree level, the TL and SL collinear limits are related
by exploiting crossing symmetry, and 
the corresponding splitting
matrix $\sp^{(0)}(p_1,p_2;{\widetilde P})$ is simply obtained by applying the
(wave function) crossing relations mentioned at the end of Sect.~\ref{sec:tree}.
At the one-loop level, we have to deal with the splitting matrix 
$\sp^{(1)}(p_1,p_2;{\widetilde P})$ in Eq.~(\ref{sp112gen}), and we can try to
proceed in an analogous way. Using crossing symmetry, the treatment of the
one-loop contribution $\sp^{(1)}_H(p_1,p_2;{\widetilde P})$ is straightforward;
$\sp^{(1)}_H(p_1,p_2;{\widetilde P})$ contains $(i)$ wave function factors,
which are treated by the corresponding crossing relations, and $(ii)$
rational functions of the collinear momenta, which are invariant under crossing.
The one-loop troubles originate from the factor 
$I_C(p_1,p_2;{\widetilde P})$, since it contains $f(\ep;z_1)$ and 
$f(\ep;z_2)$.

The function $f(\ep;x)$ (see Eq.~(\ref{fep}))
has a branch-cut singularity if the variable $x$ is real and {\em negative}.
The branch-cut singularity arises from the corresponding singularity
of the hypergeometric function
${}_2F_1(1, -\ep; 1-\ep; 1-1/x)$.
In the case of the TL collinear limit, $z_1$ and $z_2$ are both positive
(see Eq.~(\ref{tlc}) and recall that $z_1 + z_2 =1$), and the 
functions $f(\ep;z_1)$ and $f(\ep;z_2)$ are both well-defined.
In the case of the SL collinear limit, one of the two variables $z_1$ and $z_2$
necessarily has a negative value (see Eq.~(\ref{tlc})); therefore,
one of the two functions, either $f(\ep;z_1)$ or $f(\ep;z_2)$, 
in Eq.~(\ref{ictl}) is necessarily evaluated along its 
branch-cut singularity and, hence, it is ill-defined.

In summary, the issue of the TL vs. SL collinear limits is as follows.
The results in Eqs.~(\ref{fact12one}), (\ref{sp112gen}) and 
(\ref{ictl}) cannot be extended from the TL to the SL collinear limit by using
crossing symmetry, since this leads to ill-defined results (mathematical
expressions). As shown in the next section, the solution of the issue
involves not only the (mathematical) definition of the function 
$f(\ep;x)$ along (or, more precisely, in the vicinity of) its branch-cut
singularity, but also the introduction of new physical effects.

\section{One-loop amplitudes: general (including space-like) collinear limit}
\label{sec:genlim}

\subsection{Generalized factorization and violation of strict 
collinear 
factorization}
\label{sec:gencol}

The extension of the colour-space collinear formula in
Eq.~(\ref{fact12one}) to general kinematical 
configurations\footnote{In Appendix~\ref{sec:appa}, we illustrate
the SL collinear limit of colour subamplitudes for
the specific case of pure
multigluon matrix elements at the one-loop level.},
which include the two-parton collinear limit in the SL region, is
\beeq
\label{fact12onegen}
\ket{\cm^{(1)}(p_1,p_2,\dots,p_n)}
&\simeq& \sp^{(1)}(p_1,p_2;{\widetilde P};p_3,\dots,p_n) 
\;\;\ket{\cm^{(0)}(\wp,\dots,p_n)} \nn \\
&+& \sp^{(0)}(p_1,p_2;{\widetilde P}) 
\;\;\ket{\cm^{(1)}(\wp,\dots,p_n)}
\;\;.
\eeeq
The essential difference  with respect to Eq.~(\ref{fact12one})
is that the one-loop splitting matrix $\sp^{(1)}$ on the right-hand side of
Eq.~(\ref{fact12onegen}) depends not only on the collinear partons but also
on the momenta and quantum numbers of the non-collinear partons
in the original matrix element $\ket{\cm^{(1)}(p_1,p_2,\dots,p_n)}$.
Thus, $\sp^{(1)}$ is {\em no longer} (strictly) universal, 
since it retains some dependence on
the process (matrix element) from which the splitting matrix derives.
The reduced tree-level, $\ket{\cm^{(0)}(\wp,\dots,p_n)}$,
and one-loop, $\ket{\cm^{(1)}(\wp,\dots,p_n)}$, matrix elements on the 
right-hand side of Eq.~(\ref{fact12onegen}) are the same as those in 
Eq.~(\ref{fact12one}): they are still related to the original matrix element 
$\ket{\cm^{(1)}(p_1,p_2,\dots,p_n)}$ through the same factorization procedure
that is used in Eq.~(\ref{fact12one}) (i.e. in the case of the TL collinear 
limit).

The explicit form of the general one-loop splitting matrix $\sp^{(1)}$
in Eq.~(\ref{fact12onegen}) is
\beeq
\label{spgen}
&&\!\!\!\!\! \sp^{(1)}(p_1,p_2;{\widetilde P};p_3,\dots,p_n) = 
\sp^{(1)}_{H}(p_1,p_2;{\widetilde P}) + 
{\bom I}_C(p_1,p_2;p_3,\dots,p_n) \;
\sp^{(0)}(p_1,p_2;{\widetilde P}) \;, \nn \\
&&\!\!\!\!\!{}
\eeeq
where $\sp^{(1)}_{H}(p_1,p_2;{\widetilde P})$  is exactly the same (universal)
term as in Eq.~(\ref{sp112gen}). The difference with respect to the TL
expression in Eq.~(\ref{sp112gen}) arises from the replacement of 
$I_C(p_1,p_2;{\widetilde P})$ with ${\bom I}_C(p_1,p_2;p_3,\dots,p_n)$.
The term $I_C(p_1,p_2;{\widetilde P})$ is a $c$-number (i.e., colourless) 
factor, while ${\bom I}_C(p_1,p_2;p_3,\dots,p_n)$ is a {\em colour matrix}.
Moreover, $I_C(p_1,p_2;{\widetilde P})$ depends on the collinear 
variables $z_1,z_2,s_{12}$ and the flavour of the collinear partons
$A_1,A_2$ and $A$ (see Eq.~(\ref{ictl})), 
while  ${\bom I}_C(p_1,p_2;p_3,\dots,p_n)$ {\em also} 
depends on the momentum and colour of the non-collinear partons.

The expression of the colour operator ${\bom I}_C$ can be presented by
using the same notation as in Eq.~(\ref{ictl}). We can also exploit the fact
that in any kinematical configurations 
(see Eqs.~(\ref{tlc}) and (\ref{slc})) one of the two collinear variables,
$z_1$ and $z_2$,  {\em necessarily} has positive values
(recall that $z_1+z_2=1$). Therefore, with no loss of generality, we can set
(choose) 
\beeq
z_1 > 0 \;\;, \nonumber
\eeeq 
and write the following explicit expression of the colour operator:
\beeq
\label{ic12gen}
{\bom I}_C(p_1,p_2;p_3,\dots,p_n) &=& 
 \; \g^2 \; c_{\Gamma} \;
\left( \frac{-s_{12} -i0}{\mu^2} \right)^{-\ep} \\
&\times& \left\{  \; \frac{1}{\ep^2} \; \Bigl( C_{12} - C_1 - C_2 \Bigr)
+ \frac{1}{\ep} \; \Bigl( \gamma_{12} - \gamma_1 - \gamma_2 
+ \cbet0
\Bigr)
\right. \nn \\
&-&\left. \! \frac{1}{\ep} \, \left[ \Bigl(C_{12} + C_1 - C_2 \Bigr) \;f(\ep;z_1)
 - 2 \sum_{j=3}^n  \,{\bom T}_2 \cdot {\bom T}_j
 \;f(\ep;z_2- i0 s_{j2})\;
\right] \right\} \,.\nn 
\eeeq
Here, the subscript $j$ $(j=3,\dots,n)$ refers to the non-collinear parton with
momentum $p_j$, and $s_{j2}=(p_j+p_2)^2$ is
the invariant mass squared of the system formed by the $j$-th 
non-collinear parton and
the collinear parton $A_2$. The colour charge (matrix)\footnote{We use the 
same notation
as in Refs.~\cite{csdip, Catani:1998bh}:
more details about colour charges and  colour algebra
relations can be found therein.} 
of the parton with
momentum $p_k$ $(k=1,2,3,\dots,n)$ is denoted by ${\bom T}_k$,
and we define ${\bom T}_k \cdot {\bom T}_l \equiv \sum_c T^c_k T^c_l$ 
($c=1,\dots,N_c^2-1$).

On the right-hand side of Eq.~(\ref{ic12gen}), the function $f(\ep;z_1)$
is well-defined, since $z_1 > 0$.
The functional dependence on $z_2$ is also well-defined, since it is given by
either $f(\ep;z_2- i0)$ if $s_{j2} > 0$, 
or $f(\ep;z_2+ i0)$ if $s_{j2} < 0$.
Owing to the $i0$ prescription, if $z_2 < 0$, the function $f(\ep;z_2 \pm i0)$
is always evaluated either above or below its branch-cut singularity.  The presence and structure of the branch-cut singularity are
	 physical consequences of causality, as discussed in
	 Sect.~\ref{sec:caus}.

Comparing Eq.~(\ref{ictl}) with Eq.~(\ref{ic12gen}), we see that the
difference is due to a single contribution. In the TL case of Eq.~(\ref{ictl})
this contribution is proportional to the following term:
\beq
\label{d12tl}
\delta(p_1,p_2;{\widetilde P}) = 
-\; \frac{1}{\ep} \, \Bigl(C_{12} + C_2 - C_1 \Bigr) \;f(\ep;z_2) \;\;,
\eeq
while in the general case of Eq.~(\ref{ic12gen}) this term is replaced
by the following colour operator:
\beq
\label{d12gen}
{\bom \delta}(p_1,p_2;p_3,\dots,p_n) = 
+ \; \frac{2}{\ep} \,  
\sum_{j=3}^n  \,{\bom T}_2 \cdot {\bom T}_j
 \;f(\ep;z_2- i0 s_{j2})\; \;\;. 
\eeq
The `analytic' continuation from the TL collinear region to a generic
collinear region is thus achieved by the introduction of 
a colour--energy correlated $i0$ prescription.
The main new physical effect in Eq.~(\ref{d12gen}) is the presence of
{\em colour correlations} between the collinear and non-collinear partons. 
This effect produces violation of strict (\naive) collinear factorization
of the scattering amplitudes. 

A detailed derivation of the results in 
Eqs.~(\ref{fact12onegen})--(\ref{ic12gen}), including the
extension 
to the multiple collinear limit of $m$ $(m \geq 3)$ parton momenta
(see Sect.~\ref{sec:multi}),
will be presented 
in a forthcoming paper.
In Sect.~\ref{sec:irone}, we illustrate the explicit computation of 
the IR divergent 
part of the one-loop splitting matrix $\sp^{(1)}$ in Eq.~(\ref{fact12onegen}).
The result of this computation can be regarded as a consistency check of
Eqs.~(\ref{fact12onegen})--(\ref{ic12gen}).

In the following we discuss some consequences of the results in 
Eqs.~(\ref{fact12onegen})--(\ref{ic12gen}). To this purpose, we first 
present some colour algebra relations. An important relation is colour
conservation; we have 
\beq
\label{colcon}
\sum_{k=1}^n \;{\bom T}_k = 0 \;\;,
\eeq
or, equivalently,
\beq
\label{colcon12}
\sum_{j=3}^n \;{\bom T}_j = - \left( \,{\bom T}_1 + {\bom T}_2 \,\right) \;\;.
\eeq
Precisely speaking, the relations in Eqs.~(\ref{colcon}) and (\ref{colcon12})
are valid in operator form when the colour charges act onto an overall 
colour-singlet vector, with $n$ partons, in colour space. 
Such vectors are, for instance, the matrix element vector
$\ket{\cm^{(l)}(p_1,p_2,\dots,p_n)}$ $(l=0,1)$ and the vector
$\sp^{(0)}(p_1,p_2;{\widetilde P}) \;\;\ket{\cm^{(l)}(\wp,\dots,p_n)}$
in Eq.~(\ref{fact12onegen}). As a consequence of 
colour conservation
in the tree-level collinear splitting process 
$A \to A_1 A_2\,$, we have\footnote{The relation in Eq.~(\ref{colconsp}) is also
valid when replacing $\sp^{(0)}$ with the one-loop contribution $\sp^{(1)}_H$.}
\beq
\label{colconsp}
\left( \, {\bom T}_1 + {\bom T}_2 \, \right) \;\sp^{(0)}(p_1,p_2;{\widetilde P})
= \sp^{(0)}(p_1,p_2;{\widetilde P}) \;\,{\bom T}_{\wp} \;\;,
\eeq
where ${\bom T}_{\wp}\,$ denotes the colour charge of the parent collinear
parton $A$.
We also recall that ${\bom T}_k^2= C_k$, where $C_k$ is the Casimir factor
of the $k$-th parton. Therefore, Eq.~(\ref{colconsp}) implies:
\beq
\label{col2conp}
({\bom T}_1 + {\bom T}_2)^2 \;\;\sp^{(0)}(p_1,p_2;{\widetilde P})
= C_{12} \;\,\sp^{(0)}(p_1,p_2;{\widetilde P}) \;\;,
\eeq
or, equivalently,
\beq
\label{t1t2coh}
2 \;{\bom T}_2  \cdot ({\bom T}_1 + {\bom T}_2) 
\;\;\sp^{(0)}(p_1,p_2;{\widetilde P})
= \Bigl( C_{12} + C_2 - C_1 \Bigr) \;\sp^{(0)}(p_1,p_2;{\widetilde P}) \;\;,
\eeq
which follows from the identity
$({\bom T}_1 + {\bom T}_2)^2 - {\bom T}_1^2 - {\bom T}_2^2 =
2 \;{\bom T}_1 \cdot {\bom T}_2$

Using simple colour algebra relations, we can easily show that the general
results in Eqs.~(\ref{fact12onegen})--(\ref{ic12gen}) lead to the TL results
illustrated in Sect.~\ref{sec:tl}. Considering the collinear limit in the TL
region, we have $z_2 > 0$ (see Eq.~(\ref{tlc})) and, therefore, the
$i0$ prescription in $f(\ep;z_2- i0 s_{j2})$ is harmless. Removing the 
$i0$ prescription on the right-hand side of Eq.~(\ref{d12gen}), we have
\beq
\label{d12gencoh}
{\bom \delta}(p_1,p_2;p_3,\dots,p_n) = 
+ \; \frac{2}{\ep} \;f(\ep;z_2) \;\, 
 {\bom T}_2 \cdot \sum_{j=3}^n \,{\bom T}_j
  \;\;, \quad \;\; z_2 > 0 \;\;,
\eeq
and we can perform the sum over the colour charges of the non-collinear partons.
Since ${\bom \delta}$ acts onto the colour vector
$\sp^{(0)}(p_1,p_2;{\widetilde P}) \;\;\ket{\cm^{(0)}(\wp,\dots,p_n)}$
(see Eqs.~(\ref{fact12onegen})--(\ref{ic12gen})), the sum over the 
colour charges can  be carried out explicitly by using Eqs.~(\ref{colcon12})
and 
(\ref{t1t2coh})
and, hence, 
Eq.~(\ref{d12gencoh})
becomes 
\beq
\label{d12gentl}
{\bom \delta}(p_1,p_2;p_3,\dots,p_n) = 
-\; \frac{1}{\ep} \, \Bigl(C_{12} + C_2 - C_1 \Bigr) \;f(\ep;z_2) 
\;\;, \quad \; z_2 > 0 \;\;,
\eeq
which is equal to the TL expression in Eq.~(\ref{d12tl}).

In summary, the absence of evident colour correlations in 
Eq.~(\ref{d12gentl}) or, equivalently, the validity of strict collinear
factorization in the TL collinear limit is a physical consequence 
of {\em colour coherence} (and colour conservation). 
In the case of the TL collinear limit, the
non-collinear partons act coherently as a single parton, whose colour charge
is equal to the {\em total} charge of the non-collinear partons 
(see Eq.~(\ref{d12gencoh})). Owing to colour conservation, this colour charge
is equal (modulo the overall sign) to the colour charge of the parent parton; 
therefore,
the total contribution of the interactions 
(which are separately not factorized) of a collinear parton 
with the non-collinear partons is
effectively equivalent to a {\em single} interaction with the {\em parent} 
parton.
This interaction
factorizes and produces the 
colour coefficient in Eq.~(\ref{d12gentl}).

Incidentally, exploiting Eq.~(\ref{d12gentl}),
we note that we can remove the constraint $z_1 > 0$ and write
the colour operator of Eq.~(\ref{ic12gen}) in a form that has a
manifestly symmetric dependence on the variables of the two collinear partons
$A_1$ and $A_2$. This symmetric form is
 \beeq
\label{ic12gensym}
{\bom I}_C(p_1,p_2;p_3,\dots,p_n) &=& 
 \; \g^2 \; c_{\Gamma} \;
\left( \frac{-s_{12} -i0}{\mu^2} \right)^{-\ep} \\
&\times& \left\{  \; \frac{1}{\ep^2} \; \Bigl( C_{12} - C_1 - C_2 \Bigr)
+ \frac{1}{\ep} \; \Bigl( \gamma_{12} - \gamma_1 - \gamma_2 
+ \cbet0
\Bigr)
\right. \nn \\
&+&\left. \! \frac{2}{\ep} \, \sum_{j=3}^n  \,{\bom T}_j \cdot 
\Bigl[ {\bom T}_1
 \;f(\ep;z_1- i0 s_{j1})
 +   {\bom T}_2
 \;f(\ep;z_2- i0 s_{j2})\;
\Bigr] \right\} \,,\nn 
\eeeq
and it is obtained from Eq.~(\ref{ic12gen}) simply by using Eqs.~(\ref{d12gen})
and (\ref{d12gentl}), with the replacement of the subscripts $2 \leftrightarrow 1$
(i.e., $z_2 \leftrightarrow  z_1, \;{\bom T}_2 \leftrightarrow {\bom T}_1$ and so
forth).

\subsection{The collinear limit of amplitudes with $n=3$ QCD partons}
\label{sec:3part}

The simplest case in which the two-parton collinear limit can be studied occurs
when the original scattering amplitude
$\cm(p_1,p_2,\dots,p_n)$ involves only $n=3$ QCD partons and, necessarily
(because of kinematics),
additional colourless external legs
(with non-vanishing momentum). 
In this case the colour algebra of the
operator ${\bom I}_C$ (or, simply, ${\bom \delta}$) can be carried out in closed
form. Setting $n=3$ in Eq.~(\ref{d12gen}) and using the colour-charge relations
(\ref{colcon12}) and (\ref{t1t2coh}), we obtain
\beeq
{\bom \delta}(p_1,p_2;p_3) &=& 
+ \; \frac{2}{\ep} \;\,{\bom T}_2 \cdot {\bom T}_3
 \;f(\ep;z_2- i0 s_{23})
= - \; \frac{2}{\ep} \;\,{\bom T}_2 \cdot ( {\bom T}_1 + {\bom T}_2)
 \;f(\ep;z_2- i0 s_{23}) \nn \\
\label{d123p}
&=&  - \; \frac{1}{\ep} \;\Bigl(C_{12} + C_2 - C_1 \Bigr)  
\;f(\ep;z_2- i0 s_{23})\; \;\;. 
\eeeq
We see that the operator ${\bom \delta}$ is proportional to the unit matrix in
colour space. Nevertheless, the $c$-number function in Eq.~(\ref{d123p})
still retains process-dependent features that derive from the violation of 
strict collinear factorization (i.e., from the $i0$ prescription in
$f(\ep;z_2- i0 s_{23})$).

\setcounter{footnote}{2}

\begin{figure}[htb]
\begin{center}

\begin{picture}(400,380)(0,0)
\SetWidth{1.5}

\SetOffset(0,300)

\Vertex(30,50){1.8}
\Photon(-10,50)(30,50){4}{4}
\ArrowLine(30,50)(65,15)
\ArrowLine(30,50)(65,85)
\Gluon(30,50)(75,50){4}{5}
\Text(30,-10)[]{$e^+ e^-$}
\Text(75,98)[]{$q(p_1)$}
\Text(75,40)[]{$g(p_2)$}
\Text(75,5)[]{$\bar q(p_3)$}
\GCirc(30,50){12}{.5}
\GCirc(30,50){9}{1}
\Text(110,65)[]{$p_1||p_2$}
\SetWidth{3}
\LongArrow(95,50)(130,50)
\SetWidth{1.5}

\SetOffset(170,300)
\Vertex(30,50){1.8}
\Photon(-10,50)(30,50){4}{4}
\Line(30,50)(43,64)
\ArrowLine(30,50)(60,20)
\Vertex(65,85){1.8}
\Line(50,70)(65,85)
\ArrowLine(65,85)(71,109)
\Gluon(89,91)(65,85){4}{3}
\Text(52,98)[]{TL}
\GCirc(30,50){12}{.5}
\GCirc(30,50){9}{1}
\Text(100,50)[]{\color{black} \large \bf +}

\SetOffset(300,300)

\Vertex(30,50){1.8}
\Photon(-10,50)(30,50){4}{4}
\Line(30,50)(43,64)
\ArrowLine(30,50)(60,20)
\Vertex(65,85){1.8}
\Line(50,70)(65,85)
\ArrowLine(65,85)(71,109)
\Gluon(89,91)(65,85){4}{3}
\GCirc(65,85){6}{.5}
\GCirc(65,85){4}{1}
\Text(52,98)[]{TL}

\SetOffset(0,158)

\Vertex(30,50){1.8}
\Photon(-10,15)(30,50){4}{4}
\ArrowLine(30,50)(65,15)
\ArrowLine(-10,85)(30,50)
\Gluon(30,50)(65,85){4}{5}
\Text(30,-10)[]{DIS}
\Text(-5,98)[]{$\bar q(-p_1)$}
\Text(75,98)[]{$g(p_2)$}
\Text(75,5)[]{$\bar q(p_3)$}
\GCirc(30,50){12}{.5}
\GCirc(30,50){9}{1}

\Text(110,65)[]{$p_1||p_2$}
\SetWidth{3}
\LongArrow(95,50)(130,50)
\SetWidth{1.5}

\SetOffset(140,150)

\Text(28,104)[]{SL}
\Vertex(34,88){1.8}
\ArrowLine(5,88)(34,88)
\Gluon(34,88)(56,105){4}{3}
\Line(34,88)(56,75)
\Vertex(86,58){1.8}
\Line(65,70)(86,58)
\ArrowLine(86,58)(125,58)
\Photon(50,30)(86,58){4}{4}
\GCirc(86,58){12}{.5}
\GCirc(86,58){9}{1}
\Text(140,58)[]{\color{black} \large \bf +}

\SetOffset(285,150)

\Vertex(34,88){1.8}
\ArrowLine(5,88)(34,88)
\Gluon(34,88)(56,105){4}{3}
\Line(34,88)(56,75)
\Vertex(86,58){1.8}
\Line(65,70)(86,58)
\ArrowLine(86,58)(125,58)
\Photon(50,30)(86,58){4}{4}
\Text(18,104)[]{DIS (SL)} 
\Text(28,72)[]{$s_{23}>0$} 
\GCirc(34,88){6}{.5}
\GCirc(34,88){4}{1}

\SetOffset(0,8)

\Vertex(30,50){1.8}
\Photon(30,50)(65,15){4}{4}
\ArrowLine(-10,15)(30,50)
\ArrowLine(-10,85)(30,50)
\Gluon(30,50)(65,85){4}{5}
\Text(30,-10)[]{DY}
\Text(-5,98)[]{$\bar q(-p_1)$}
\Text(75,98)[]{$g(p_2)$}
\Text(-5,5)[]{$q(-p_3)$}
\GCirc(30,50){12}{.5}
\GCirc(30,50){9}{1}

\Text(110,65)[]{$p_1||p_2$}
\SetWidth{3}
\LongArrow(95,50)(130,50)
\SetWidth{1.5}

\SetOffset(140,0)

\Text(28,104)[]{SL}
\Vertex(34,88){1.8}
\ArrowLine(5,88)(34,88)
\Gluon(34,88)(56,105){4}{3}
\Line(34,88)(56,75)
\Vertex(86,58){1.8}
\Line(65,70)(86,58)
\ArrowLine(50,30)(86,58)
\Photon(86,58)(125,58){4}{4}
\GCirc(86,58){12}{.5}
\GCirc(86,58){9}{1}
\Text(140,58)[]{\color{black} \large \bf +}

\SetOffset(285,0)

\Vertex(34,88){1.8}
\ArrowLine(5,88)(34,88)
\Gluon(34,88)(56,105){4}{3}
\Line(34,88)(56,75)
\Vertex(86,58){1.8}
\Line(65,70)(86,58)
\ArrowLine(50,30)(86,58)
\Photon(86,58)(125,58){4}{4}
\Text(18,104)[]{DY (SL)}
\Text(28,72)[]{$s_{23}<0$}
\GCirc(34,88){6}{.5}
\GCirc(34,88){4}{1}

\end{picture}
\end{center}
\caption{\label{fig:3part}
{\em Two-parton collinear factorization of the one-loop amplitude
$\cm(q(p_1),g(p_2),{\overline q}(p_3);\gamma^*)$ with $n=3$ QCD partons 
in different kinematical configurations 
($e^+e^-$, DIS, DY)
related by analytic continuation.
The tree-level (pointlike vertices) collinear splitting subprocesses 
$q \to q g$ in the TL ($e^+e^-$) and SL (DIS or DY) regions
are related by crossing symmetry.
In the SL region, the one-loop (annular vertices)
splitting subprocesses depend on the process and are different in the DIS and DY
configurations.}
}
\end{figure}
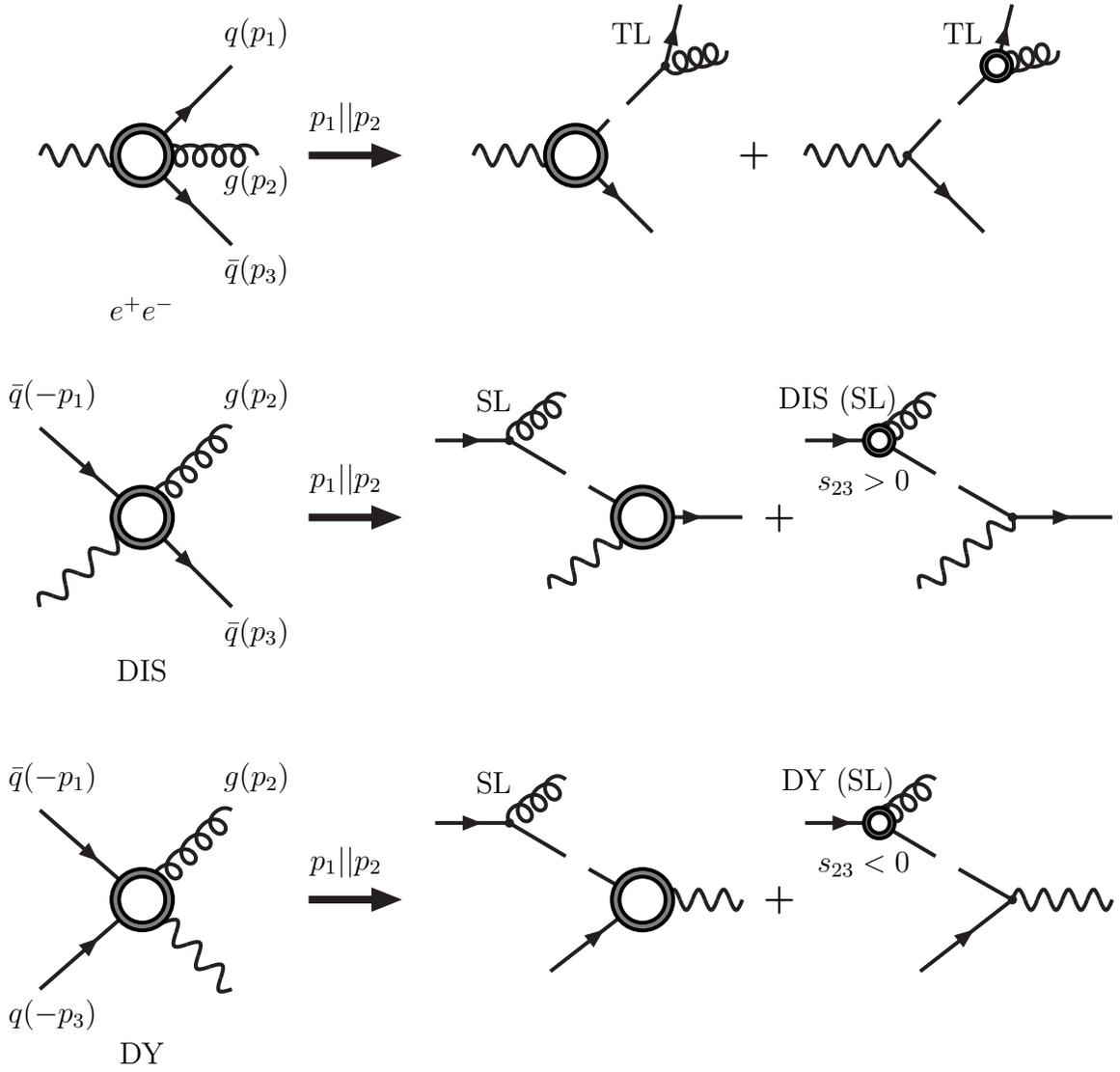

To remark these process-dependent features, we consider the specific 
example\footnote{A completely analogous example is obtained by replacing
the off-shell photon with a Higgs boson (in this case, the $q{\bar q}$ pair
can also be replaced by two gluons).} in which the external legs of the 
matrix element $\cm(p_1,p_2,p_3)= \cm(q(p_1),g(p_2),
{\overline q}(p_3);\gamma^*)$
are a gluon, a quark--antiquark pair and an off-shell photon $\gamma^*$.
The off-shell photon can be coupled to a lepton pair, thus leading to the
partonic subprocess of different physical processes, such as, hadron production
in $e^+e^-$ annihilation ($e^+e^-$) or in lepton--hadron deep-inelastic
scattering (DIS), and the production of lepton pairs through the
Drell--Yan (DY) mechanism in hadron--hadron collisions. These different
processes simply require the analytic continuation of the
{\em same} matrix element, $\cm(q(p_1),g(p_2),
{\overline q}(p_3);\gamma^*)$,
to different kinematical regions, which are specified by the sign of the
`energies' $p_i^0$ of the outgoing momenta  $p_i$. To be definite, we fix 
$p_2^0 > 0$ and we examine the physical partonic processes
(see Fig.~\ref{fig:3part})
\beeq
\label{eepro}
&&\;\;\;\; \gamma^* \;\to\; q(p_1) + g(p_2) + 
{\overline q}(p_3) \;\;,
\quad \;\;\;p_1^0>0, \,p_3^0>0  \;\;\;\;\;(e^+e^-) \,,\\
\label{dispro}
&&\;\; \gamma^* + 
{\overline q}(-p_1)
\;\to\; g(p_2) + 
{\overline q}(p_3) \;\;,
\quad \;\;p_1^0<0<p_3^0  \;\;\;\;\;\;\;\;\; ({\rm DIS}) \,,\\
\label{dypro}
&&q(-p_3) + 
{\overline q}(-p_1) 
\;\to\; \gamma^* + g(p_2) \;\;,
\quad \;\,p_1^0<0, \, p_3^0<0  \;\;\;\;\;({\rm DY}) \,,
\eeeq
in the limit where the momenta $p_1$ and $p_2$ become collinear.
Using the notation of Eqs.~(\ref{tls}) and (\ref{sls}),
the collinear splitting 
$q \to q(p_1) \,g(p_2)$
formally describes two different
physical subprocesses: in $e^+e^-$ annihilation  
(Eq.~(\ref{eepro})), we are dealing with the TL subprocess 
$q^* \to q(z) \,g(1-z)$, where the {\em final-state} gluon is collinear to the
{\em final-state} quark; in DIS (Eq.~(\ref{dispro})) and DY (Eq.~(\ref{dypro}))
production, we are dealing with the SL subprocess
${\overline q} \to {\overline q}^*(z) \,g(1-z)$, where the {\em final-state} 
gluon is collinear to the
{\em initial-state} antiquark.
At the tree-level, these two physical splitting subprocesses are related 
by crossing symmetry,
namely, by the exchange of the final-state quark with the 
initial-state antiquark.

The generalized factorization formula that describes the collinear limit
of the one-loop scattering amplitude of the processes in 
Eqs.~(\ref{eepro})--(\ref{dypro}) involves the `operator' 
${\bom \delta}$ of Eq.~(\ref{d123p}).
Using Eq.~(\ref{d123p}) in the specific kinematical regions of
Eqs.~(\ref{eepro})--(\ref{dypro}), we have
\begin{itemize}
\item $q^* \to q(z) \,g(1-z) \quad\;\;\;(0 < z < 1 \;, \;\; z_2=1-z > 0 )$:
\end{itemize}
\beq
\label{tlsee}
{\bom \delta}^{(e^+e^-)}(p_1,p_2;p_3) =
- \; \frac{1}{\ep} \;C_A  \;f(\ep;z_2) \;\;,
\eeq
\begin{itemize}
\item ${\overline q} \to {\overline q}^*(z) \,g(1-z)
\quad\;\;\;(0 < z < 1 \;, \;\; z_2=1-1/z < 0 )$:
\end{itemize}
\beeq
\label{sldis}
{\bom \delta}^{({\rm DIS})}(p_1,p_2;p_3) &=&
- \; \frac{1}{\ep} \;C_A  \;f(\ep;z_2 - i0) \;\;, \\
\label{sldy}
{\bom \delta}^{({\rm DY})}(p_1,p_2;p_3) &=&
- \; \frac{1}{\ep} \;C_A  \;f(\ep;z_2 + i0) \;\;.
\eeeq
The differences between the expressions in Eqs.~(\ref{tlsee}), (\ref{sldis})
and (\ref{sldy})
highlight the effect of the violation of strict collinear factorization
at the one-loop level.
In going from the TL expression (\ref{tlsee}) to the related 
SL expressions (Eqs.~(\ref{sldis}) and (\ref{sldy})), it is not sufficient to
use the replacement $z \to 1/z$: in fact, the crossing relation
$z \leftrightarrow 1/z$ has to be supplemented with an appropriate $i0$
prescription. More importantly, the two SL expressions in Eqs.~(\ref{sldis}) and 
(\ref{sldy}) are different: although they are dealing with the `{\em same}'
SL splitting subprocess (radiation of a final-state gluon collinearly to an
initial-state antiquark), the 
singular collinear factors
of the scattering amplitude are different since they refer to an 
initial-state antiquark in two different physical processes (the DIS and DY
processes).

\clearpage


\subsection{The collinear limit of multiparton 
amplitudes}
\label{sec:multipart}

If the original matrix element $\cm(p_1,p_2,\dots,p_n)$
has $n$ external QCD partons with $n \geq 4$, the colour algebra involved
in the collinear limit cannot be explicitly carried out in general form. 
The action of the colour operator ${\bom I}_C$ (or ${\bom \delta}$)
onto the colour vector $\sp^{(0)} \times \ket{\cm^{(0)}}$
indeed depends on the colour-flow structure of the colour vector:
this structure has to be specified in order to explicitly perform
the colour algebra. 
Some general features of the collinear limit, which are independent of the 
specific colour structure of $\cm(p_1,p_2,\dots,p_n)$, are illustrated below.

The function $f(\ep;z)$ in Eq.~(\ref{fep}) is an analytic function 
of the complex variable $z$. When the argument $z$ approaches the real axis,
$f(\ep;z)$ has a real and an imaginary part. We write
\beq
\label{fepri}
f(\ep;x-i0s) = f_R(\ep;x) - i \;{\rm sign}(s) \;f_I(\ep;x) \;\;, 
\quad (x \;{\rm real}) \;\;,
\eeq
where the real $(f_R)$ and imaginary $(f_I)$ parts are defined as
\beq
\label{fepr}
f_R(\ep;x) \equiv {\rm Re} \Bigl[ f(\ep;x \pm i0) \Bigr] = 
\f{1}{2} \Bigl[ f(\ep;x+i0) + f(\ep;x-i0) \Bigr] \;\;,
\eeq
\beq
\label{fepi}
f_I(\ep;x) \equiv {\rm Im} \Bigl[ f(\ep;x+i0) \Bigr]
= \f{1}{2 i} \Bigl[ f(\ep;x+i0) - f(\ep;x-i0) \Bigr] \;\;.
\eeq
We thus consider Eq.~(\ref{d12gen}) and we apply the decomposition in
Eq.~(\ref{fepri}) to $f(\ep;z_2- i0 s_{j2})$. We obtain
\beq
\label{d12genri}
{\bom \delta}(p_1,p_2;p_3,\dots,p_n) = 
+ \; \frac{2}{\ep}  \;{\bom T}_2 \cdot  
 \sum_{j=3}^n \, {\bom T}_j
 \;\Bigl[
f_R(\ep;z_2) - i \;{\rm sign}(s_{j2}) \;f_I(\ep;z_2) \Bigr]
  \;\;. 
\eeq
Note that the term proportional to $f_R(\ep;z_2)$ does not depend on
$s_{j2}$, and it 
can be treated 
as the right-hand side of 
Eq.~(\ref{d12gencoh});
we can perform the sum over the colour charges of the non-collinear partons
and, using Eqs.~(\ref{colcon12})
and (\ref{t1t2coh}), we obtain
\beq
\label{d12ri}
{\bom \delta}(p_1,p_2;p_3,\dots,p_n) = \delta_R(p_1,p_2;\wp)
+ i \;{\bom \delta}_I(p_1,p_2;p_3,\dots,p_n)
 \;\;, 
\eeq
where
\beq
\label{d12r}
\delta_R(p_1,p_2;\wp) = 
- \; \frac{1}{\ep} \;\Bigl(C_{12} + C_2 - C_1 \Bigr)  
\;f_R(\ep;z_2) 
 \,, 
\eeq
\beq
\label{d12i}
{\bom \delta}_I(p_1,p_2;p_3,\dots,p_n) = 
- \; \frac{2}{\ep}  \;{\bom T}_2 \cdot  
\left( \sum_{j=3}^n \, {\bom T}_j
 \;
  {\rm sign}(s_{j2})  
 \right)\;f_I(\ep;z_2) \;\;. 
\eeq

The expressions in Eqs.~(\ref{d12gen}) and (\ref{d12ri}) are equivalent.
The latter explicitly shows that the `real' (more precisely, hermitian) 
part, $\delta_R$, of the colour operator $\bom \delta$ is proportional to the
unit matrix 
in colour space. Incidentally, the form of $\delta_R$ is analogous
to that of the corresponding term, $\delta$, in the TL case 
(see Eq.~(\ref{d12tl})); the only difference is that $z_2$  
can have negative values in the general case of Eq.~(\ref{d12r}).
The `imaginary' (more precisely, antihermitian) part, $i {\bom \delta}_I$,
of the colour operator $\bom \delta$ has instead a non-trivial dependence on the
colour charges of the non-collinear partons; this part is responsible for
violation of strict collinear factorization.

For practical computational applications of the factorization formula
(\ref{fact12onegen}), it is useful to expand the one-loop splitting matrix 
$\sp^{(1)}$ in powers of $\ep$. This eventually requires the corresponding
expansion of the function $f(\ep;z_2- i0 s_{j2})$ in Eqs.~(\ref{ic12gen}) and
(\ref{d12gen}) or, equivalently, the real functions $f_R(\ep;z_2)$ and 
$f_I(\ep;z_2)$ in Eqs.~(\ref{d12r}) and (\ref{d12i}).
If $z_2$ is positive (TL collinear limit), the $i0$ prescription is not 
needed, and Eq.~(\ref{fepex}) explicitly gives the expansion of 
$f(\ep;z_2)$. In the case of the SL collinear limit, $z_2$ is negative
and the polylogarithms in Eq.~(\ref{fepex}) are evaluated close to (either
above or below) their branch-cut singularity. To simplify the $\ep$ expansion
in the SL case,
we can use the relation between the hypergeometric functions 
${}_2F_1$ of complex argument $z$ and $1/z$; Eq.~(\ref{fep}) can thus be written
as
\beq
\label{fexpgen}
f(\ep;x-i0s) = f(- \ep;1-x) + \f{1}{\ep} 
\left[ \Gamma(1+\ep) \, \Gamma(1-\ep) \left( \f{x-i0s}{1-x} \right)^{-\ep}
-1 \right] \;,   \;(x \;{\rm real, \;and} \; x < 1) \;.
\eeq
If $x$ is negative, 
$f(- \ep;1-x)$  can safely be expanded as in 
Eq.~(\ref{fepex}). Therefore, the right-hand side of Eq.~(\ref{fexpgen})
gives a simple $\ep$ expansion of $f(\ep;z_2- i0 s_{j2})$ in the SL case.
The formula (\ref{fexpgen}) can also be used to evaluate 
the real and imaginary
parts, $f_R$ and $f_I$, and their expansion in powers of $\ep$; we have
\beq
\label{frex}
f_R(\ep;x) = \Theta(x) \;f(\ep;x) + \Theta(-x) \,
\left\{ \;f(-\ep;1-x) + \f{1}{\ep} 
\left[ \f{\pi \ep}{\tan (\pi \ep)} \left( \f{-x}{1-x} \right)^{-\ep}
-1 \right]
\right\} \;\;,
\eeq
\beq
\label{fiex}
f_I(\ep;x) = - \;\Theta(-x) \;\, \pi \; \left( \f{-x}{1-x} \right)^{-\ep} 
\;\;.
\eeq

We explicitly report the expansion of the functions $f_R(\ep;x)$  and 
$f_I(\ep;x)$ up to order $\ep^2$:
\beeq
\label{frex2}
\Theta(-x) \;f_R(\ep;x) \!\!\!&=&\!\! \Theta(-x) \,
\left\{ \;- \ln(-x) + \ep \left[ \;{\rm Li}_2\left(\f{-x}{1-x}\right)
+ \f{1}{2} \,\ln^2\left(\f{-x}{1-x}\right) - \f{\pi^2}{3}
\right] \right. \nn \\
&-&\!\!\!\! \left. \ep^2 \!\left[ \;{\rm Li}_3\left(\f{-x}{1-x}\right)
+ \f{1}{6} \,\ln^3\left(\f{-x}{1-x}\right)
- \f{\pi^2}{3} \,\ln\left(\f{-x}{1-x}\right)
\right]\! + {\cal O}(\ep^3)
\right\} ,
\eeeq
\beq
\label{fiex2}
f_I(\ep;x) = - \;\Theta(-x) \;\, \pi \;\left[ 
1 - \ep \,\ln\left(\f{-x}{1-x}\right) + 
\ep^2 \;\f{1}{2} \,\ln^2\left(\f{-x}{1-x}\right) + {\cal O}(\ep^3)
\right] \;.
\eeq
If $x > 0$, the $\ep$ expansion of $f_R(\ep;x)$ is given in 
Eq.~(\ref{fepex}).

\subsection{
The SL collinear limit in lepton--hadron and hadron--hadron 
collisions}
\label{sec:SL4}

We present some additional general comments\footnote{A related discussion,
limited to the specific case of amplitudes with $n=3$ QCD partons, has been
presented in Sect.~\ref{sec:3part}.}
on the SL collinear limit 
in the kinematical configuration\footnote{Related comments
apply to the SL configuration in which $p_2^0 < 0 < p_1^0$ and $\wp^0 > 0$
(see the final part of Sect.~\ref{sec:tree}).}
of Eq.~(\ref{sls}).
This kinematical configuration occurs in the case of hard-scattering 
observables in lepton--hadron and hadron--hadron collisions.

In lepton--hadron 
DIS, the 
partonic
matrix elements $\cm(p_1,p_2,p_3,\dots,p_n)$ involve the collision of a lepton
and a parton in the initial state. The initial-state physical parton 
with `outgoing' momentum $p_1$ and energy $p_1^0 < 0$ can radiate the collinear
parton with momentum $p_2$ in the final state ($p_2^0 > 0$); then,
the accompanying
(`parent') parton with `outgoing' momentum $\wp$ and energy ${\wp}^{\,0} < 0$
acts as initial-state physical parton in the hard-scattering subprocess
controlled by the `reduced' matrix elements $\cm(\wp,p_3,\dots,p_n)$.
All the other partons (i.e. the non-collinear partons)
are produced in the final state and, thus, $s_{j2} > 0$ (with $j\geq 3$).
In this kinematical configuration the colour operator 
${\bom \delta}(p_1,p_2;p_3,\dots,p_n)$ is 
\beq
\label{d12dis}
{\bom \delta}(p_1,p_2;p_3,\dots,p_n) = 
- \; \frac{1}{\ep} \;\Bigl(C_{12} + C_2 - C_1 \Bigr)  
\;f(\ep;z_2- i0) \;, \quad \;\;s_{j2} > 0  \;(j\geq 3)
 \,. 
\eeq
This expression is obtained from Eq.~(\ref{d12gen}) by following
the same steps as in Eqs.~(\ref{d12gencoh}) and (\ref{d12gentl}).
Actually, the DIS expression in Eq.~(\ref{d12dis}) differs from the
TL expression in Eq.~(\ref{d12gentl}) only because 
of the replacement $f(\ep;z_2) \to f(\ep;z_2- i0)$,
which, roughly speaking, simply produces an additional
imaginary part. 
In Eq.~(\ref{d12dis}) we note the absence of explicit dependence on 
the colours and momenta of the non-collinear partons.
This implies that the two-parton SL collinear limit in DIS configurations
effectively takes (mimics) a strictly-factorized form.
This `effective' strict collinear factorization 
is eventually the result of a colour-coherence phenomenon, analogously to
what happens for the TL collinear limit. 
The interactions 
(which are separately not factorized) of the collinear parton $p_2$
with the non-collinear partons produce imaginary parts; however, since all these
interactions involve {\em final-state} partons, the imaginary parts
combine coherently to mimic a single effective interaction with the parent
parton.
This global final-state effect is embodied in the definite $i0$ prescription
of Eq.~(\ref{d12dis}).

\vspace{0.5cm}
 \begin{figure}[htb]
 \begin{center}
 \begin{tabular}{c}
 \epsfxsize=8truecm
 \epsffile{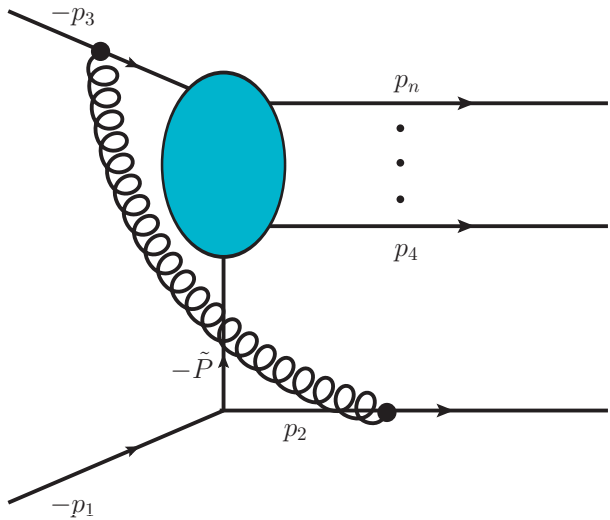}\\
  \end{tabular}
 \end{center}
 \caption{\label{fig:t2t3}
 {\em Two-parton SL collinear limit $(p_i \simeq z_i \wp, \;i=1,2)$
 of the scattering amplitude
 $\cm(p_1,p_2,p_3,\dots,p_n)$ with $n \geq 4$ QCD partons in parton--parton
 hard-scattering configurations. Typical (irreducible) colour structure 
 of factorization breaking correlations at the one-loop level.} 
 }
 \end{figure}

In hadron--hadron collisions (see Fig.~\ref{fig:t2t3}),
the SL collinear splitting of Eq.~(\ref{sls})
takes place in a partonic environment that differs from that of  
lepton--hadron collisions. The main difference is due to the fact that the
matrix element $\cm(p_1,p_2,p_3,\dots,p_n)$ involves the initial-state
collision of {\em two} QCD partons, the
collinear parton with momentum $p_1$ and another parton, and both partons carry
colour charge. The real part $\delta_R$ of the colour operator $\bom \delta$
is the same as in Eq.~(\ref{d12dis}) (see also Eq.~(\ref{d12r})), 
while the imaginary part 
${\bom \delta}_I$ is sensitive
to the colour charge of the initial-state non-collinear parton. 
We assign the `outgoing' momentum $p_3$  and the energy $p_3^0 < 0$ 
to this parton, and we can write the following explicit form of 
${\bom \delta}_I$:
\beq
\label{d12had}
{\bom \delta}_I(p_1,p_2;p_3,..,p_n) = 
+ \; \frac{1}{\ep}    
\left\{ \Bigl(C_{12} + C_2 - C_1 \Bigr) + 
4 \;{\bom T}_2 \cdot {\bom T}_3  
 \right\}\;f_I(\ep;z_2) \,, \;\;  
s_{23} < 0 < s_{j2} \;(j \geq 4) \,.
\eeq
Combining this imaginary part with the real part (see Eq.~(\ref{d12r})),
the colour operator ${\bom \delta}$ can be written in the following form: 
\beq
\label{d12hadt}
{\bom \delta}(p_1,p_2;p_3,\dots,p_n) = 
- \; \frac{1}{\ep} \;\Bigl(C_{12} + C_2 - C_1 \Bigr)  
\;f(\ep;z_2- i0) + \; \frac{i}{\ep} \;4 \;{\bom T}_2 \cdot {\bom T}_3
\;f_I(\ep;z_2) \;\;, 
\eeq
which clearly exhibits the difference with respect to the DIS expression in 
Eq.~(\ref{d12dis}).

The expression in Eq.~(\ref{d12had}) is obtained from Eq.~(\ref{d12i})
by using simple colour algebra relations. Since $s_{23} < 0$ and 
$s_{j2} > 0 \; (j \geq 4)$, the colour charge factor in Eq.~(\ref{d12i})
gives
\beeq
\label{d12icc}
- \;{\bom T}_2 \cdot  
\left\{ \sum_{j=3}^n \, {\bom T}_j
 \;{\rm sign}(s_{j2})  
 \right\} &=& {\bom T}_2 \cdot  \left\{ {\bom T}_3 - \sum_{j=4}^n \, {\bom T}_j
  \right\}
= {\bom T}_2 \cdot  \left( {\bom T}_1 + {\bom T}_2 + 2 \,
{\bom T}_3 
   \right) \nn \\
&=& \f{1}{2}  \Bigl(C_{12} + C_2 - C_1 \Bigr) + 2 \;{\bom T}_2 \cdot {\bom T}_3
 \;\;, 
\eeeq
where we have used Eq.~(\ref{colcon}) (namely, $\sum_{j=4}^n \, {\bom T}_j =
- ( {\bom T}_1 + {\bom T}_2 + {\bom T}_3)$) and Eq.~(\ref{t1t2coh}). Inserting
Eq.~(\ref{d12icc}) in Eq.~(\ref{d12i}), we get the result in Eq.~(\ref{d12had}).

The initial-state (hence, SL) collinear splitting in physical
parton--parton scattering amplitudes $\cm(p_1,p_2,\dots,p_n)$
with $n \geq 4$ QCD partons necessarily involves colour correlations 
between the collinear
and non-collinear partons. 
A representation of these correlations in minimal form
is presented in the expression (\ref{d12had}) of the imaginary part of the
colour charge operator~${\bom \delta}$ (see also Fig.~\ref{fig:t2t3}).

\subsection{Factorization, colour coherence and causality}
\label{sec:caus}

The factorized structures of Eqs.~(\ref{fact12one}) and (\ref{fact12onegen}) 
and, in particular,
the violation of strict factorization in the case of
SL collinear configurations\footnote{A related effect, in the context of 
the computation of logarithmically-enhanced contributions
to gaps-between-jets cross sections, was 
described 
as a `breakdown of \naive\ coherence' 
\cite{Forshaw:2006fk, Seymour:2007vw}.}
deserve some physical (though qualitative)
interpretation. We offer some comments about that.

The universal factorization of scattering amplitudes in the collinear limit 
is expected on the basis of a simple physical picture. When the two
partons\footnote{The qualitative discussion of this subsection can
straightforwardly be extended to the multiple collinear limit of
three or more parton momenta.} $A_1$ and  $A_2$
with momenta $p_1$ and $p_2$ become collinear, their invariant mass $s_{12}$
is much smaller than $s_{j2} \sim s_{j1}$, where the subscript $j$ generically
denotes a non-collinear parton $(j=3,4,\dots,n)$. Therefore, we are dealing
with a two-scale process. Interactions between the collinear partons
take place at the small energy scale $s_{12}$ and, hence, at large space-time
distances; whereas, interactions between the non-collinear partons and
interactions between the collinear and non-collinear partons require a large
energy scale and, hence, they take place at small space-time distances.
This space-time separation of large and small distances produces factorization.
The (large-distance) physical subprocess $A \to A_1 A_2$, which generates the
collinear partons,
is factorized from the (short-distance) scattering amplitude that involves the 
parent collinear parton $A$ and the 
non-collinear partons.

This simple space-time factorized picture is expressed, at the tree level,
by the universal factorization formula in Eq.~(\ref{fact12t}). At the one-loop
(or higher-loop) level, the same space-time picture is too \naive, at least in
the case of gauge theories. Owing to the long-range nature of gauge 
interactions, partonic fluctuations with arbitrarily-large 
wavelength (i.e. arbitrarily-small momentum) can propagate over 
widely-separated space-time
regions, thus spoiling the factorization between the small-distance
and large-distance subprocesses involved in the collinear limit. 

To be more precise, we refer to one-loop interactions due to a gluon with
soft momentum $q$ ($q \to 0)$. The soft gluon produces   
{\em pairwise} interactions between a collinear and a non-collinear parton.
In the kinematical regions where the
soft-gluon momentum has a very small angle with respect to 
the momentum of one of the external partons (either a collinear or a
non-collinear parton), the one-loop interaction produces factorized
(though IR divergent) contributions\footnote{ 
These contributions are eventually factorized since the small-angle 
restriction, namely, the restriction to be collinear
to the external leg, effectively bounds the soft-gluon interaction to act
in the space-time region of the physical subprocess that involves 
the corresponding external leg.}. Therefore, we are left to consider 
interactions due to a soft {\em wide-angle} gluon.
Each of these pairwise interactions between a non-collinear $(j=3,4,\dots,n)$
and a collinear $(i=1,2)$ parton is 
proportional to the colour-charge term  
${\bom T}_j \cdot {\bom T}_i\,$, and it is not factorized (it depends
on the colour charge of both the non-collinear and collinear partons). 
These 
interactions {\em separately} spoil collinear
factorization. 
Nonetheless, factorization can be recovered because of {\em
colour coherence}.

The colour coherence mechanism that leads to factorization is as follows.
We consider the kinematical region of {\em wide-angle}
interaction. It is the region where $\theta_{jq}$
($\theta_{jq}$ is the angle between $p_j$ and the soft momentum $q$)  
is large and, specifically, the region where
$\theta_{jq}$ can reach values that are 
parametrically\footnote{To be slightly more precise, this wide-angle region is
specified by the boundary $\theta_{jq} \ltap \theta_{ji} - \theta_{12}$.
Indeed, from the viewpoint of the ensemble of the two collinear partons,
there is a different wide-angle region. This is the region where $\theta_{2q}$
(or $\theta_{1q}$) is parametrically of the order of $\theta_{12}$. In this
region, considering the collinear limit $\theta_{12} \to 0$, the system of the
non-collinear partons coherently interacts with the collinear parton $p_2$
(or $p_1$), as discussed below Eq.~(\ref{d12gentl}).} 
of the order of $\theta_{ji}$ ($\theta_{ji}$ is the angle
between $p_j$ and the collinear momentum $p_i$). After integration over the 
soft-gluon momentum, the wide-angle interaction
proportional to ${\bom T}_j \cdot {\bom T}_1\,$ depends on $\theta_{j1}$ and,
analogously, the wide-angle interaction with the other collinear parton is
proportional to ${\bom T}_j \cdot {\bom T}_2\,$  and depends on $\theta_{j2}$.
In the collinear limit we have  $\theta_{j1}=\theta_{j2}=\theta_{j\wp}$ 
($\wp$ is the momentum of the parent parton)
and, therefore, these two interaction contributions
are combined in a single contribution that depends on $\theta_{j\wp}$ and
is proportional to the colour-charge term
${\bom T}_j \cdot ({\bom T}_1 + {\bom T}_2)={\bom T}_j \cdot {\bom T}_{\wp}\,$.
This single contribution is exactly the contribution of the 
wide-angle interaction between the non-collinear parton $p_j$ and the parent
parton $\wp$: this contribution is factorized in the one-loop scattering
amplitude $\ket{\cm^{(1)}(\wp,\dots,p_n)}$ on the right-hand side of 
Eqs.~(\ref{fact12one}) and (\ref{fact12onegen}).

In summary, in the collinear limit the system of the two collinear partons 
acts coherently with respect to non-factorizable wide-angle interactions 
with each of the non-collinear partons. Owing to this coherent action, these
interactions are removed from the collinear splitting matrix 
$\sp^{(1)}$ and re-factorized in the matrix element 
$\ket{\cm^{(1)}(\wp,\dots,p_n)}$ of the factorization formulae 
(\ref{fact12one}) and (\ref{fact12onegen}).

The colour coherence mechanism that we have just illustrated is completely
analogous to the mechanism that acts by considering the radiation 
of two collinear
partons and a soft gluon in tree-level scattering amplitudes. In this mixed
soft--collinear limit, the collinear splitting process is factorized from the
soft-gluon current of the parent parton (see Sects.~3.4 and 3.5 in 
Ref.~\cite{Catani:1999ss}). However, there is an essential difference 
between the radiation of a {\em real} gluon in tree-level amplitudes 
and the interactions of a {\em virtual} gluon in one-loop amplitudes.
This difference is eventually responsible for the violation of strict
factorization in the SL collinear limit
of one-loop amplitudes.
 
A real gluon with soft momentum $q$ is always on-shell ($q^2=0$).
A virtual gluon instead produces both a {\em radiative} (roughly speaking, real)
and an {\em absorptive} (roughly speaking, imaginary) contribution
to the one-loop amplitude. The radiative contribution is produced when 
the virtual
soft gluon is on-shell ($q^2=0$), while the absorptive contribution is 
produced when the soft wide-angle gluon is slightly off-shell ($q^2 \sim - {\bf
q}_\perp^2$, where ${\bf q}_\perp$ is the gluon transverse momentum with respect
to the direction of the momenta of the pair of interacting external partons). 

The colour coherence phenomenon discussed so far in this subsection refers 
to the radiative part of the one-loop interactions. The absorptive (imaginary)
part of the one-loop wide-angle interaction between the 
non-collinear parton $p_j$ and the collinear parton $p_i$ is very similar to its
radiative part (it is still proportional to 
${\bom T}_j \cdot {\bom T}_i\,$  and it simply depends on $\theta_{j\wp}$ in the
collinear limit), but it is non-vanishing {\em only} if $s_{ji} > 0$
(as recalled below, this constraint follows from {\em causality}). 
Combining the absorptive part of the interactions of $p_j$ with the two collinear
partons (as we did previously by combining the radiative part), 
we obtain a contribution that is proportional to
${\bom T}_j \cdot [ \,{\bom T}_1 \,\Theta(s_{j1}) + 
{\bom T}_2 \,\Theta(s_{j2}) \,]$, and we see that the two collinear partons can
act coherently {\em only} if their energies $p_1^0$ and $p_2^0$ have the 
same sign (i.e., in the case of the TL collinear limit). In the absence of this 
coherent action, the collinear splitting subprocess retains 
absorptive interactions
with the non-collinear partons, and strict collinear factorization is violated in
the SL collinear limit. In the generalized factorization formula 
(\ref{fact12onegen}), the amplitude $\ket{\cm^{(1)}(\wp,\dots,p_n)}$ includes
the absorptive part of the interactions with the parent parton $\wp$
(namely, the terms ${\bom T}_j \cdot \,{\bom T}_{\wp} \;\Theta(s_{j \wp})$), and
the remaining absorptive contribution effectively included in 
$\sp^{(1)}$ is proportional to 
\beeq
&&{\bom T}_j \cdot \Bigl[ \,{\bom T}_1 \,\Theta(s_{j1}) + 
{\bom T}_2 \,\Theta(s_{j2}) \,\Bigr] - 
{\bom T}_j \cdot \,{\bom T}_{\wp} \;\Theta(s_{j \wp}) \nn \\ 
&&= {\bom T}_j \cdot 
\Bigl[ \,{\bom T}_1 \,\Bigl( \Theta(s_{j1}) - \Theta(s_{j \wp}) \Bigr)
+ \,{\bom T}_2 \,\Bigl( \Theta(s_{j2}) - \Theta(s_{j \wp}) \Bigr) \,\Bigr]
\;\;.
\nn
\eeeq
If, for example, we consider the SL case with $z_2 <0$ (thus, $z_1 >0$), 
the energies 
$s_{j1}$ and $s_{j \wp}$ have the same sign, while the energies 
$s_{j2}$ and $s_{j \wp}$ have opposite sign; the absorptive contribution to 
$\sp^{(1)}$ is thus proportional to 
${\bom T}_j \cdot \,{\bom T}_2 
\,\Bigl( \Theta(s_{j2}) - \Theta(s_{j \wp}) \Bigr)
= {\bom T}_j \cdot \,{\bom T}_2 \;{\rm sign}(s_{j2})$, which is exactly the
colour-charge factor that appears in the right-hand side of Eq.~(\ref{d12i}).

The one-loop interaction between the partons $j$ and $i$ has a non-vanishing 
absorptive part only if the parton energies $p_j^0$ and $p_i^0$ have the same
sign (i.e. only if $s_{ji} > 0)$. This requires that the two partons belong to
either the physical initial state or the physical final state of the scattering
amplitude. In other words, the absorptive part has a definite causal structure 
(and origin): it arises from long-range interactions that takes place at large
asymptotic times $t \to \pm \infty$, either in the past (considering 
initial-state partons) or in the future (considering final-state partons), with
respect to the short time intervals that control the hard-scattering 
subprocess. 

In gauge theories, factorization is potentially spoiled by the long-range nature
of gauge interactions. Colour coherence can effectively restore the 
space-time factorization of small-distance and large-distance subprocesses. 
Colour coherence argument requires no distinctions between large space distances
and large time distances. Causality does make distinctions between
large distances at $t \to -\infty$ and $t \to +\infty$. Therefore, if the 
large-distance subprocess involves interactions at both 
$t \to -\infty$ and $t \to +\infty$ (as is the case of the two-parton collinear
splitting in the SL region), the factorization power of colour coherence
is {\em limited} by causality. This limitation leads to violation of `strict'
factorization.

The absorptive part of the one-loop splitting matrix 
$\sp^{(1)}$ in Eq.~(\ref{fact12onegen}) is responsible for violation of strict
factorization in the SL collinear region. We observe that this absorptive part is
IR divergent. Its IR divergent contribution (see e.g. Eq.~(\ref{ic12gen}) 
and remember that $f(\ep;z - i0) = - \ln (z -i0)  + {\cal O}(\ep)$) is proportional
to 
$- i \pi \,{\bom T}_j \cdot \,{\bom T}_2/\ep$,
which exactly corresponds to
the contribution of the non-abelian analogue of the QED Coulomb phase. It is
well known (see, e.g., Refs.~\cite{Ralston:1982pa, Catani:1985xt, 
Bonciani:2003nt, Forshaw:2006fk, Bomhof:2004aw, 
Bacchetta:2005rm, Rogers:2010dm, Aybat:2008ct, Bauer:2010cc, DelDuca:2011xm} 
and references
therein) that `Coulomb
gluons' lead to non-trivial QCD effects, especially in relation to various
factorization issues and in connection with resummations of
logarithmically-enhanced radiative corrections.
Our study of the one-loop splitting matrix 
$\sp^{(1)}(p_1,p_2;{\widetilde P};p_3,\dots,p_n)$ (including its absorptive part) 
is performed to all orders in the $\ep$ expansion. We thus note that our results 
on the SL collinear limit of two partons are not limited to the treatment of 
Coulomb gluon effects to leading IR (or logarithmic) accuracy.


\setcounter{footnote}{2}

\section{Multiparton collinear limit and generalized 
factorization at one-loop order and beyond}
\label{sec:multiall}

\subsection{Multiparton collinear limit of tree-level and one-loop amplitudes}
\label{sec:multi}

The definition of the collinear limit of a set $\{ p_1, \dots, p_m \}$ of
$m$ ($m \geq 3$) parton momenta follows the corresponding definition of 
the two-parton
collinear limit (see Eqs.~(\ref{kin2})--(\ref{ptil})).
The multiparton
collinear limit is approached when the momenta
of the $m$ partons become parallel.
This implies that all the parton subenergies
\beq
s_{i \ell}=(p_i+p_\ell)^2 \;\;, 
\quad \quad {\rm with} \;\;\;\;\; i,\ell \in C= \{\,1,\dots,m \,\} \;\;, 
\eeq
are of the {\em same} order and vanish {\em simultaneously}. 
In analogy with the two-parton collinear limit, 
we introduce the light-like momentum\footnote{To be precise, a more
appropriate notation should be used to distinguish the two vectors in
Eqs.~(\ref{ptil}) and (\ref{ptilm}). We do not introduce such a distinction,
since we always use Eqs.~(\ref{ptil}) and (\ref{ptilm}) in connection with the
corresponding collinear limit of $2$ and $m$ partons, respectively.} 
${\wp}^\mu$:
\beq
\label{ptilm}
{\wp}^\mu = 
(p_1 + \dots + p_m)^\mu   
- \frac{(p_1 + \dots + p_m)^2 \; n^\mu}{2 \, n \cdot (p_1 + \dots + p_m)} 
\;\;. 
\eeq
In the multiparton collinear limit, the vector ${\wp}^\mu$
approaches the collinear direction and we have
$p_i^\mu \to z_i {\wp}^\mu$, where the longitudinal-momentum 
fractions $z_i$ are
\beq
\label{zim}
z_i = \frac{n \cdot p_i}{n \cdot {\wp}} =
\frac{n \cdot p_i}{n \cdot (p_1 + \dots + p_m)} \;\;, \quad \;\;\; 
i \in C \;\;,
\eeq
and they fulfill the constraint $\sum_{i=1}^m z_i =1$.
More details on the kinematics of the 
multiparton collinear limit can be found in
Ref.~\cite{Catani:1999ss}.

As in the case of the two-parton collinear limit,
the dynamics of the 
multiparton
collinear limit of scattering amplitudes is
different in the TL and SL regions. The TL region is specified by the constraints
$s_{i \ell} = (p_i+p_\ell)^2 > 0 $, where $\{p_i,p_\ell\}$ refers to {\em any} pair
of collinear-parton momenta; note that these contraints imply $1> z_i > 0$.
If these contraints are not fulfilled, we are dealing with the SL region.

According to this definition, in the TL case, {\em all} the partons in the
collinear set are either final-state partons
(i.e., physically outgoing partons with `energies' $p_i^0 > 0$)
or initial-state partons
(i.e., physically incoming partons with `energies' $p_i^0 < 0$).
In the SL case, the collinear set involves 
at least one parton
in the initial state and, necessarily, one or more partons in the final state.

In the kinematical configuration where the $m$ parton momenta 
$p_1, \dots, p_m$ become 
simultaneously parallel, the matrix element 
$\cm=\cm(p_1,\dots,p_m,p_{m+1},\dots,p_n)$
becomes singular.
The {\em dominant} singular behaviour is
$\cm(p_1,\dots,p_m,p_{m+1},\dots,p_n)\sim (1/{\sqrt s})^{m-1}\, 
{\rm mod\,}(\ln^k s)$ (see Eq.~(\ref{mscale12}) for comparison),
where the logarithmic enhancement is due to scaling violation that occurs through
loop radiative corrections.
Here $s$ denotes a generic two-parton (i.e., $s_{i \ell}$) or multiparton
(e.g., $(p_1+p_2+p_3)^2$) subenergy of the system of the $m$ collinear partons.

The extension of the collinear-factorization formulae 
(\ref{fact12t}), (\ref{fact12one}) and (\ref{fact12onegen}) to the case of the
multiparton collinear
limit is 
\beq
\label{factmt}
\ket{\cm^{(0)}}
\simeq \sp^{(0)}(p_1,\dots,p_m;{\widetilde P}) 
\;\;\ket{\cmbar^{(0)}}
\;\;,
\eeq
\beeq
\label{factmonegen}
\ket{\cm^{(1)}}
&\simeq& \sp^{(1)}(p_1,\dots,p_m;{\widetilde P};p_{m+1},\dots,p_n) 
\;\;\ket{\cmbar^{(0)}} \nn \\
&+& \sp^{(0)}(p_1,\dots,p_m;{\widetilde P}) 
\;\;\ket{\cmbar^{(1)}}
\;\;.
\eeeq
On the right-hand side of Eqs.~(\ref{factmt}) and (\ref{factmonegen}),
we have neglected 
contributions that are subdominant (though,
still singular) in the collinear
limit, and we have denoted the `reduced' matrix element by introducing the 
following shorthand notation:
\beq
\label{mbardef}
\cmbar = \cm(\wp,p_{m+1},\dots,p_n) \;\;.
\eeq
The reduced matrix element $\cmbar$ is obtained from the original matrix 
element  $\cm$
by replacing the $m$ collinear partons $A_1,\dots,A_m$ (whose momenta are 
$p_1,\dots,p_m$) with a single parton $A$, which carries the momentum
$\wp$. The flavour of the parent parton $A$ is determined by flavour
conservation in the splitting subprocess $A \to A_1 \dots A_m$.

The process dependence of the tree-level factorization formula
(\ref{factmt}) is entirely embodied
in the matrix elements $\cm$ and $\cmbar$.
The splitting matrix $\sp^{(0)}(p_1,\dots,p_m;{\widetilde P})$,
which
captures the dominant singular behaviour in the multiparton collinear limit,
is universal (process independent).
It depends on the momenta and quantum
numbers (flavour, spin, colour) of the sole partons that are involved
in the collinear
splitting $A \to A_1 \dots A_m$. 
The colour dependence can explicitly be denoted as
(see Eq.~(\ref{colsp02}) in the case of $m=2$ collinear partons)
\beq
\label{colsp0m}
Sp^{(0)\;(c_1,\dots,c_m;\,c)}(p_1,\dots,p_m;{\widetilde P}) \equiv
\bra{c_1,\dots,c_m} \;\sp^{(0)}(p_1,\dots,p_m;{\widetilde P})
\; \ket{c} \;\;,
\eeq
where $c_1,\dots,c_m$ are the colour indices of the partons $A_1,\cdots,A_m$,
and $c$ is the colour index of the parent parton $A$.

At the tree level, the TL and SL collinear limits have the same structure 
and are simply related by crossing symmetry relations. In going from the TL
to the SL regions, the multiparton splitting matrix 
$\sp^{(0)}(p_1,\dots,p_m;{\widetilde P})$ in Eq.~(\ref{factmt})
only varies because of the replacement of the wave function factors of the
collinear partons (see the final part of Sect.~\ref{sec:tree}).
The dependence of $\sp^{(0)}$ on the colours and momenta of the collinear 
partons is unchanged in the TL and SL regions.

At the one-loop order, the TL and SL collinear limits have a different structure.
As a consequence of the violation of strict collinear factorization,
Eq.~(\ref{factmonegen}) is presented in the form of generalized 
collinear factorization (see Eq.~(\ref{fact12onegen})).
In the SL collinear region, the multiparton one-loop splitting matrix
$\sp^{(1)}$ also depends on the  
momenta and colour charges of the
non-collinear partons in the matrix elements $\cm$ and $\cmbar$.
Introducing the colour dependence in explicit form, we have
\beeq
\label{colsp1m}
&&\!\!\!\!\!\!\!\!Sp^{(1)\;(c_1,\dots,c_n;
\,c,c_{m+1}^\prime,\dots,c_n^\prime)}(p_1,\dots,p_m;
{\widetilde P};p_{m+1},\dots,p_n) 
\nn \\
&&\!\!\equiv \bra{c_1,\dots,c_n} 
\;\sp^{(1)}(p_1,\dots,p_m;{\widetilde P};p_{m+1},\dots,p_n)
\; \ket{c,c_{m+1}^\prime,\dots,c_n^\prime} \;\;,
\eeeq
where $c_1,\dots,c_n$ are the colour indices of {\em all} (collinear and
non-collinear) the external partons in the original matrix element $\cm$,
and $c,c_{m+1}^\prime,\dots,c_n^\prime$ are the colour indices of {\em all}
the external partons (i.e. the parent collinear parton and 
the non-collinear partons)
in the reduced matrix element $\cmbar$. We remark that $\sp^{(1)}$ 
does not depend on
the spin polarization states of the non-collinear partons,
since the violation of strict collinear factorization  
originates from
{\em soft}
interactions between the non-collinear and collinear 
partons (see Sect.~\ref{sec:caus}).
This origin of the violation of strict collinear factorization
also implies that the one-loop multiparton splitting matrix
$\sp^{(1)}$ has factorization breaking terms with a {\em linear}
dependence on the colour matrix of the non-collinear partons
(see Eq.~(\ref{ic12gensym}) and also Sect.~\ref{sec:irone}). 
In the TL collinear region \cite{Kosower:1999xi, Catani:2003vu}
strict collinear factorization is recovered,
and $\sp^{(1)}$ is universal (i.e., independent of the non-collinear partons);
thus, we can write: 
\beq
\label{colsp1mtl}
\sp^{(1)}(p_1,\dots,p_m;{\widetilde P};p_{m+1},\dots,p_n) =
\sp^{(1)}(p_1,\dots,p_m;{\widetilde P})  \;\;, 
\quad ({\rm TL \,\;coll. \,\;lim.}) \;\;, 
\eeq
where, precisely speaking, the notation means that $\sp^{(1)}$ is proportional to
the unit matrix in the colour subspace of the non-collinear partons.

As recalled in Sect.~\ref{sec:tree}, collinear factorization of QCD amplitudes
is usually presented in a colour-stripped formulation,
which is obtained upon decomposition  of the matrix element $\cm$
in colour subamplitudes. In this formulation, the singular behaviour of the
colour subamplitudes in the multiparton collinear limit is described
by splitting amplitudes 
${\rm Split}^{(0)}(p_1,\dots,p_m;{\widetilde P})$
(at the tree level) and ${\rm Split}^{(1)}$ (at the one-loop order).
The splitting amplitudes can be regarded as matrices in helicity space, since
they depend on the helicity states of the collinear partons.
The splitting matrix $\sp$ in Eqs.~(\ref{factmt}) and (\ref{factmonegen})
is a generalization of the splitting amplitude, since $\sp$ describes collinear
factorization in colour+helicity space.
In the case of the two-parton collinear limit, there is a 
straightforward direct proportionality (through a single colour matrix) between
the splitting matrix $\sp^{(0)}$ and the splitting amplitude 
${\rm Split}^{(0)}$ (see Eq.~(\ref{spvssplit})).
Considering the collinear limit of $m$ partons, 
with $m \geq 3$, the corresponding splitting matrix $\sp^{(0)}$ can get
contributions from different colour structures.
In general,
$\sp^{(0)}$ can be expressed as a linear combination of colour-matrix factors
whose coefficients are
kinematical splitting amplitudes ${\rm Split}^{(0)}$. Equivalently, the 
splitting amplitudes ${\rm Split}^{(0)}$ can be regarded (defined) 
as colour-stripped components of the splitting
matrix $\sp^{(0)}$. This correspondence between 
$\sp$ and ${\rm Split}$ functions extends from the tree-level to one-loop
order in the TL collinear region. One-loop splitting amplitudes can be 
introduced also in the SL region (see Appendix~\ref{sec:appa}),
by properly taking into account 
the violation of strict collinear factorization
and the ensuing colour entanglement between collinear and non-collinear partons.

At the tree level, the multiparton $(m \geq 3)$ collinear limit has
been extensively studied in the literature. In the case of
$m=3$ collinear partons, explicit results for all QCD splitting processes
$A \to A_1 A_2 A_3$ are known for both squared amplitudes
\cite{Campbell:1997hg, Catani:1998nv, Catani:1999ss}
and amplitudes 
\cite{DelDuca:1999ha, Birthwright:2005ak, Birthwright:2005vi}.
At the amplitude level, the multiparton collinear limit
is explicitly known also in the cases of
$m=4$ \cite{DelDuca:1999ha, Birthwright:2005ak}
and $m=5$~and~6 \cite{Birthwright:2005ak} gluons.
Considering some specific classes of helicity configurations
of the collinear partons, the authors of 
Refs.~\cite{Birthwright:2005ak, Birthwright:2005vi} derived general results 
for the splitting amplitude  ${\rm Split}^{(0)}$ that are valid for 
an arbitrary number $m$ of gluons and of gluons plus up to four fermions
$(q, q{\bar q}, qq{\bar q}, qq{\bar q}{\bar q})$.

At the one-loop level, the multiparton collinear limit in the TL region
was studied in Ref.~\cite{Catani:2003vu}: we explicitly computed the one-loop
splitting matrix for the triple collinear limit 
$q \to q_1 q_2^{\,\prime} {\bar q}_3^{\,\prime}$
($q$ and $q^{\,\prime}$ are quarks with different flavour), and we presented
the general structure of the IR and ultraviolet divergences of the one-loop
splitting matrices. The latter result is recalled in Sect.~\ref{sec:irone}, 
where it is also extended to the SL collinear region.

\subsection{Generalized collinear factorization at all orders}
\label{sec:allorder}

The TL collinear limit of all-order QCD amplitudes is studied in 
Ref.~\cite{Kosower:1999xi}, by using the unitarity sewing method 
\cite{Bern:1993qk, Bern:1996je}. The detailed analysis 
of Ref.~\cite{Kosower:1999xi}
is limited to the leading-colour terms, but it can be extended to 
subleading-colour contributions as shown by the explicit computations of 
splitting amplitudes at one-loop \cite{Kosower:1999rx} 
and two-loop \cite{Bern:2004cz}
orders. The collinear limit can be studied by using a different method
\cite{Catani:1999ss}
that relies on the factorization properties of Feynman diagrams in light-like
axial gauges. By exploiting colour-coherence of QCD radiation, this method,
which directly applies in colour space, can be extended \cite{Catani:2003vu}
from tree-level to one-loop amplitudes in the TL collinear region.
In Eq.~(\ref{factallL}), we propose an all-order factorization formula that is
valid in both the TL and SL collinear regions. In the TL collinear limit, 
Eq.~(\ref{factallL}) represents a colour-space generalization of the all-order
results of Ref.~\cite{Kosower:1999xi} or, 
similarly,
an all-order
generalization of the colour-space factorization of 
Refs.~\cite{Catani:1999ss, Catani:2003vu}. The extension from the TL to the SL
collinear regions is based on the generalized factorization structure (and the
physical insight) that we have uncovered at the one-loop order
(see Sects.~\ref{sec:genlim} and \ref{sec:multi}).
In Sect.~\ref{sec:bey2}, we show that Eq.~(\ref{factallL}) is consistent 
with the
all-order factorization structure of the IR divergences of QCD amplitudes.

The generalized factorization formula for the multiparton collinear limit
of the {\em all-order} matrix element $\cm$ is
\beq
\label{factallL}
\ket{\cm} \;
\simeq \;\sp(p_1,\dots,p_m;{\widetilde P};p_{m+1},\dots,p_n) 
\;\;\ket{\cmbar} \;\;,
\eeq
where $\sp$ is the all-order splitting matrix. The loop expansion of the
unrenormalized splitting matrix is
\beeq
\label{loopexsp}
&&\!\!\!\!\!\!\!\!\!\!\!\!\!\sp(p_1,\dots,p_m;{\widetilde P};p_{m+1},\dots,p_n) = 
\sp^{(0)}(p_1,\dots,p_m;{\widetilde P}) \nn \\
&&\!\!\!\!+ \;\sp^{(1)}(p_1,\dots,p_m;{\widetilde P};p_{m+1},\dots,p_n) 
+ \sp^{(2)}(p_1,\dots,p_m;{\widetilde P};p_{m+1},\dots,p_n)  + \dots \;,
\eeeq
where $\sp^{(1)}$ is its one-loop contribution, 
$\sp^{(2)}$ is the two-loop splitting matrix, and so forth.
The loop expansion of the matrix element $\cm$ is defined in 
Eq.~(\ref{loopexnog}),
and an analogous expression (it is obtained by simply replacing
$\cm^{(k)}$ with $\cmbar^{(k)}$) applies to the reduced matrix element
$\cmbar$. Inserting these expansions in Eq.~(\ref{factallL}), we obtain
factorization formulae that are valid order-by-order in the number of loops.
At the tree level and one-loop order we recover Eqs.~(\ref{factmt})
and (\ref{factmonegen}), respectively. The explicit factorization 
formula for the
{\em two-loop} matrix element $\cm^{(2)}$ is 
\beeq
\label{factwogen}
\ket{\cm^{(2)}}
&\simeq& \sp^{(2)}(p_1,\dots,p_m;{\widetilde P};p_{m+1},\dots,p_n) 
\;\;\ket{\cmbar^{(0)}} \nn \\
&+& \sp^{(1)}(p_1,\dots,p_m;{\widetilde P};p_{m+1},\dots,p_n) 
\;\;\ket{\cmbar^{(1)}} \nn \\
&+& \sp^{(0)}(p_1,\dots,p_m;{\widetilde P}) 
\;\;\ket{\cmbar^{(2)}}
\;\;.
\eeeq

In the case of the TL collinear limit, strict factorization is valid and
the splitting matrix is process independent at each order in the loop
expansion; we can remove the reference to the non-collinear partons, and we can
simply write
\beq
\label{colsptlallL}
\sp(p_1,\dots,p_m;{\widetilde P};p_{m+1},\dots,p_n) =
\sp(p_1,\dots,p_m;{\widetilde P})  \;\;, 
\quad ({\rm TL \,\;coll. \,\;lim.}) \;\;. 
\eeq
Beyond the tree level, in the case of the SL collinear limit,
the splitting matrix $\sp$ acquires also a dependence on the external
{\em non-collinear} partons of $\cm$ and $\cmbar$, although it is 
(expected to be) independent of the spin polarization states of these
partons.

The all-order matrix elements $\cm$ and $\cmbar$ in Eq.~(\ref{factallL})
are invariant under the renormalization procedure, since they are scattering
amplitudes whose external partons are on-shell and with physical polarization
states. Therefore, the splitting matrix $\sp$ (the remaining ingredient in
Eq.~(\ref{factallL})) is also invariant. The renormalization of 
$\cm, \cmbar$ and $\sp$ simply amounts to replace the bare coupling constant
with the renormalized QCD coupling. 
The perturbative
(loop) expansion with respect to the renormalized coupling is denoted as
follows:
\beq
\label{loopexmren}
\cm = 
\cm^{(0, R)} 
+ \cm^{(1, R)} + \cm^{(2, R)}  + \dots\;\;,
\eeq
\beq
\label{loopexspren}
\sp = 
 \sp^{(0, R)} 
+ \sp^{(1, R)} + \sp^{(2, R)}  + \dots\;\;,
\eeq
where the superscripts $(k, R)$ ($k=0,1,2,\dots$) refer to the renormalized
expansions, whereas the superscripts $(k)$ refer to the 
unrenormalized expansions in the corresponding
Eqs.~(\ref{loopexnog}) and (\ref{loopexsp}).
Since perturbative renormalization {\em commutes} with the collinear limit, the
perturbative factorization formulae 
(\ref{factmt}), (\ref{factmonegen}) and (\ref{factwogen})
can equivalently be written by using the renormalized expansion. We simply have:
\beeq
\label{facttren}
\ket{\cm^{(0,R)}} &\simeq& \sp^{(0,R)} \;\ket{\cmbar^{(0,R)}} 
\;\;, \\
\label{factoren}
\ket{\cm^{(1,R)}} &\simeq& \sp^{(1,R)} \;\ket{\cmbar^{(0,R)}} +
\sp^{(0,R)} \;\ket{\cmbar^{(1,R)}}
\;\;, \\
\label{facttworen}
\ket{\cm^{(2,R)}} &\simeq& \sp^{(2,R)} \;\ket{\cmbar^{(0,R)}} 
+ \sp^{(1,R)} \;\ket{\cmbar^{(1,R)}} + \sp^{(0,R)} 
\;\ket{\cmbar^{(2,R)}} \;\;.
\eeeq

The main features of the results that we present in the following sections 
do not depend on the specific renormalization procedure. To avoid possible
related ambiguities, 
we specify 
the renormalization (and regularization) procedure that we actually use 
in writing explicit expressions.
The unrenormalized quantities are evaluated in $d=4-2\ep$ dimensions by 
using\footnote{The relation between
different regularization schemes within dimensional regularization can be found
in Refs.~\cite{Kunszt:1993sd, Catani:1996pk, Bern:2002zk}.}
{\em conventional dimensional regularization} (CDR);
in particular, the parton momenta are $d$ dimensional, and the gluon has 
$d-2=2-2\ep$ physical polarization states.
The renormalized QCD coupling at the scale $\mu_R$ is denoted by
$\g(\mu_R^2)$ ($\as=\g^2/(4\pi)$), and it is obtained from the 
unrenormalized (bare) coupling $\g$ by a modified minimal subtraction
($\msbar$) procedure.
We use the following explicit relation:
\beq
\label{msren}
\mu^{2\,\ep} \,\g^2  \;S_{\ep} = \mu_R^{2\,\ep} \;\g^2(\mu_R^2)  
\left[ 1 -  \f{\as(\mu_R^2)}{2\pi} \;\frac{\cbet0}{\ep}
+ \left( \f{\as(\mu_R^2)}{2\pi} \right)^2
\left( \frac{\cbet0^2}{\ep^2} - \frac{\bone^2}{2\ep}
\right) + {\cal O}(\as^3(\mu_R^2)) \right] \;\,,
\eeq 
where $S_{\ep}$ is the customary $\msbar$ factor 
($\gamma_E= - \psi(1) = 0.5772\dots$ is the Euler number),
\beq
\label{spher}
S_{\ep} = \exp \left[ \,\ep \,(\ln 4\pi + \psi(1)) \right] \;\;,
\eeq
and $\cbet0$ (see Eq.~(\ref{norbeta0})) 
and $\bone$ are the first two coefficients of 
the QCD $\beta$ function, with
\beq
\label{b1}
\bone = 
( 17\,C_A^2 - 5\,C_A N_f - 3\,C_F N_f )/6 \;\;.
\eeq

In the following, all the renormalized expressions refer to the perturbative
expansion with respect to
$\as(\mu^2)$ (i.e., the renormalization scale $\mu_R$ is always set to be equal
to the dimensional-regularization scale $\mu$).
For instance, considering the splitting matrix $\sp$ of $m$ collinear partons,
the tree-level and one-loop relations between the renormalized 
(see Eq.~(\ref{loopexspren}))
and unrenormalized (see Eq.~(\ref{loopexsp})) contributions are as follows:
\beq
\sp^{(0,R)} = 
\left[ \sp^{(0)} \right]_{\g  = \g(\mu^2) S_{\ep}^{\,-\f{1}{2}}} \;\;,
\eeq
\beq
\label{sp1ren}
\sp^{(1,R)} = 
- \,\f{\as(\mu^2)}{2\pi} \;\f{m-1}{2} \;\frac{\cbet0}{\ep} 
\;\sp^{(0,R)} \,+ \left[ \sp^{(1)} 
\right]_{\g  = \g(\mu^2) S_{\ep}^{\,-\f{1}{2}}} \;\;.
\eeq
The first term on the right-hand side of Eq.~(\ref{sp1ren}) originates from the
fact that the corresponding tree-level contribution $\sp^{(0)}$ is proportional
to $\g^{m-1}$.


In the case of the two-parton collinear limit, for later purposes,
we can write the renormalized
version of Eq.~(\ref{spgen}) in the following form:
\beq
\label{sp2oneren}
\sp^{(1,R)}(p_1,p_2;{\widetilde P};p_3,\dots,p_n) = 
\itc^{(1)}(\ep) \;
\sp^{(0,R)}(p_1,p_2;{\widetilde P}) 
+ \sp^{(1,R)}_{H}(p_1,p_2;{\widetilde P}) \;,
\eeq
\beeq
\label{it2cone}
\itc^{(1)}(\ep) &=& \f{\as(\mu^2)}{2\pi} \;\f{1}{2} \;\,{\widetilde c}_{\Gamma}
\left( \frac{-s_{12} -i0}{\mu^2} \right)^{-\ep} \nn \\
&\times&
\left\{  \; \frac{1}{\ep^2} \; \Bigl( C_{12} - C_1 - C_2 \Bigr)
+ \frac{1}{\ep} \; \Bigl( \gamma_{12} - \gamma_1 - \gamma_2 \Bigr)
\right. \nn \\
&-& \left. \; \frac{1}{\ep} \, \left[ \Bigl(C_{12} + C_1 - C_2 \Bigr) 
\;f_R(\ep;z_1)
 +  \Bigl(C_{12} + C_2 - C_1 \Bigr) \;f_R(\ep;z_2)\;
\right] \right.  \nn \\
&-& \;\;i \left. \,\frac{2}{\ep} \; \sum_{j=3}^n  \,\sum_{i=1,2}
\,{\bom T}_j \cdot {\bom T}_i \;\,{\rm sign}(s_{ij}) \;\,\Theta(- z_i)
\;\,f_I(\ep;z_i) \;\right\} \;\;,
\eeeq
where the coefficient ${\widetilde c}_{\Gamma}= 1 +{\cal O}(\ep^2)$ is related 
to the volume factor $c_{\Gamma}$ in Eq.~(\ref{cgammaf}) and it is defined as
\beq
\label{ctgamma}
{\widetilde c}_{\Gamma} \equiv \left( 4\pi \right)^2 \;S_{\ep}^{-1} \;c_{\Gamma}
= \frac{\Gamma(1+\epsilon)\, 
\Gamma^2(1-\epsilon)}{\Gamma(1-2\epsilon)} \;e^{- \ep \,\psi(1)} \;\;\;.
\eeq

The colour operator $\itc^{(1)}(\ep)$ is directly related to the corresponding 
unrenormalized operator ${\bom I}_C(p_1,p_2;p_3,\dots,p_n)$
(the functional dependence on the parton momenta is not explicitly 
recalled in the argument of $\itc^{(1)}$)
in Eq.~(\ref{ic12gensym}). In the expression on the right-hand side of
Eq.~(\ref{it2cone}), the colour correlation contributions proportional to
${\bom T}_j \cdot {\bom T}_i$ are written as in 
Eqs.~(\ref{d12genri})--(\ref{d12i}),
by using the decomposition of
$f(\ep;z)$ in its real and imaginary parts.
We also note that the contribution proportional to $\cbet0$ in ${\bom I}_C$
and the renormalization counterterm of $\sp^{(1,R)}$ (i.e. the first term
on the right-hand side of Eq.~(\ref{sp1ren})) have been included in the
definition of $\sp^{(1,R)}_{H}$. Thus, in Eq.~(\ref{sp2oneren}),  
$\sp^{(1,R)}_{H}$ has been defined as
\beeq
\sp^{(1,R)}_{H}(p_1,p_2;{\widetilde P}) &\equiv&
\f{\as(\mu^2)}{2\pi} \;\frac{\cbet0}{2\ep}
\left[ \,{\widetilde c}_{\Gamma} \left( \frac{-s_{12} -i0}{\mu^2} \right)^{-\ep}
-1 \right]
\;\sp^{(0,R)}(p_1,p_2;{\widetilde P}) \nn \\
&+& \left[ \sp^{(1)}_{H}(p_1,p_2;{\widetilde P}) 
\right]_{\g  = \g(\mu^2) S_{\ep}^{\,-\f{1}{2}}} \;\;\;\;.
\eeeq
The first term on the right-hand side is finite when $\ep \to  0$,
since the ultraviolet divergence of ${\bom I}_C$ is cancelled by the
renormalization of the one-loop splitting matrix. The remaining $\ep$ poles 
of $\sp^{(1,R)}$ are of IR origin, and they are included in the colour operator
$\itc^{(1)}(\ep)$.

\subsection{The IR divergences of the one-loop splitting matrix}
\label{sec:irone}

Analogously to the case of two-collinear partons, 
the one-loop splitting matrix $\sp^{(1, R)}$ of the multiparton collinear limit
in Eq.~(\ref{factoren}) (or Eq.~(\ref{factmonegen})) 
has IR divergences. They show up as double $(1/\ep^2)$ and single $(1/\ep)$
poles in the expansion around the point $\ep=0$.
To make the IR behaviour explicit, we separate the divergent and finite terms
as follows:
\beq
\label{sp1df}
\sp^{(1, R)} = \sp^{(1) \,{\rm div.}} + \sp^{(1) \,{\rm fin.}} \;\;,
\eeq
where $\sp^{(1) \,{\rm div.}}$ contains the $\ep$ poles, while
$\sp^{(1) \,{\rm fin.}}$ is finite when $\ep \to 0$. Then, we can present
our general result for the IR divergent part.

The computation of the divergent part shows that it can be written in a
colour-space factorized form:
\beq
\label{sp1div}
\sp^{(1) \,{\rm div.}}(p_1,\dots,p_m;{\widetilde P};p_{m+1},\dots,p_n) 
= \imc(\ep) \;\,\sp^{(0, R)}(p_1,\dots,p_m;{\widetilde P}) \;\;,
\eeq
where $\sp^{(0, R)}$ is the tree-level splitting matrix, 
{\em including}\footnote{The double pole $1/\ep^2$ in $\imc$ interferes with
the term of ${\cal O}(\ep)$ in $\sp^{(0, R)}$, thus contributing to the single
pole $1/\ep$ in $\sp^{(1) \,{\rm div.}}$.}
its complete (i.e. exact) dependence on $\ep$.
The colour space operator $\imc$ depends on the collinear partons and, in the SL
region, it also depends on the momenta and colour charges of the
non-collinear partons in the original matrix
element $\cm$. Analogously to the notation in Eq.~(\ref{sp2oneren}), the
functional dependence on all partons is not explicitly denoted in the 
argument of $\imc$. We present $\imc$ in a general form, which is valid in both
the SL and TL collinear limits. Considering the collinear splitting process
$A \to A_1 \dots A_m$, the explicit expression of $\imc$ 
is
\beeq
\label{i1mcexp}
\imc(\ep) 
&=&\f{\as(\mu^2)}{2\pi} \;\frac{1}{2} \;
\left\{ \left( \,\frac{1}{\ep^2} \,C_{\wp} + \frac{1}{\ep} \,\gamma_{\wp} \right)
- \sum_{\substack{i \,\in \, C \\ {}}}
\left( \,\frac{1}{\ep^2} \,C_i + \frac{1}{\ep} \,\gamma_i\right)
\right. \nn \\
&-& 
\frac{1}{\ep} \sum_{\substack{i,\ell \,\in C \\ i \,\neq\, \ell}} 
{\bom T}_i \cdot {\bom T}_\ell
\, \ln\left(\f{-s_{i \ell} - i0}{\mu^2}\right)
\left.
- \;\frac{2}{\ep} \sum_{\substack{i \,\in C \\ j \,\in NC}} 
{\bom T}_i \cdot {\bom T}_j
 \, \ln\bigl( z_i - i0 s_{ij} \bigr) 
 \right\} \;,
\eeeq
where the subscript ${\wp}$  refers to the parent collinear parton $A$.
The sets
\beq
	C= \{1, \dots, m \} \quad {\rm and} \quad  NC=\{m+1, \dots, n \}
	\nn
	\eeq
denote the collinear and non-collinear partons, respectively.

The expression (\ref{i1mcexp}) can be written in the following equivalent form:
\beeq
\label{i1mcexd}
\imc(\ep) 
&=&\f{\as(\mu^2)}{2\pi} \;\frac{1}{2} \;
\left\{ \left( \,\frac{1}{\ep^2} \,C_{\wp} + \frac{1}{\ep} \,\gamma_{\wp} \right)
- \sum_{\substack{i \,\in \, C \\ {}}}
\left( \,\frac{1}{\ep^2} \,C_i + \frac{1}{\ep} \,\gamma_i
- \,\frac{2}{\ep} \,C_i \,\ln |z_i| \right)
\right. \nn \\
&-& 
\frac{1}{\ep} \left.\sum_{\substack{i,\ell \,\in C \\ i \,\neq\, \ell}} 
{\bom T}_i \cdot {\bom T}_\ell
\, \ln\left(\f{- s_{i \ell} - i0}{|z_i| \, |z_{\ell}| \, \mu^2}\right)
 \right\} + \dmc(\ep) \;\;,
\eeeq
where
\beq
\label{del1}
\dmc(\ep) = \f{\as(\mu^2)}{2\pi} \,\;\f{i \pi}{\ep} \;
\sum_{\substack{i \,\in C \\ j \,\in NC}} 
{\bom T}_i \cdot {\bom T}_j \;\Theta(-z_i) \;{\rm sign}(s_{ij}) \;\;.
\eeq
The result in Eq.~(\ref{i1mcexp}) (or Eq.~(\ref{i1mcexd})) contains double and
single poles in
$1/\ep$. For the specific case of $m=2$ collinear partons, this result agrees
with the $\ep$ poles of the complete one-loop result in Eqs.~(\ref{sp2oneren})
and (\ref{it2cone}). In the general case of $m$ collinear partons, we can see
that $\imc$ embodies colour-correlation terms that produce violation of strict
collinear factorization. In the expression (\ref{i1mcexd}), these correlations
are fully taken into account by the colour operator $\dmc(\ep)$.
The expression (\ref{del1}) explicitly shows that the operator $\dmc$
is {\em antihermitian}.
In particular, $\dmc(\ep)$ is proportional to a single pole $1/\ep$ and
exactly corresponds to the colour-correlation terms that are produced by the
non-abelian Coulomb phase (see Sect.~\ref{sec:caus}).

The computation of the finite part of Eq.~(\ref{sp1df}),
in the case of the SL collinear limit of $m=3$ partons, is in progress.
In general, we can anticipate that $\sp^{(1) \,{\rm fin.}}$ also contains
factorization breaking contributions, and they are of the same type as those
in $\dmc$, namely, they have the form of two-parton
correlations ${\bom T}_i \cdot {\bom T}_j$ with $i \in C$ and $j \in NC$.

In the TL collinear region, the longitudinal-momentum fractions
$z_i$ are positive. Therefore, $\dmc$ vanishes and the result in 
Eq.~(\ref{i1mcexd}) has a strict-factorization form, which agrees 
with the result in Eq.~(11) of Ref.~\cite{Catani:2003vu}.
To be precise, Eq.~(11) of Ref.~\cite{Catani:2003vu} contains an additional
single pole
of UV origin (that expression refers to the unrenormalized splitting matrix)
and additional IR finite terms\footnote{The separation in Eq.~(\ref{sp1df})
is, in part, a matter of definition, since some IR finite terms can always be moved
from $\sp^{(1) \,{\rm fin.}}$ to $\sp^{(1) \,{\rm div.}}$.},
which include the ${\cal O}(\ep^0)$ terms of $\sp^{(1)}$ that explicitly depend
on the regularization scheme. In the TL collinear region, the expression 
(\ref{i1mcexd}) can also be rewritten in a slightly-simpler form, as follows:
\beeq
\label{i1mcexptl}
\!\!\!\!\!\!\!\!\!\!\!\!
\imc(\ep) 
&=&\f{\as(\mu^2)}{2\pi} \;\frac{1}{2} \;
\left\{ \left( \,\frac{1}{\ep^2} \,C_{\wp} + \frac{1}{\ep} \,\gamma_{\wp} \right)
- \sum_{\substack{i \,\in \, C \\ {}}}
\left( \,\frac{1}{\ep^2} \,C_i + \frac{1}{\ep} \,\gamma_i
- \,\frac{2}{\ep} \,C_i \,\ln z_i \right)
\right. \nn \\
&-& 
\frac{1}{\ep} \sum_{\substack{i,\ell \,\in C \\ i \,\neq\, \ell}} 
{\bom T}_i \cdot {\bom T}_\ell
\, \ln\left(\f{s_{i \ell}}{z_i z_{\ell} \,\mu^2}\right)
\left.
+ \;\frac{i \pi}{\ep}  
\left( \,C_{\wp} - \sum_{i \,\in \, C} \,C_i \right)
 \right\} \;,
\;\; ({\rm TL \,\;coll. \,\;lim.}) \,.
\eeeq
We can see that the antihermitian part of $\imc$ is simply an imaginary
$c$-number. Therefore, in the TL collinear limit, the non-abelian Coulomb phase
effectively takes an abelian form.

The derivation of the results in 
Eqs.~(\ref{sp1div}) and (\ref{i1mcexp}) is presented below.
As for the rewriting in Eqs.~(\ref{i1mcexd}) and (\ref{i1mcexptl}), it simply
follows from colour conservation properties.
Considering the last term on the right-hand side of Eq.~(\ref{i1mcexp}),
we can write 
$\ln\bigl( z_i - i0 s_{ij} \bigr)= 
\ln |z_i| - i \pi \Theta(-z_i) \;{\rm sign}(s_{ij})$ and then, using the
conservation of the {\em total} colour charge,
\beq
\sum_{j \, \in NC} \;{\bom T}_j = - \sum_{i \, \in C} \;{\bom T}_i \;\;,
\eeq
we directly obtain Eqs.~(\ref{i1mcexd}) and (\ref{del1}).
Considering the last term in the curly bracket on the right-hand side of 
Eq.~(\ref{i1mcexd}), we can write 
$\ln(-s_{i \ell} - i0)= \ln s_{i \ell} - i\pi$ ($s_{i \ell} > 0$ in the TL
collinear region), and then Eq.~(\ref{i1mcexptl}) is directly obtained by
using
$\;\sum_{i,\ell \,\in C} {\bom T}_i \cdot {\bom T}_\ell = C_{\wp}$
or, more precisely, by using the relation
\beq
\label{collchargecon}
\Bigl( \;\sum_{i \,\in C} {\bom T}_i \;\Bigr) \;\sp^{(0, R)}
= \sp^{(0, R)} \;{\bom T}_{\wp} \;\;.
\eeq
This relation (which generalizes Eq.~(\ref{colconsp}) to the case of
$m$ collinear partons) follows from the fact that the colour charge 
${\bom T}_{\wp}$ of the parent parton $A$ is conserved in the tree-level
collinear splitting $A \to A_1 \dots A_m$. 

Note that, in the case of the TL collinear
splitting, Eq.~(\ref{collchargecon}) remains valid when replacing 
the tree-level splitting matrix $\sp^{(0, R)}$
with the all-order splitting matrix $\sp$. 
We remark that this all-order generalization does not apply 
to the SL collinear limit. In the SL case,
owing to the violation of strict factorization, $\sp^{(1)}$, $\sp^{(2)}$
and so forth, contain charge interactions between collinear and non-collinear
partons: therefore, only their {\em total} charge is conserved
(and the equality in (\ref{collchargecon}) is not valid
beyond the tree level).

The results in Eqs.~(\ref{sp1div}) and (\ref{i1mcexp}) can be derived in a
simple way by exploiting the known IR structure of generic one-loop QCD
amplitudes \cite{Giele:1991vf, csdip, Catani:1998bh}.
Considering the scattering amplitude $\cm$ with $n$ external QCD partons (and
any number of colourless external legs), the renormalized one-loop contribution
can be written in the following factorized form \cite{Catani:1998bh}:
\beq
\label{m1ir}
\ket{\cm^{(1, R)}} = {\bom I}^{(1)}_M(\ep) \;
\ket{\cm^{(0, R)}} +  \;\ket{\cm^{(1) \,{\rm fin.}}} 
\;\;,
\eeq
where
\beq
\label{i1m}
{\bom I}^{(1)}_M(\ep) 
=  \f{\as(\mu^2)}{2\pi} \;\frac{1}{2} 
\;\Bigl\{ \;
- \sum_{i=1}^n \left( \frac{1}{\ep^2} \,C_i 
+ \frac{1}{\ep} \,\gamma_i\right)
- \frac{1}{\ep} \sum_{\substack{i,j=1 \\ i \neq j}}^n 
\;{\bom T}_i \cdot {\bom T}_j
\, \ln\left(\f{-s_{ij} - i0}{\mu^2}\right) 
\; \Bigr\}
\;\;,
\eeq
and the one-loop term $\ket{\cm^{(1) \,{\rm fin.}}}$ is finite when 
$\ep \to 0$. The singular dependence on $\ep$ is embodied in the factor 
${\bom I}^{(1)}_M(\ep)$ that acts as a colour-charge operator onto the
tree-level matrix element $\ket{\cm^{(0, R)}}$ (which retains its complete
dependence on $\ep$). 
The colour operator ${\bom I}^{(1)}_M$ is 
equivalent, though it is not exactly equal, to the operator ${\bom I}^{(1)}$
introduced in Ref.~\cite{Catani:1998bh}; the expression (\ref{i1m})
is obtained from ${\bom I}^{(1)}$ by removing all terms that are finite at
$\ep = 0$ (these terms are absorbed in the definition of 
$\ket{\cm^{(1) \,{\rm fin.}}}$).

The universality structure of the IR factorization formulae (\ref{m1ir})
and (\ref{i1m}) has direct consequences on the collinear limit of the 
scattering amplitudes. Indeed, Eq.~(\ref{m1ir}) can be applied to the matrix
element $\cm$ before performing the collinear limit, and an analogous relation
(which is obtained by the replacements $\cm \to \cmbar$ and 
${\bom I}^{(1)}_{M} \to {\bom I}^{(1)}_{\Mbar}\,$) applies to the reduced 
matrix element $\cmbar$. Therefore, we have:
\beeq
\label{facto1}
&&\!\!\!\!\!\!\!\!\!\!\!\!\!\!\!\!\!\!\!\!\!\!\! \sp^{(1,R)} 
\; \ket{\cmbar^{(0,R)}} \simeq 
\;\ket{\cm^{(1,R)}}
- \,\sp^{(0,R)} \;\ket{\cmbar^{(1,R)}} \\
\label{facto2} 
\!\!\!\! =&&\!\!\!\!\!\! {\bom I}^{(1)}_M(\ep) \;
\ket{\cm^{(0, R)}}  - \,\sp^{(0,R)} 
\;{\bom I}^{(1)}_{\Mbar}(\ep) \;
\ket{\cmbar^{(0, R)}} 
+ 
\;\ket{\cm^{(1) \,{\rm fin.}}}
- \,\sp^{(0,R)} \;\ket{\cmbar^{(1) \,{\rm fin.}}} \\
\label{facto3}
\!\!\!\! \simeq&&\!\!\!\!\!\! 
\left( {\bom I}^{(1)}_M(\ep) \;\sp^{(0,R)}
- \,\sp^{(0,R)} \;{\bom I}^{(1)}_{\Mbar}(\ep) \right) \;
\ket{\cmbar^{(0, R)}} 
+ \;\ket{\cm^{(1) \,{\rm fin.}}}
- \,\sp^{(0,R)} \;\ket{\cmbar^{(1) \,{\rm fin.}}} 
\,. 
\eeeq
Equation (\ref{facto1}) is just a rewriting of the one-loop collinear 
factorization in Eq.~(\ref{factoren}). Then, Eq.~(\ref{facto2}) is obtained by
using Eq.~(\ref{m1ir}) for both $\cm^{(1,R)}$ and $\cmbar^{(1,R)}$. Finally,
Eq.~(\ref{facto3}) is obtained by applying the tree-level factorization formula
(\ref{facttren}) to $\cm^{(0,R)}$. Performing these steps, we have only
neglected subdominant collinear terms (as denoted by the approximate 
equalities in Eq.~(\ref{facto1}) and (\ref{facto3})).
The last terms, $\ket{\cm^{(1) \,{\rm fin.}}}$ and 
$\sp^{(0,R)} \;\ket{\cmbar^{(1) \,{\rm fin.}}}$, on the right-hand side of
Eq.~(\ref{facto3}) do not contain any $\ep$ poles, and they are finite if 
$\ep \to 0$. Thus, in the collinear limit, we can write
\beq
\label{divone}
\sp^{(1,R)}  \;\simeq \;
 {\bom I}^{(1)}_M(\ep) \;\sp^{(0,R)}
- \,\sp^{(0,R)} \;{\bom I}^{(1)}_{\Mbar}(\ep)  
\;+  \;{\cal O}(\ep^0) 
\;\;. 
\eeq

The IR divergent contributions in Eqs.~(\ref{sp1div}) and (\ref{i1mcexp})
directly derive from the relation (\ref{divone}), by simply using colour-charge
conservation. To illustrate the derivation in detail, we note that, according 
to Eq.~(\ref{i1m}), the operator $\;{\bom I}^{(1)}_{M}$ on the right-hand side
of Eq.~(\ref{divone}) contains three classes of contributions: 
$(a)$ terms that only depend on the non-collinear partons; 
$(b)$ terms that only depend on the collinear partons in $\cm$;
$(c)$ terms that depend on both the collinear and non-collinear partons.
Correspondingly, the operator $\;{\bom I}^{(1)}_{\Mbar}$
contains the following classes of contributions:
$({\overline a})$ terms that only depend on the non-collinear partons; 
$({\overline b})$ terms that only depend on the parent collinear parton 
$A$ in $\cmbar$;
$({\overline c})$ terms that depend on both the parent parton
and the non-collinear partons.
The terms of the classes $(a)$ and $({\overline a})$ are exactly equal
and commute with $\sp^{(0,R)}$: therefore, they cancel on the right-hand side 
of Eq.~(\ref{divone}). 
The terms of the class $({\overline b})$ are proportional to either
$C_{\wp}$ or $\gamma_{\wp}$; they commute with $\sp^{(0,R)}$ and combine
with the terms of the class $(b)$, thus leading to the first three 
contributions in the curly bracket of Eq.~(\ref{i1mcexp}).
The last contribution (which depends on ${\bom T}_i \cdot {\bom T}_j$)
in the curly bracket of Eq.~(\ref{i1mcexp})
is due to the terms of the remaining classes $(c)$ and $({\overline c})$.
Indeed, these terms give the following contribution to the right-hand side of
Eq.~(\ref{divone}) (we omit an overall factor of $\as/(4\pi)$ and use the
notation $s_{j\wp}= 2p_j\cdot{\wp}$):
\beeq
\label{divcbarc}
&&- \, \frac{2}{\ep} \sum_{\substack{i\,\in C \\ j \,\in NC}}
 \;{\bom T}_j \cdot {\bom T}_i
\, \ln\left(\f{-s_{ji} - i0}{\mu^2}\right) \;\sp^{(0,R)}
+ \sp^{(0,R)} \;\frac{2}{\ep} \sum_{j \,\in NC}
 \;{\bom T}_j \cdot {\bom T}_{\wp}
\, \ln\left(\f{-s_{j\wp} - i0}{\mu^2}\right) \nn \\
&&\\
\label{divcbarc1}
&&\simeq
- \, \frac{2}{\ep} \sum_{\substack{i\,\in C \\ j \,\in NC}}
 \;{\bom T}_j \cdot {\bom T}_i
\,\ln \left(z_i- i0 s_{ji}\right)  \;\sp^{(0,R)} \nn \\
&&\;\;\;\; - \, \frac{2}{\ep} \sum_{j \,\in NC}
\; \ln\left(\f{-s_{j\wp} - i0}{\mu^2}\right) 
\left[
\;\sum_{i\,\in C}\;{\bom T}_j \cdot {\bom T}_i
\;\sp^{(0,R)}
- \sp^{(0,R)} 
 \;{\bom T}_j \cdot {\bom T}_{\wp} \right]
 \;\;.
\eeeq
In going from Eq.~(\ref{divcbarc}) to  Eq.~(\ref{divcbarc1}), we have performed
the collinear limit $p_i \simeq z_i \wp$ ($i \in C$), and we have thus used the
following collinear approximation:
\beeq
\label{logcol}
\ln \left(\f{-s_{ji} -i0}{\mu^2}\right) &\simeq&
\ln \left(\f{-z_i s_{j\wp} -i0}{\mu^2}\right) =
\ln \left(z_i+ i0 s_{j\wp}\right) + 
\ln \left(\f{- s_{j\wp} -i0}{\mu^2}\right) \nn \\
&\simeq& \ln \left(z_i- i0 s_{ji}\right) + 
\ln \left(\f{- s_{j\wp} -i0}{\mu^2}\right) \;\;.
\eeeq
The term in the square bracket of Eq.~(\ref{divcbarc1}) vanishes
(we recall that the non-collinear charge ${\bom T}_j$ commutes with 
$\sp^{(0,R)}$) because of the conservation of the collinear charge 
${\bom T}_{\wp}$ of the parent parton (see Eq.~(\ref{collchargecon})).
Therefore,  
the expression (\ref{divcbarc1})
exactly corresponds to the
contribution to Eq.~(\ref{sp1div})
from the last term in the curly bracket of Eq.~(\ref{i1mcexp}).
This completes the derivation of the results in 
Eqs.~(\ref{sp1div}) and (\ref{i1mcexp}).

The structure of Eqs.~(\ref{sp1df})--(\ref{i1mcexp})
and 
its derivation agree with the discussion in Sect.~\ref{sec:caus}.
In particular, the pairwise interaction terms of the classes 
$(c)$ (i.e. the first contribution in Eq.~(\ref{divcbarc}))
and $({\overline c})$ (i.e. the second contribution in Eq.~(\ref{divcbarc}))
include the IR divergent part of the one-loop contributions due to the soft and
wide-angle region of the loop momentum.
The terms of the class $({\overline c})$ correspond to one-loop 
contributions to $\cm$ that have been removed from the one-loop
splitting matrix $\sp^{(1,R)}$ and re-factorized
in $\cmbar$ (see Eqs.~(\ref{facto1}) and (\ref{divone})). 
These terms, which are rewritten as second contribution in the square bracket of
Eq.~(\ref{divcbarc1}), are effectively equal to part of the contribution of the
class $(c)$: this part (i.e. the first contribution in the square bracket of
Eq.~(\ref{divcbarc1})) is due to the one-loop interactions at wide angle with
respect to the direction of the system of collinear partons in $\cm$.
The vanishing of the square bracket term in Eq.~(\ref{divcbarc1}) is produced by
the coherent action of the collinear-parton system with respect to
non-factorizable wide-angle interactions with each of the non-collinear partons.
This colour-coherence mechanism guarantees strict factorization in the TL
collinear limit. 
The remaining part of the terms of the class $(c)$
(i.e. the first term in Eq.~(\ref{divcbarc1})) includes 
an absorptive (imaginary) component and 
a radiative (real) component.
Owing to their causality structure, these remaining absorptive contributions
are non-vanishing only in the SL collinear limit (i.e., $z_i < 0$): 
the absorptive part of the one-loop interactions between collinear 
and non-collinear partons
produces the violation of strict collinear factorization.
The remaining radiative contributions do not violate 
strict collinear factorization,
since the non-collinear partons act coherently as a single parton, whose colour
charge is equal (modulo the overall sign) to the total colour charge of the
collinear partons.

To conclude this section, we rewrite the steps in 
Eqs.~(\ref{divone})--(\ref{logcol})
by using a slightly different and more compact notation.
This rewriting anticipates the notation that we use in some of the following
sections 
(e.g., Sect.~\ref{sec:IRall} and Appendix~\ref{sec:appb}).
We consider the right-hand side of Eq.~(\ref{divone}),
and we write it in the following form:
\beeq
\label{divonere}
{}\!\!\!\!\!\!\!\!\!\!\!\! {\bom I}^{(1)}_M(\ep) \;\sp^{(0,R)}
-\,\sp^{(0,R)} \;{\bom I}^{(1)}_{\Mbar}(\ep)  
\;+  \;{\cal O}(\ep^0) &\simeq&
 \;{\bom I}^{(1)}(\ep) \;\sp^{(0,R)}
- \,\sp^{(0,R)} \;{\bom {\overline I}}^{(1)}(\ep)
\\
\label{divone2}
&=& \Bigl[ \;{\bom I}^{(1)}(\ep) - {\bom {\overline I}}^{(1)}(\ep) \;
\Bigr] \;\sp^{(0,R)}
\;\;. 
\eeeq
The right-hand side of Eq.~(\ref{divonere}) is obtained by neglecting IR finite
contributions of
${\cal O}(\ep^0)$, performing the collinear limit of the operator 
${\bom I}^{(1)}_M$ and defining the operator ${\bom {\overline I}}^{(1)}$;
our notation is
\beq
\label{ionedef}
\quad \quad \quad \quad  \quad
{\bom I}^{(1)}_M(\ep) \;\simeq \;{\bom I}^{(1)}(\ep) \;\;,
\quad \quad \;(s_{ij} \simeq z_i \,s_{j\wp}, \;\;i \in C, \; j \in NC) \;\;,
\eeq
\beq
\label{ibaronedef}
\!\!\!\!\!\!\!\!\!
\Bigl[ \;{\bom I}^{(1)}_{\Mbar}(\ep) \;
\Bigr]_{{\bom T}_{\wp}= \,- \sum_{j \in NC} {\bom T}_j}
 \equiv \;\;{\bom {\overline I}}^{(1)}(\ep) \;\;.
\eeq
The operator ${\bom I}^{(1)}$ in Eq.~(\ref{ionedef}) is obtained from 
${\bom I}^{(1)}_M$ in Eq.~(\ref{i1m})
by implementing the collinear approximation
$p_i \simeq z_i \wp$ $\;(i \in C)$ in all the terms of 
${\bom I}^{(1)}_M$ that are not singular in the collinear limit. The explicit
expression of ${\bom I}^{(1)}(\ep)$ is derived by using collinear relations as
in Eqs.~(\ref{divcbarc1}) and (\ref{logcol}), and it is presented in 
Eq.~(\ref{ionefor}) of
Appendix~\ref{sec:appb}. The IR operator 
${\bom I}^{(1)}_{\Mbar}$ of the reduced matrix element
$\cmbar$ depends on the colour charges ${\bom T}_j$ of the 
non-collinear partons and on the colour charge ${\bom T}_{\wp}$
of the parent collinear parton. The definition of 
${\bom {\overline I}}^{(1)}$ in Eq.~(\ref{ibaronedef})
simply amounts to the implementation of the colour conservation relation
${\bom T}_{\wp}= \,- \sum_{j \in NC} {\bom T}_j$.
The operator ${\bom {\overline I}}^{(1)}$ is obtained from 
${\bom I}^{(1)}_{\Mbar}$ by replacing
the colour matrix  
$- {\bom T}_{\wp}$ with the sum of the non-collinear charges.
This replacement produces the explicit expression of 
${\bom {\overline I}}^{(1)}$ that is presented in Eq.~(\ref{ibaronefor})
of Appendix~\ref{sec:appb}. According to this definition,
the matrix structure of
${\bom {\overline I}}^{(1)}$ only depends on the colour charges of 
the non-collinear partons: therefore, ${\bom {\overline I}}^{(1)}$
commutes with $\sp^{(0,R)}$ (since $\sp^{(0,R)}$ does not depend on the
non-collinear partons), and the commutation leads to Eq.~(\ref{divone2}).
The IR divergent part 
$\sp^{(1) \,{\rm div.}}$ of the one-loop splitting matrix
$\sp^{(1,R)}$ in Eq.~(\ref{sp1df})
is obtained by performing the collinear limit of Eq.~(\ref{divone})
and removing the IR finite terms of ${\cal O}(\ep^0)$.
These steps are carried out in Eqs.~(\ref{divonere}) and (\ref{divone2}).
Equating the relations (\ref{sp1div}) and (\ref{divone2}),
we thus obtain the following representation of the IR divergent operator 
$\imc$:
\beq
\label{i1mc}
\imc(\ep) = {\bom I}^{(1)}(\ep) - {\bom {\overline I}}^{(1)}(\ep) \;\;.
\eeq
The explicit expressions (\ref{ionefor}) and (\ref{ibaronefor})
of ${\bom I}^{(1)}$ and ${\bom {\overline I}}^{(1)}$ can be inserted in 
Eq.~(\ref{i1mc}), and we can directly check that we reobtain the result in
Eq.~(\ref{i1mcexp}).

\setcounter{footnote}{2}
\section{
All-order structure of the collinear limit and 
\\
space-like two-loop results}
\label{sec:bey2}

\subsection{The IR structure of the splitting matrix
}
\label{sec:IRall}

The IR structure of multiloop QCD amplitudes is not independent of their
collinear behaviour. This fact can be exploited to extract non-trivial
information \cite{Catani:2003vu, Bern:2004cz, Becher:2009qa, Dixon:2009ur}.
In Sect.~\ref{sec:irone}, 
the IR divergences of the one-loop splitting matrix 
$\sp^{(1,R)}$ were extracted from the known IR structure of the one-loop
QCD amplitudes. In this section, our study of the collinear limit is extended
beyond the one-loop order.

We consider a generic scattering amplitude $\cm$ with $n$ external QCD partons 
(and any number of colourless external legs). The IR structure of $\cm$
at two-loop order is known in explicit form, and it is given by the following
colour-space factorization formula \cite{Catani:1998bh}:
\beq
\label{m2ir}
\ket{\cm^{(2, R)}} = {\bom I}^{(2)}_M(\ep) \;\ket{\cm^{(0, R)}}
+ {\bom I}^{(1)}_M(\ep) \;
\ket{\cm^{(1, R)}} +  \;\ket{\cm^{(2) \,{\rm fin.}}} 
\;\;,
\eeq
where the contributions 
$\cm^{(k, R)}$
($k=0,1,2$) refer to the
renormalized expansion of the scattering amplitude $\cm$
(see Eq.~(\ref{loopexmren})).
The two-loop term 
$\cm^{(2) \,{\rm fin.}}$ is finite when $\ep \to 0$.
The colour-charge operator ${\bom I}^{(1)}_M$ is the IR divergent operator that
controls the one-loop factorization formula (\ref{m1ir}).
The IR divergent two-loop operator ${\bom I}^{(2)}_M$ can be written in terms 
of ${\bom I}^{(1)}_M$ in the following form \cite{Catani:1998bh, Aybat:2006wq}:
\beeq
\label{i2m}
{\bom I}^{(2)}_M(\ep) 
&=&  -\, \f{1}{2} \left[ {\bom I}^{(1)}_M(\ep) \right]^2
+ \f{\as(\mu^2)}{2\pi} 
\left\{ + \frac{1}{\ep} \,\cbet0 
\left[ {\bom I}^{(1)}_M(2\ep) - {\bom I}^{(1)}_M(\ep)\right]
+ K \;{\bom I}^{(1)}_M(2\ep) \right\} \nn \\
&+& 
\left(\f{\as(\mu^2)}{2\pi}\right)^2  \,\frac{1}{\ep} \;
\sum_{i=1}^{n} 
\; H^{(2)}_i \;\;.
\eeeq
Note that ${\bom I}^{(1)}_M$ appears on the right-hand side with two different
arguments, namely, ${\bom I}^{(1)}_M(\ep)$ and ${\bom I}^{(1)}_M(2\ep)$.
The expression (\ref{i2m}) includes IR poles of the type $1/\ep^n$, with
$n=4,3,2,1$. The dominant pole terms, $1/\ep^4$ and  $1/\ep^3$, are fully
controlled by ${\bom I}^{(1)}_M$. The double-pole terms,  $1/\ep^2$, also
depend on the value of the coefficient $K$ \cite{Catani:1998bh}:
\beq
\label{kcoef}
K = \left( \frac{67}{18} - \frac{\pi^2}{6} \right) C_A - \frac{5}{9} \,N_f
\;\;.
\eeq
The control of the single-pole terms, $1/\ep$, also requires the knowledge
of the $c$-number coefficients $H^{(2)}_i$  \cite{Aybat:2006wq}, which
depend on the flavour $i$ (quark, antiquark or gluon) of the parton with 
momentum $p_i$. The value of $H^{(2)}_i$ can be extracted 
\cite{Catani:1998bh, Aybat:2006wq} from explicit computations of 
two-loop amplitudes;
the quark (antiquark) coefficient $H^{(2)}_q$ is
\cite{Aybat:2006wq, Catani:1998bh, Anastasiou:2000kg} 
\beeq
\label{h2q}
H^{(2)}_q = H^{(2)}_{\bar q} &=& \frac{1}{4}
\left(\frac{\pi^2}{2}-6 ~\zeta_3 
-\frac{3}{8}\right) C_F^2
+\frac{1}{8}
\left( 13 ~\zeta_3 +\frac{245}{108}-\frac{17}{12} ~\pi^2 \right) C_A C_F
\nonumber \\
&+& 
\frac{1}{8} \left(-\frac{25}{54}+\frac{\pi^2}{6} \right) N_f C_F \;\; , 
\eeeq
and the gluon coefficient $H^{(2)}_g$ is
\cite{Aybat:2006wq, Anastasiou:2001sv} 
\beeq
\label{h2g}
H^{(2)}_g &=&  \frac{5}{108} \, N_f^2
+ \frac{1}{8} \, C_F N_f
+ \frac{1}{24}\left(\frac{ \pi^2}{6}-\frac{58}{9} \right)  N_f C_A
\nonumber \\ &&
+\frac{1}{24}\left(3 \zeta_3 +\frac{5}{2}
-\frac{11}{12} ~\pi^2 \right) C_A^2 \;\; . 
\eeeq

The expression (\ref{i2m}) of the two-loop operator ${\bom I}^{(2)}_M(\ep)$
is similar (and equivalent) to that of the operator ${\bom I}^{(2)}$ in
Ref.~\cite{Catani:1998bh}: the differences eventually amount to IR finite
contributions to Eq.~(\ref{m2ir}) that are included in the definition of 
$\cm^{(2) \,{\rm fin.}}$.
The essential difference is due to the fact that ${\bom I}^{(2)}_M(\ep)$
{\em and} ${\bom I}^{(1)}_M(\ep)$ include {\em only} contributions from IR
poles $1/\ep^n$ ($n \geq 1)$, with no additional IR finite contributions.
The key remark of Ref.~\cite{Aybat:2006wq} is that this `minimal form' 
of the IR divergent operators ${\bom I}^{(1)}_M$ and ${\bom I}^{(2)}_M$
highly simplifies the expression of ${\bom I}^{(2)}_M$. The single-pole
coefficients $H^{(2)}_i$ in Eqs.~(\ref{i2m}) are $c$-numbers 
\cite{Aybat:2006wq}, whereas the form of ${\bom I}^{(2)}$ in 
Ref.~\cite{Catani:1998bh} includes more complex colour-matrix correlations at
${\cal O}(1/\ep)$ (see Appendix~A.3 in Ref.~\cite{Bern:2004cz}).
We note that the coefficients $H^{(2)}_i$ in Eqs.~(\ref{i2m}), (\ref{h2q})
and (\ref{h2g}) are directly related to the coefficients 
$E^{\,[i]\,(2)}_1$
in the second paper of Ref.~\cite{Aybat:2006wq} (see Eq.~(3.9) therein);
the precise relation is
\beq
\label{hvse}
H^{(2)}_i = 4 \, E^{\,[i]\,(2)}_1 + \frac{1}{4}\, K \gamma_i \;\;.
\eeq

The colour-space factorization formula (\ref{m2ir}) can directly be exploited 
to extract explicit information (see Sects.~\ref{sec:2part2loop} and 
\ref{sec:mpart2loop})
on the collinear limit of two-loop QCD amplitudes. The procedure
(which is illustrated in Appendix~\ref{sec:appb}) 
is similar to that used at one-loop order in
Sect.~\ref{sec:irone}. This procedure can be extended beyond the two-loop 
order, as discussed below.

The one-loop and two-loop IR factorization formulae (\ref{m1ir}) and
(\ref{m2ir}) can be iterated to higher-loop orders. The {\em all-order}
generalization is
\beq
\label{mallir}
\ket{\cm } = {\bom I}_M(\ep) \;
\ket{\cm } +  \;\ket{\cm^{\,{\rm fin.}}} 
\;\;,
\eeq
where the operator ${\bom I}_M(\ep)$ is IR divergent, while the matrix element
term $\cm^{\,{\rm fin.}}$ is IR finite. The perturbative (loop) expansions of 
${\bom I}_M$ and  $\cm^{\,{\rm fin.}}$ in terms of the renormalized QCD coupling
are
\beq
\label{imall}
{\bom I}_M(\ep) = \;{\bom I}^{(1)}_M(\ep) + {\bom I}^{(2)}_M(\ep) +
{\bom I}^{(3)}_M(\ep) + \cdots \;\;,
\eeq
\beq
\label{mfinall}
\cm^{\,{\rm fin.}} = \;\cm^{(0,R)} + \cm^{(1)\,{\rm fin.}} +
\cm^{(2)\,{\rm fin.}} + \cm^{(3)\,{\rm fin.}} + \cdots \;\;,
\eeq
where $\cm^{(0,R)}$ is the (complete) tree-level expression of the matrix
element $\cm$.
Inserting Eqs.~(\ref{loopexmren}),
(\ref{imall}) and (\ref{mfinall})
in Eq.~(\ref{mallir}) and performing the loop expansion,
we recover Eqs.~(\ref{m1ir}) and (\ref{m2ir}),
and we obtain corresponding IR factorization formulae at three-loop order and 
higher-order levels.

The recursive structure of Eq.~(\ref{mallir}) can be rewritten in the following
form:
\beq
\label{mallfact}
\ket{\cm } = {\bf V}_M(\ep) 
\;\ket{\cm^{\,{\rm fin.}}} 
\;\;,
\eeq
where the all-order IR factor ${\bf V}_M(\ep)$ only depends on ${\bom I}_M(\ep)$;
the inverse operator ${\bf V}_M^{-1}$ simply is
\beq
\label{vvsi}
{\bf V}_M^{-1}(\ep) = 1 - {\bom I}_M(\ep) \;\;.
\eeq 

\setcounter{footnote}{2}
We can also express the IR factor ${\bf V}_M$ in exponential
form:
\beq
\label{vexp}
{\bf V}_M(\ep) = \exp \Bigl\{ \,{\bom I}_{M,\,{\rm cor}}(\ep) \Bigr\} \;\;,
\eeq
where the relation between ${\bom I}_{M,\,{\rm cor}}$ and ${\bom I}_M$ is
\beq
\label{icorvsi}
\exp \Bigl\{ \,-\, {\bom I}_{M,\,{\rm cor}}(\ep) \Bigr\} \equiv
 1 - {\bom I}_M(\ep) \;\;.
\eeq
The perturbative expansion of the operator ${\bom I}_{M,\,{\rm cor}}$
is
\beq
\label{imcorall}
{\bom I}_{M,\,{\rm cor}}(\ep) = \;{\bom I}^{(1)}_{M,\,{\rm cor}}(\ep) 
+ {\bom I}^{(2)}_{M,\,{\rm cor}}(\ep) +
{\bom I}^{(3)}_{M,\,{\rm cor}}(\ep) + \cdots \;\;,
\eeq
and the perturbative contributions ${\bom I}^{(k)}_{M,\,{\rm cor}}$
are directly related to the corresponding contributions in Eq.~(\ref{imall})
(in particular, ${\bom I}^{(1)}_{M,\,{\rm cor}}= {\bom I}^{(1)}_M$).
Owing to the definition (\ref{icorvsi}), the perturbative terms of the
exponentiated operator $-\,{\bom I}_{M,\,{\rm cor}}$ give the
irreducible-correlation component 
(in a statistical language)
of the perturbative
expansion of the operator  $-{\bom I}_M$.
 
In the context of the IR structure of multiparton scattering amplitudes, the
exponentiated representation in Eq.~(\ref{vexp}) exists and is particularly
suitable in view of the physical property of exponentiation of the IR divergent
contributions to QCD scattering amplitudes with {\em multiple} 
(i.e. $n \geq 3$) external
legs \cite{ leadingIR, Sterman:2002qn}.
IR exponentiation means that the {\em dominant} IR divergences 
at high perturbative
order are directly captured by simply exponentiating the IR divergent terms that
appear at lower orders. This also implies that the exponent function is less IR
divergent than the exponential function. Using dimensional regularization,
the dominant IR divergence of the operator ${\bom I}_M$ at the $n$-th
perturbative order (see Eq.~(\ref{imall})) is
${\bom I}_M^{(n)}(\ep) \sim \as^n/\ep^{2n}$, whereas the perturbative expansion
(\ref{imcorall}) of the exponentiated operator 
${\bom I}_{M,\,{\rm cor}}$ has a less singular IR behaviour of the type
${\bom I}^{(n)}_{M,\,{\rm cor}}(\ep) \sim \as^n/\ep^{n+1}$.
The operator ${\bom I}_{M,\,{\rm cor}}(\ep)$ has a compact all-order integral
representation that is given in terms of a perturbatively-computable kernel of
soft and collinear anomalous dimensions 
\cite{Sterman:2002qn, Aybat:2006wq, Dixon:2008gr, Becher:2009cu, Gardi:2009qi,
Becher:2009qa, Dixon:2009ur}.
We do not use this integral representation in the present paper.

We have briefly recalled some results on the IR structure
of the QCD scattering amplitudes.
These results are sufficient for the following discussion of the multiparton
collinear limit. Considering the collinear limit of $m$ parton momenta 
$\{p_1, \dots, p_m\}$ of the all-order matrix element $\cm$ with $n$ external QCD
partons, we obtain
\beeq
\label{mallfin}
\ket{\cm^{\,{\rm fin.}}} &=& 
{\bf V}_M^{-1}(\ep)
\;\ket{\cm }
\simeq {\bf V}_M^{-1}(\ep)
\;\,\sp
\;\,\ket{\cmbar } \\
\label{mallfin2}
&=& {\bf V}_M^{-1}(\ep)
\;\,\sp
\;\,{\bf V}_{\Mbar}(\ep)
\;\,\ket{\cmbar^{\,{\rm fin.} }} \;\;.
\eeeq
The first equality on the line (\ref{mallfin}) is just a rewriting of the IR
factorization formula (\ref{mallfact}). Then, we have applied the collinear
formula (\ref{factallL})
and, finally, in the expression (\ref{mallfin2}) we have used 
Eq.~(\ref{mallfact}) by replacing $\cm$ with the reduced matrix element
$\cmbar$.
The collinear limit performed in Eqs.~(\ref{mallfin}) and (\ref{mallfin2})
relates the matrix element 
$\cm^{\,{\rm fin.}}$ with the reduced matrix element
$\cmbar^{\,{\rm fin.}}$. Since both $\cm^{\,{\rm fin.}}$ and 
$\cmbar^{\,{\rm fin.}}$ are IR finite, the 
colour matrix 
${\bf V}_M^{-1} \;\,\sp \;\,{\bf V}_{\Mbar}$ in Eq.~(\ref{mallfin2})
must be IR finite in the collinear limit.
The collinear limit of this colour matrix is denoted by 
$\sp^{\,{\rm fin.}}$, and we can write
\beq
\label{spfinapp}
\sp^{\,{\rm fin.}} \simeq {\bf V}_M^{-1}(\ep)
\;\,\sp
\;\,{\bf V}_{\Mbar}(\ep) \;\;.
\eeq
Note that we have not yet implemented the collinear limit in the operator
${\bf V}_M$. We thus introduce the IR divergent operators
${\bf V}$ and ${\bf {\overline V}}$
as follows:
\beq
\label{vdef}
{\bf V}_M(\ep) \;\simeq \;{\bf V}(\ep) \;\;,
\eeq
\beq
\label{vbardef}
\Bigl[ \;{\bf V}_{\Mbar}(\ep) \;
\Bigr]_{{\bom T}_{\wp}= \,- \sum_{j \in NC} {\bom T}_j}
 \equiv \;{\bf {\overline V}}(\ep) \;\;.
\eeq
The relations (\ref{vdef}) and (\ref{vbardef})
represent the all-order generalization of the one-loop relations in
Eqs.~(\ref{ionedef}) and (\ref{ibaronedef}).
The relation (\ref{vdef}) defines ${\bf V}(\ep)$ through the collinear limit
of ${\bf V}_M(\ep)$, which is obtained by using the 
approximation 
$p_i \simeq z_i \wp$ $\;(i=1,\dots,m)$ for all the terms of 
${\bf V}_M$ that are not singular in the collinear limit.
The operator ${\bf V}_{\Mbar}$ depends on the colour charges of the partons
in $\cmbar$: the non-collinear partons with momentum $p_j$ $\;(j \in NC)$
and the parent collinear parton with momentum $\wp$.
These colour charges are related by colour conservation
$({\bom T}_{\wp} + \sum_{j \in NC} {\bom T}_j = 0)$. The relation 
(\ref{vbardef}) defines ${\bf {\overline V}}(\ep)$ through the implementation
of colour conservation: in all the terms of ${\bf V}_{\Mbar}$
that depends on ${\bom T}_{\wp}$, the colour matrix  
$- {\bom T}_{\wp}$ is replaced by the sum of the non-collinear charges.
According to this definition, the colour matrix structure of 
${\bf {\overline V}}$ only depends on the colour charges of the non-collinear
partons; therefore, the colour operator ${\bf {\overline V}}$ has a well
defined action onto both $\cmbar$ and $\cm$, or, equivalently, onto both 
the right-hand and left-hand sides of the colour matrices 
$\sp$ and $\sp^{\,{\rm fin.}}$.

Using the definitions (\ref{vdef}) and (\ref{vbardef}),
Eq.~(\ref{spfinapp}) gives
\beq
\label{spfindef}
\sp = {\bf V}(\ep) \;\,\sp^{\,{\rm fin.}} 
\;\,{\bf {\overline V}}^{\,-1}(\ep)
\;\;,
\eeq
where the all-order IR finite splitting matrix 
$\sp^{\,{\rm fin.}}$ has the renormalized perturbative expansion:
\beq
\label{spfinex}
\sp^{\,{\rm fin.}} = \sp^{(0,R)} + \sp^{(1) \,{\rm fin.}} +
\sp^{(2) \,{\rm fin.}} + \dots \;\;\;.
\eeq
The relation (\ref{spfindef}) presents the structure of the IR divergences of
the all-order splitting matrix $\sp$ for the multiparton collinear limit.
The IR divergences are embodied in the operators 
${\bf V}(\ep)$ and  ${\bf {\overline V}}(\ep)$.
The operator ${\bf V}$ depends on the colour charges and momenta of the
collinear {\em and} non-collinear partons in $\cm$. The 
operator ${\bf {\overline V}}$ depends on the momentum $\wp$ of the parent
collinear parton and on the colour charges and momenta of the
{\em non-collinear} partons. The IR dependence on the 
non-collinear partons {\em implies} that, in general (actually, in the SL
collinear case), the splitting matrix $\sp$ violates strict collinear
factorization.
We also note
that the IR factorization formula (\ref{spfindef}) has a
non-abelian structure (e.g., 
${\bf V} \,\sp^{\,{\rm fin.}} \,{\bf {\overline V}}^{\,-1}
\neq {\bf V} \,{\bf {\overline V}}^{\,-1} \,\sp^{\,{\rm fin.}}$
and ${\bf V} \,{\bf {\overline V}}^{\,-1} \neq
{\bf {\overline V}}^{\,-1} \,{\bf V}$).
In the SL collinear region, this structure produces strict-factorization
breaking effects with distinctive non-abelian features
(see Sects.~\ref{sec:2part2loop} and 
\ref{sec:mpart2loop}).

We have derived Eq.~(\ref{spfindef}) by exploiting the generalized collinear
factorization in Eq.~(\ref{factallL})
and the IR factorization property (Eq.~(\ref{mallir}) or,
equivalently, Eq.~(\ref{mallfact}))
of the multiparton QCD amplitudes. According to the structure of 
Eqs.~(\ref{vvsi}) and (\ref{icorvsi}), the all-order splitting matrix operators 
${\bf V}$ and ${\bf {\overline V}}$ have the following equivalent
representations:
\beq
\label{vm1}
{\bf V}^{-1}(\ep) = 1 - {\bom I}(\ep) = 
\exp \Bigl\{ \,-\, {\bom I}_{{\rm cor}}(\ep) \Bigr\} \;\;,
\eeq
\beq
\label{vbarm1}
{\bf {\overline V}}^{-1}(\ep) = 1 - {\bom {\overline I}}(\ep) = 
\exp \Bigl\{ \,-\, {\bom {\overline I}}_{{\rm cor}}(\ep) \Bigr\} \;\;,
\eeq
where the all-order operators ${\bom I}$ and ${\bom {\overline I}}$ 
(${\bom I}_{{\rm cor}}$ and ${\bom {\overline I}}_{{\rm cor}}$) are obtained 
from
the corresponding amplitude operators 
${\bom I}_M$  and ${\bom I}_{\Mbar}$ by using relations that are analogous
to those in Eqs.~(\ref{vdef}) and (\ref{vbardef}).
The one-loop contribution 
${\bom I}^{(1)}$  (${\bom {\overline I}}^{(1)}$) to 
${\bom I}$ (${\bom {\overline I}}$) has already been introduced in 
Eq.~(\ref{ionedef})
(Eq.~(\ref{ibaronedef})).
The perturbative expansion of Eqs.~(\ref{spfindef}), 
(\ref{vm1}) and (\ref{vbarm1}) is explicitly worked out in
Appendix~\ref{sec:appb}, and the ensuing two-loop results are presented and
discussed in the following Sects.~\ref{sec:2part2loop} and 
\ref{sec:mpart2loop}.

As discussed in Sect.~\ref{sec:irone} and at the beginning of this subsection,
the structure of the one-loop and two-loop IR factorization formulae
(\ref{m1ir}) and (\ref{m2ir}) does not specify in a unique way the explicit form of
the IR operators 
${\bom I}^{(1)}_{M}(\ep)$ and ${\bom I}^{(2)}_{M}(\ep)$.
Indeed, an IR finite redefinition  of
$\cm^{(1)\,{\rm fin.}}$ and $\cm^{(2)\,{\rm fin.}}$
can be compensated by a corresponding redefinition of 
${\bom I}^{(1)}_{M}(\ep)$ and ${\bom I}^{(2)}_{M}(\ep)$
(at two-loop order this redefinition can modify ${\bom I}^{(2)}_{M}(\ep)$
even at ${\cal O}(1/\ep)$).
This kind of invariance (and the ensuing arbitrariness) applies
to the all-order formulae (\ref{mallir}) and (\ref{mallfact}).
The IR factorization invariance is particularly evident in Eq.~(\ref{mallfact}): 
this equation is invariant under the transformations (redefinitions)
$\ket{\cm^{{\rm fin.}}} \to {\bom U}_{\rm fin.} \;\ket{\cm^{{\rm fin.}}}$
and ${\bom V}_{M}(\ep)\to {\bom V}_{M}(\ep) 
\;\left( {\bom U}_{\rm fin.}\right)^{-1}$, where ${\bom U}_{\rm fin.}$ is an
invertible IR finite operator. In the explicit expressions 
(\ref{i1m}) and (\ref{i2m}),
we have used a `minimal form' of 
${\bom I}^{(1)}_{M}$ and ${\bom I}^{(2)}_{M}$, namely, a form in which
${\bom I}^{(1)}_{M}(\ep)$ and ${\bom I}^{(2)}_{M}(\ep)$ include only terms
proportional to the IR poles $1/\ep^k$ $(k\geq 1)$. However, the other relations and
derivations presented in this subsection are independent of this minimal form.
In particular, the IR factorization formula (\ref{spfindef})
for the splitting matrix does not require (or necessarily imply) that the IR
divergent operators ${\bf V}(\ep)$ and ${\bf {\overline V}}(\ep)$ have a minimal
form.
Incidentally, we note that, in the one-loop expression (\ref{sp2oneren})
of the two-parton splitting matrix $\sp^{(1,R)}$,
the IR divergent operator $\itc^{(1)}(\ep)$ (see Eq.~(\ref{it2cone}))
does not have a minimal form, whereas the one-loop multiparton operator
$\imc(\ep)$ in Eqs.~(\ref{sp1div}) and (\ref{i1mcexp})
has a minimal form. An analogous comment applies to the two-loop results
discussed in Sects.~\ref{sec:2part2loop} and 
\ref{sec:mpart2loop}.

In the TL collinear region, strict factorization is valid and, therefore,
the result in Eq.~(\ref{spfindef}) takes a simplified form.
In Appendix~\ref{sec:tlircon}, we show that 
the IR structure of the splitting matrix $\sp$ for the multiparton
TL collinear limit can be presented as follows:
\beq
\label{spallirtl}
\sp(p_1,\dots,p_m;{\widetilde P})  = {\bf V}_{\rm TL}(\ep) \;\,
\sp^{\,{\rm fin.}}(p_1,\dots,p_m;{\widetilde P})\;\;, 
\quad \quad \quad \;\;\; ({\rm TL \,\;coll. \,\;lim.}) \;\;, 
\eeq
or, in the equivalent iterative form:
\beq
\label{spallittl}
\sp(p_1,\dots,p_m;{\widetilde P})  = {\bf I}_{\rm TL}(\ep) 
\;\, \sp(p_1,\dots,p_m;{\widetilde P}) \, + \,
\sp^{\,{\rm fin.}}(p_1,\dots,p_m;{\widetilde P})\;\;, 
\quad  ({\rm TL \,\;coll. \,\;lim.}) \;\;, 
\eeq
where $\sp^{\,{\rm fin.}}$ and the IR divergent operator ${\bf V}_{\rm TL}$
(or ${\bf I}_{\rm TL}$),
\beq
\label{itldef}
{\bf V}^{-1}_{\rm TL}(\ep) = 
\exp \Bigl\{ \,-\,{\bom I}_{\rm TL, \,cor}(\ep) \Bigr\} 
\equiv 1 - \,{\bf I}_{\rm TL}(\ep) \;\;, 
\eeq
are strictly factorized (completely independent of the 
non-collinear partons). The explicit perturbative expression of 
${\bf I}_{\rm TL}(\ep)$ up to two-loop order is given in 
Appendix~\ref{sec:tlircon}.

The all-order IR structure of the TL collinear limit of $m=2$ partons
was discussed in Refs.~\cite{Becher:2009qa, Dixon:2009ur}.
Our discussion in Appendix~\ref{sec:tlircon} and the 
expressions in
Eqs.~(\ref{spallirtl})--(\ref{itldef}) generalize the corresponding results 
of Refs.~\cite{Becher:2009qa, Dixon:2009ur}
to the case of $m \geq 3$ collinear partons.
Both the operators ${\bf V}$ and ${\bf {\overline V}}$ in Eq.~(\ref{spfindef})
depend on the non-collinear partons, while the operator 
${\bf V}_{\rm TL}$ in Eq.~(\ref{spallirtl}) is independent of the 
non-collinear partons. Since ${\bf V}_{\rm TL}$ eventually originates from 
${\bf V}$ and ${\bf {\overline V}}$, the strictly-factorized form
of  ${\bf V}_{\rm TL}$ implies a non-trivial cancellation of the combined
dependence of ${\bf V}$ and ${\bf {\overline V}}$ on the 
non-collinear partons. This cancellation constrains the form of 
${\bf V}$ and ${\bf {\overline V}}$ and, therefore, it also constrains
the general colour and kinematical structure of the scattering amplitude
operator ${\bf V}_M(\ep)$ in the IR factorization formula (\ref{mallfact})
(we recall that ${\bf V}$ and ${\bf {\overline V}}$ derive from 
${\bf V}_{M}$
through the collinear-limit procedure 
in Eqs.~(\ref{vdef}) and (\ref{vbardef})).
This constraint, which is a consistency requirement between 
strict collinear factorization and IR factorization, is particularly
sharp if the IR divergent operator ${\bf V}_M(\ep)$ is expressed in its minimal
form
(see Refs.~\cite{Becher:2009qa, Dixon:2009ur} and Appendix~\ref{sec:tlircon}).

\subsection{Two-parton collinear limit at two-loop order}
\label{sec:2part2loop}

The TL collinear limit of two-loop QCD amplitudes was studied in 
Refs.~\cite{Bern:2004cz, Badger:2004uk}. The authors of Ref.~\cite{Bern:2004cz}
considered the two-parton collinear splitting $g \to g \,g$ and, using the
unitarity sewing method, they performed a direct (process-independent) 
computation of the corresponding two-loop splitting amplitude 
${\rm Split}^{(2)}$. The authors of Ref.~\cite{Badger:2004uk}
exploited the universality of collinear factorization to extract 
${\rm Split}^{(2)}$ by taking the collinear limit of explicit two-loop results
of scattering amplitudes. Considering various scattering amplitudes with three
external QCD partons (and one additional colourless external particle), 
the splitting amplitudes of all the QCD subprocesses $A \to A_1  \,A_2$
were computed in Ref.~\cite{Badger:2004uk}. 
In Refs.~\cite{Bern:2004cz, Badger:2004uk}, the computation of
${\rm Split}^{(2)}$ was explicitly carried out up to ${\cal O}(\ep^0)$,
i.e. by neglecting terms that vanish in the limit $\ep \to 0$. 

In this subsection we examine the two-parton collinear limit in both the TL and
SL regions. Since we use factorization in colour space, we consider the two-loop
splitting matrix $\sp^{(2)}$.
The study of 
Refs.~\cite{Bern:2004cz, Badger:2004uk} mostly refers to the splitting 
amplitude ${\rm Split}^{(2)}$, which controls the collinear behaviour of colour
subamplitudes. In the TL collinear region, the relation between
$\sp^{(2)}$ and ${\rm Split}^{(2)}$ is exactly the same as at the tree level 
and at one-loop order; we can simply consider Eq.~(\ref{spvssplit}) and
perform the replacements  $\sp^{(0)} \to \sp^{(2)}$
and ${\rm Split}^{(0)} \to {\rm Split}^{(2)}$. The two-loop validity of 
the proportionality relation in Eq.~(\ref{spvssplit}) follows from the fact
that $\sp^{(2)}$ only involves a {\em single} (and unique)
colour matrix for {\em each} 
flavour configuration of the splitting process $A \to A_1 \, A_2$.
Indeed, in the case of the subprocesses $q \to q \,g$, 
${\bar q} \to {\bar q} \,g$ and $g \to q \,{\bar q}$, the colour matrix
$t^a_{\alpha \, \alpha'}$ (see Eqs.~(\ref{qqg0})--(\ref{gqqbar0}))
is the sole colour structure that is allowed by colour conservation
(this conclusion is independent of the perturbative order).
In the case of the subprocess $g \to g_1 \,g_2$, two different
colour matrices, namely,
$f_{a_1 a_2 a}$ (see Eq.~(\ref{ggg0})) and $d_{a_1 a_2 a}$ ($d_{a b c}$
is the fully-symmetrized trace of $t^a t^b t^c$), 
are allowed by colour conservation. However, the corresponding two-loop
splitting amplitude ${\rm Split}^{(2)}(p_1,p_2;{\widetilde P})$
turns out to be antisymmetric with respect to the exchange
$1 \leftrightarrow 2$ and, therefore, $\sp^{(2)}$ is necessarily proportional
to $f_{a_1 a_2 a} \,{\rm Split}^{(2)}(p_1,p_2;{\widetilde P})$ 
\cite{Bern:2004cz} (the presence of $d_{a_1 a_2 a}$ is excluded, since
$\sp^{(2)}(p_1,p_2;{\widetilde P})$ is symmetric with respect to the exchange
$1 \leftrightarrow 2$).

The general collinear limit of two-loop QCD amplitudes $\cm^{(2)}$ is 
controlled by the generalized factorization formula in Eq.~(\ref{factwogen}) 
(or Eq.~(\ref{facttworen})). The two-loop splitting matrix $\sp^{(2)}$
is the new (irreducible) contribution to collinear factorization.
Considering the generic two-parton collinear splitting $A \to A_1 \,A_2\,$, 
we can
write the two-loop renormalized splitting matrix in the following general
(i.e., valid in both the TL and SL collinear regions) form:
\beq
\label{sp2twoc}
\sp^{(2,R)} = 
\itc^{(2)}(\ep) \;\,
\sp^{(0,R)}  
+ \itc^{(1)}(\ep) \;\,
\sp^{(1,R)} 
+ \,{\widetilde \sp}^{(2) \,{\rm fin.}}\;,
\eeq
where $\sp^{(0,R)}$ and $\sp^{(1,R)}$ are the tree-level and one-loop 
renormalized splitting matrices, respectively.
The one-loop operator $\itc^{(1)}(\ep)$ is given in Eq.~(\ref{it2cone}),
and the two-loop colour-space operator $\itc^{(2)}(\ep)$ is
\beeq
\label{i22part}
\itc^{(2)}(\ep) &=& - \, \f{1}{2} \left[ \itc^{(1)}(\ep) \right]^2
+ \f{\as(\mu^2)}{2\pi} 
\left\{ \, \frac{1}{\ep} \,\cbet0
\left[ \, \itc^{(1)}(2\ep) - \itc^{(1)}(\ep) \right]
+ K \;\itc^{(1)}(2\ep) 
\right. \nn \\
&+& \, \f{\as(\mu^2)}{2\pi} \left(\f{-s_{12} -i0}{\mu^2} \right)^{\!-2\ep}
\left.
 \,\frac{1}{\ep} \;
\Bigl( 
\,H^{(2)}_1 + H^{(2)}_2 
\Bigr. 
\Bigl. \left.
- H^{(2)}_{12}  \right. \Bigr)
\right\}
+ {\widetilde {\bf \Delta}}_{C}^{(2)}(\ep) \;\;,
\eeeq
where the coefficients $K$ and $H^{(2)}_i$  are given in
Eqs.~(\ref{kcoef})--(\ref{h2g}).
The last term on the right-hand side of Eq.~(\ref{i22part}) has the following
explicit 
expression:
\beeq
\label{d2til}
{\widetilde {\bf \Delta}}_{C}^{(2)}(\ep) 
&=& \left(\f{\as(\mu^2)}{2\pi}\right)^2  
\;\left( \f{- s_{12}}{\mu^2} \right)^{\!\!-2\ep}
\;\pi \; f_{abc} \,
\sum_{i=1,2} 
\;\sum_{\substack{j, k = 3 \\ j \,\neq\, k}}^n
\,T_i^a \,T_j^b \, T_k^c \;\Theta(-z_i) \;
{\rm sign}(s_{ij}) \;\Theta(-s_{jk}) \nn \\
&\times& \ln\left(- 
\,\f{ s_{j\wp} \; s_{k\wp} \,z_1 z_2}{s_{jk} \,s_{12}} -i0 \right)
\;\left[ \,- \, \f{1}{2 \,\ep^{\,2}} \, +  \f{1}{\ep} 
\;\ln\left( \f{ - z_i }{1-z_i} \right)  \right]
\;\;. 
\eeeq

In Eqs.~(\ref{sp2twoc}) and (\ref{i22part}), the terms $\sp^{(0,R)}$, 
$\sp^{(1,R)}$ and $\itc^{(1)}$ retain their complete dependence on $\ep$.
In the limit $\ep \to 0$, the two-loop splitting matrix has IR divergences that
lead to $\ep$-poles of the types $1/\ep^4,\, 1/\ep^3,\, 1/\ep^2$ and $1/\ep$.
The $\ep$-poles of $\sp^{(2,R)}$ are entirely embodied in the first two
contributions, $\itc^{(2)} \times \sp^{(0,R)}$ and 
$\itc^{(1)} \times \sp^{(1,R)}$, on the right-hand side of Eq.~(\ref{sp2twoc}). 
The third contribution, ${\widetilde \sp}^{(2) \,{\rm fin.}}$, 
still depends on $\ep$, but it is {\em finite} in the limit 
$\ep \to 0$. 

In the TL collinear region, the expressions in Eqs.~(\ref{sp2twoc}) and 
(\ref{i22part}) agree
with the QCD results of 
Refs.~\cite{Badger:2004uk, Bern:2004cz}. We only note that 
Refs.~\cite{Badger:2004uk, Bern:2004cz} use the 't Hooft--Veltman (HV) variant
of dimensional regularization, whereas we use the CDR scheme throughout this
paper. The comparison between these two schemes poses no difficulties, since the
results in the HV scheme are obtained from those in the CDR scheme
by simply replacing the formal wave functions $u(p), v(p), \varepsilon(p)$
of the external collinear partons in $\sp$ with 
spin polarization states of definite (positive and negative) helicity.

The derivation of the results in Eqs.~(\ref{sp2twoc})--(\ref{d2til})
is presented in Appendix~\ref{sec:appb}.
We recall that the derivation is based on two input results:
the IR factorization formula (\ref{spfindef})
(and the knowledge in explicit form \cite{Catani:1998bh, Aybat:2006wq}
of the {\em two-loop} IR factorization formula (\ref{m2ir})
of the multiparton QCD amplitudes) and the explicit knowledge
(see Sect.~\ref{sec:gencol})
of the one-loop splitting matrix $\sp^{(1,R)}$ to {\em all orders} in $\ep$ in 
{\em both} the TL and SL collinear regions.
In particular 
(see Eq.~(\ref{d2single}) and the related discussion),
the coefficients of the single-pole terms, $1/\ep$,
in Eqs.~(\ref{sp2twoc}) and (\ref{d2til}) depend on the terms of
${\cal O}(\ep^0)$ in $\sp^{(1,R)}$.

In the SL collinear region (i.e., $s_{12} < 0$), strict collinear factorization
is violated, and all the contributions (with the sole exception of 
$\sp^{(0,R)}$) to Eq.~(\ref{sp2twoc}) depend on the non-collinear partons.
We recall (see Eqs.~(\ref{sp2oneren}) and (\ref{it2cone})) that, at one-loop
order, the factorization breaking terms are due to {\em two-parton} colour
correlations of the type ${\bom T}_i \cdot {\bom T}_j$, where $i$ 
($i=1$ or 2)
labels the collinear parton with $z_i < 0$ and the label $j$
refers to a non-collinear parton.
Owing to the {\em iterative} structure of Eq.~(\ref{sp2twoc}),
$\sp^{(2,R)}$ contains factorization breaking terms of the same type as at the
one-loop level, and it also contains the `square' of these terms (see the
contributions $\itc^{(1)} \times \sp^{(1,R)}$ in Eq.~(\ref{sp2twoc})
and $( \itc^{(1)} )^2$ in Eq.~(\ref{i22part})).
Moreover, 
the result in Eq.~(\ref{i22part}) shows new aspects 
of the violation of strict collinear factorization. These aspects
are clearly illustrated by the main features of the two-loop colour operator
${\widetilde {\bf \Delta}}_{C}^{(2)}(\ep)$.

The operator ${\widetilde {\bf \Delta}}_{C}^{(2)}(\ep)$, which contains
double and single poles in $1/\ep$, 
produces violation of strict collinear factorization since it depends on the
non-collinear partons. The expression (\ref{d2til}) shows that 
${\widetilde {\bf \Delta}}_{C}^{(2)}$ is non-vanishing only in the SL collinear
region (it requires $z_i < 0$), and that 
${\widetilde {\bf \Delta}}_{C}^{(2)}$
is definitely non-abelian (it is
proportional to $f_{abc}$).

The factorization breaking terms of $\sp^{(1)}$ are due to the absorptive part
of one-loop contributions that are present in both 
abelian
and non-abelian gauge theories. The non-abelian character of 
${\widetilde {\bf \Delta}}_{C}^{(2)}$ originates from a two-loop interference
effect (see, e.g., Eqs.~(\ref{d22mcgen}) and (\ref{d22mcnew})).
The absorptive part of the interactions in a loop interferes
with the radiative and absorptive 
parts\footnote{The absorptive correlations in a loop spoil colour coherence of
the interactions in the other loop, so that {\em both} the radiative and
absorptive parts of these interactions contribute to factorization breaking
effects. As a consequence, ${\widetilde {\bf \Delta}}_{C}^{(2)}$ has a
hermitian and an antihermitian component.} 
of the interactions in the other loop.
Owing to the causality structure  of the absorptive part, the interferences
between the two loops occur at different (asymptotic) times: 
non-abelian 
interactions  with different (time) orderings do not 
commute, and this produces
new
factorization breaking terms at the two-loop level.

\begin{figure}[htb]
 \begin{center}
 \begin{tabular}{c}
 \epsfxsize=8truecm
 \epsffile{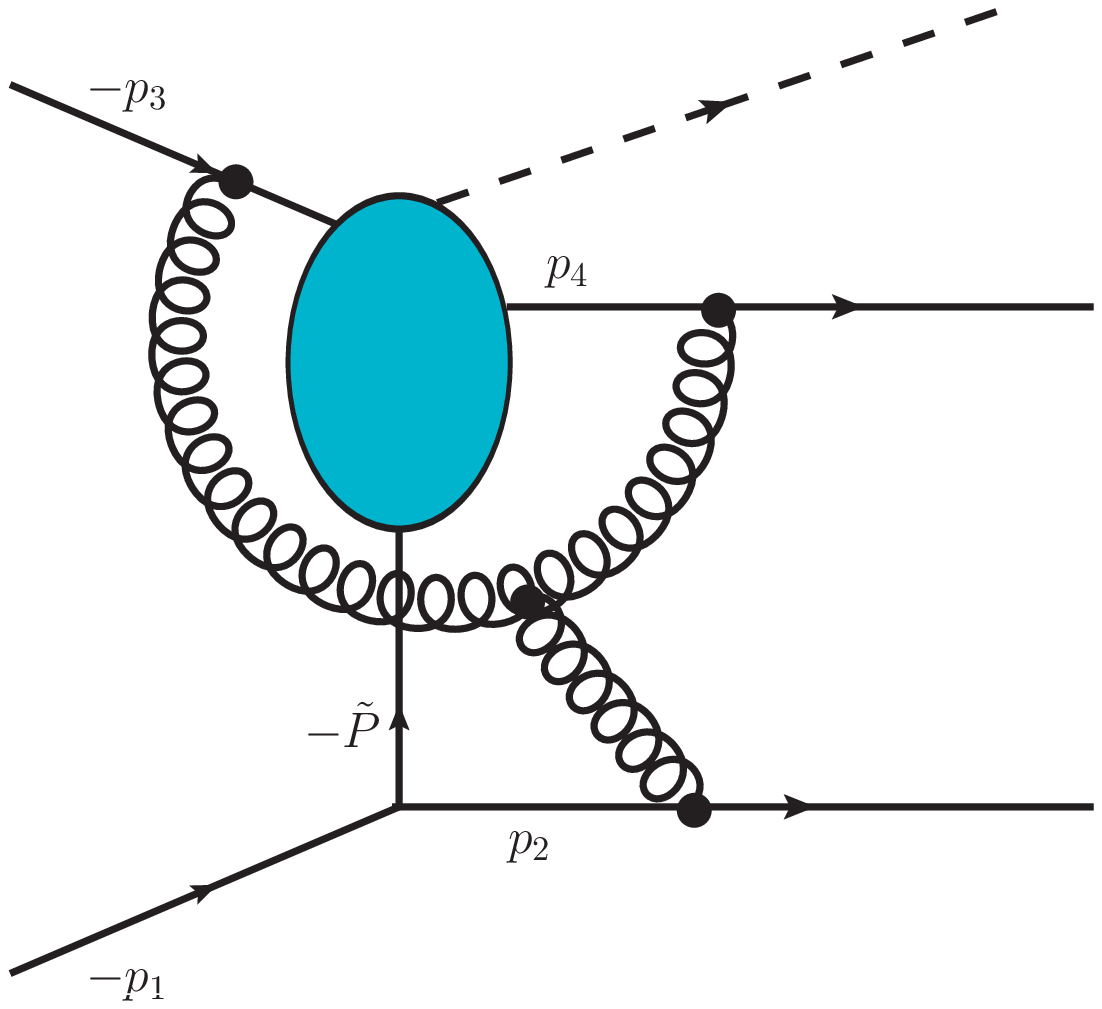}
 \epsfxsize=8truecm
 \epsffile{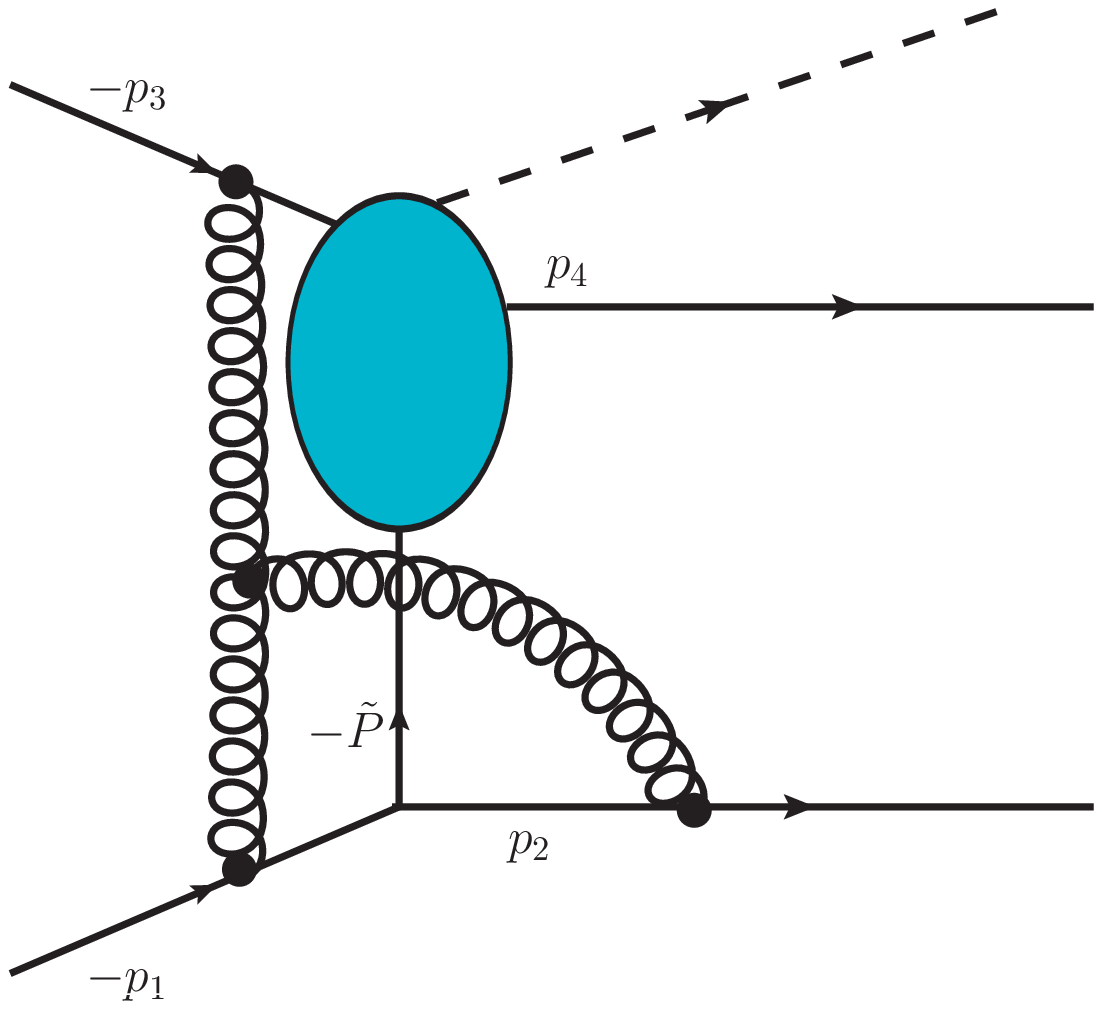} \\
 \end{tabular}
 \end{center}
 \caption{\label{fig:t2t3t4}
 {\em 
 Two-parton SL collinear limit $(p_i \simeq z_i \wp, \;i=1,2)$ of 
 two-loop QCD
 amplitudes with $n=4$ partons and additional colourless external legs
 (denoted by the dashed line) in parton--parton hard-scattering configurations.
 Representative colour structure of non-abelian factorization breaking
 correlations according to 
 (left) Eq.~(\ref{d2til4p}) and (right) Eq.(\ref{d2til4prime}).
 }}
 \end{figure}

The colour structure of ${\widetilde {\bf \Delta}}_{C}^{(2)}$ involves {\em
three-parton} correlations and, specifically, correlations between a collinear
parton and two non-collinear partons (see, e.g., Fig.~\ref{fig:t2t3t4}--left). 
Since there are two collinear partons,
to detect the effect of ${\widetilde {\bf \Delta}}_{C}^{(2)}$ we have
to consider the collinear limit of amplitudes $\cm(p_1,p_2,\dots,p_n)$
with $n \geq 4$ QCD partons (if $n=4$, $\cm$ necessarily involves additional
colourless external legs, otherwise the corresponding reduced matrix element
$\cmbar$ vanishes). We also note that the two non-collinear partons $j$ and
$k$ that are colour-correlated by ${\widetilde {\bf \Delta}}_{C}^{(2)}$
must have energies with opposite signs (i.e., $s_{jk} < 0$):
this pair of non-collinear partons consists of a physical initial-state
parton and a physical final-state parton. In particular, this energy constraint
implies that ${\widetilde {\bf \Delta}}_{C}^{(2)}$ does not contribute
to the two-parton SL collinear limit of the amplitudes that are involved
in lepton--hadron DIS. The operator ${\widetilde {\bf \Delta}}_{C}^{(2)}$
typically contributes to the SL collinear limit in hadron--hadron
hard-scattering processes. A simple example with $n=4$ QCD partons is the
collinear limit of the scattering amplitude of the 
process 
`parton + parton $\to$ vector boson + 2 partons',
where one of the final-state partons is collinear to one of the 
initial-state partons.

\vspace{0.5cm}

We present the explicit expression of ${\widetilde {\bf \Delta}}_{C}^{(2)}$
for a generic matrix element with $n=4$ QCD partons (see Fig.~\ref{fig:t2t3t4}).
We consider only the SL kinematical configurations in which 
${\widetilde {\bf \Delta}}_{C}^{(2)} \neq 0$. With no loss of generality, we can
limit ourselves to the following kinematical region (the other kinematical
regions are obtained by re-labeling the parton momenta):
\beq
z_2 < 0 \;\;, \quad s_{34} < 0 \;\;, \quad s_{23} < 0 \;\;, \nn
\eeq
and, using Eq.~(\ref{d2til}), we have
\beeq
\label{d2til4p}
{\widetilde {\bf \Delta}}_{C}^{(2)}(\ep) 
&=& - \;\f{\as^2(\mu^2)}{2\pi}  
\;\left( \f{- s_{12}}{\mu^2} \right)^{\!\!-2\ep}
\; f_{abc} 
\; T_2^a \,T_3^b \, T_4^c 
 \nn \\
&\times& \ln\left(- 
\,\f{ s_{3\wp} \; s_{4\wp} \,z_1 z_2}{s_{34} \,s_{12}} -i0 \right)
\;\left[ \,- \, \f{1}{2 \,\ep^{\,2}} \, +  \f{1}{\ep} 
\;\ln\left( \f{ - z_2 }{1-z_2} \right)  \right] \\
\label{d2til4prime}
&\simeq& \;\f{\as^2(\mu^2)}{2\pi}  
\;\left( \f{- s_{12}}{\mu^2} \right)^{\!\!-2\ep}
\; f_{abc} 
\; T_1^a \,T_2^b \, T_3^c 
 \nn \\
&\times& \ln\left(- 
\,\f{ s_{31} \; s_{42} }{s_{34} \,s_{12}} -i0 \right)
\;\left[ \,- \, \f{1}{2 \,\ep^{\,2}} \, +  \f{1}{\ep} 
\;\ln\left( - \f{z_2 }{z_1} \right)  \right] 
\;\;. 
\eeeq
Note that the expressions (\ref{d2til4p}) and (\ref{d2til4prime})
are equivalent in the collinear limit. To obtain Eq.~(\ref{d2til4prime}),
we have implemented the collinear approximation $p_i \simeq z_i \wp \;(i=1,2)$,
and we have used the $n=4$ equality
\beq
f_{abc} \; T_2^a \,T_3^b \, T_4^c = \, -
\; f_{abc} \; T_1^a \,T_2^b \, T_3^c \;\;, \quad \quad \quad (n=4) \;\;.
\eeq
This equality derives from charge conservation 
(i.e., ${\bom T}_4 = - ( {\bom T}_1 + {\bom T}_2 + {\bom T}_3 )\,$)
and from colour algebra, namely, from the general algebraic identity
\beq
\label{tipj}
i \,f_{abc} \; T_i^a \,T_j^b \,( T_i^c + T_j^c ) = - \,\delta_{ij} \;C_A 
\;{\bom T}_i^2 \;\;.
\eeq

The factorization breaking correlations depend on the momenta of the
non-collinear partons. At the one-loop order (see Eq.~(\ref{it2cone})),
this dependence only involves the sign of the energy of the non-collinear parton
(what matter is simply the physical distinction between initial-state and
final-state partons). The two-loop operator 
${\widetilde {\bf \Delta}}_{C}^{(2)}$ instead depends also on the actual size of
the momenta of the non-collinear partons. This dependence appears (see
Eq.~(\ref{d2til})) through the scale
\beq
\label{qtp}
\f{ s_{j\wp} \; s_{k\wp}}{s_{jk}} \equiv {\bom q}^2_{\perp \wp, j k} \;\;.
\eeq
We notice that $q^{\mu}_{\perp \wp, j k}\;$ ($q^{\mu}_{\perp} q_{\perp \mu}
= -\,{\bom q}^2_{\perp})$ is the component of the momentum $\wp^\mu$ that is
transverse to the direction of the momenta $p_j$ and $p_k$ in the reference
frame where these two momenta are back-to-back. Since in the collinear limit we
have $p_i \simeq z_i \wp\;$ ($i=1,2$), the scale (\ref{qtp}) can be expressed 
in terms of the transverse momentum $q^{\mu}_{\perp i, j k}\;$ of one of the
collinear partons:
\beq
\label{qti}
{\bom q}^2_{\perp i, j k} \equiv \f{ s_{ji} \; s_{ki}}{s_{jk}} 
\simeq z_i^2 \;{\bom q}^2_{\perp \wp, j k} \;\;.
\eeq

The `non-collinear' scale in Eq.~(\ref{qtp}) and the `collinear' scale
$s_{12}/(z_1 z_2)$ are always positive quantities. Thus, we have
\beq
\label{logP}
\ln\left(- 
\,\f{ s_{j\wp} \; s_{k\wp} \,z_1 z_2}{s_{jk} \,s_{12}} -i0 \right)
= \,\ln\left( 
\,\f{ s_{j\wp} \; s_{k\wp} \,z_1 z_2}{s_{jk} \,s_{12}} \right) - i \pi \;\;,
\eeq
where the logarithm on the right-hand side is a real number (with no imaginary
part). Inserting Eq.~(\ref{logP}) in Eq.~(\ref{d2til}), we see that 
${\widetilde {\bf \Delta}}_{C}^{(2)}$ has a {\em hermitian} and an 
{\em antihermitian} part (we recall that at one-loop order the factorization
breaking term is purely antihermitian). The hermitian part depends on the
non-collinear scale (\ref{qtp}). In the antihermitian part, the momenta of the
non-collinear partons appear only through the sign of their energies, and we
explicitly have 
\beeq
\label{d2tilanti}
{\widetilde {\bf \Delta}}_{C}^{(2)}(\ep) - 
\left[ {\widetilde {\bf \Delta}}_{C}^{(2)}(\ep) \right]^\dagger
&=& -\,\f{i}{2} \;\as^2(\mu^2)  
\;\left( \f{- s_{12}}{\mu^2} \right)^{\!\!-2\ep} \; f_{abc} \,
\sum_{i=1,2} 
\;\sum_{\substack{j, k = 3 \\ j \,\neq\, k}}^n
\,T_i^a \,T_j^b \, T_k^c \;\Theta(-z_i) 
\nn \\
&\times& {\rm sign}(s_{ij}) \;\Theta(-s_{jk})
\;\left[ \,- \, \f{1}{2 \,\ep^{\,2}} \, +  \f{1}{\ep} 
\;\ln\left( \f{ - z_i }{1-z_i} \right)  \right]
\\
\label{d2tilantib}
&=& +\,\f{i}{2} \;\as^2(\mu^2)  
\;\left( \f{- s_{12}}{\mu^2} \right)^{\!\!-2\ep} \; f_{abc} \,
\;\sum_{j= 3}^n
\,T_1^a \,T_2^b \, T_j^c \;{\rm sign}(s_{2j}) 
\nn \\
&\times& \sum_{i=1,2} \;\Theta(-z_i) 
\;\left[ \,- \, \f{1}{2 \,\ep^{\,2}} \, +  \f{1}{\ep} 
\;\ln\left( \f{ - z_i }{1-z_i} \right)  \right]
\;\;,
\eeeq
where the equivalence between the expressions (\ref{d2tilanti}) and
(\ref{d2tilantib}) follows from the conservation of the colour charge (see 
Eq.~(\ref{3vs3})).
We note that the expression (\ref{d2tilantib}) involves colour correlations
between a non-collinear parton and the two collinear partons. This type
of three-parton colour correlations (see, e.g., Fig.~\ref{fig:t2t3t4}--right)
completes the class of factorization breaking
structures that can appear in the two-loop splitting matrix $\sp^{(2, R)}$.

We briefly comment on the finite contribution
${\widetilde \sp}^{(2) \,{\rm fin.}}$ in Eq.~(\ref{sp2twoc}).
In the TL collinear region, the explicit expression of 
${\widetilde \sp}^{(2) \,{\rm fin.}}$ at ${\cal O}(\ep^0)$
can be extracted from the direct comparison with the results of 
Refs.~\cite{Badger:2004uk, Bern:2004cz}.
We have not computed 
${\widetilde \sp}^{(2) \,{\rm fin.}}$ in the SL collinear region.
Its expression contains terms that violate strict collinear factorization. 
These terms produce
all types of colour correlations (including  
three-parton correlations as in Eq.~(\ref{d2tilantib}))
that contribute to the IR
divergent part of $\sp^{(2, R)}$.

\subsection{Multiparton collinear limit at two-loop order}
\label{sec:mpart2loop}

To our knowledge, no results on the multiparton collinear limit 
(with $m \geq 3$ collinear partons) of the two-loop
QCD amplitudes $\cm^{(2)}$ are available in the literature.
To illustrate some main features of the corresponding two-loop splitting matrix
$\sp^{(2)}$ in the factorization formula (\ref{factwogen}) 
(or (\ref{facttworen})), we proceed analogously to Sect.~\ref{sec:irone}.
Considering the renormalized splitting matrix, we introduce the following
decomposition in IR divergent and IR finite terms:
\beq
\label{sp2divfin}
\sp^{(2, R)} = \sp^{(2) \,{\rm div.}} + \sp^{(2) \,{\rm fin.}} \;\;.
\eeq
All the $\ep$-pole contributions to $\sp^{(2, R)}$ are included in 
$\sp^{(2) \,{\rm div.}}$, which can be written as
\beq
\label{sp2mcdiv}
\sp^{(2) \,{\rm div.}} =  \imc(\ep) \;\sp^{(1, R)} +
\imctwo(\ep) \;\sp^{(0, R)} +  
{\overline \sp}^{(2) \,{\rm div.}} \;\;,
\eeq
where $\sp^{(1, R)}$ is the one-loop splitting matrix, 
and $\imc(\ep)$ is given in Eq.~(\ref{i1mcexp}).
The new two-loop colour operator $\imctwo(\ep)$ contains $\ep$ poles,
and additional IR divergent terms are included in 
${\overline \sp}^{(2) \,{\rm div.}}$. Note, however, that 
${\overline \sp}^{(2) \,{\rm div.}}$ is purely non-abelian, 
it is non-vanishing only in the SL collinear
regions, and it contains
only single poles (i.e., its IR divergences appear at 
${\cal O}(1/\ep)$). All the higher-order poles (starting at order
${\cal O}(1/\ep^4)$) are included in the first two terms on the right-hand side of
Eq.~(\ref{sp2mcdiv}).

The result in Eq.~(\ref{sp2mcdiv}) is derived in Appendix~\ref{sec:appb}.
The derivation is analogous to that of the corresponding result 
(see Eq.~(\ref{sp2twoc}))
for the two-parton collinear limit. In the case of the multiparton collinear
limit, we do not know the explicit expression of the one-loop splitting matrix 
$\sp^{(1, R)}$ at ${\cal O}(\ep^0)$ in the SL region: this prevent us from
computing the explicit expression of 
${\overline \sp}^{(2) \,{\rm div.}}$ through the method used in 
Appendix~\ref{sec:appb}.

The explicit expression of the colour operator $\imctwo$ is
\beeq
\label{i2mexpl}
\imctwo(\ep) &=& - \, \f{1}{2} \left[ \imc(\ep) \right]^2
+ \f{\as(\mu^2)}{2\pi} 
\left\{ + \frac{1}{\ep} \,\cbet0 
\left[ \imc(2\ep) - \imc(\ep) \right]
+ K \;\imc(2\ep) 
\right. \nn \\
&+& \, \f{\as(\mu^2)}{2\pi} 
\left.
 \,\frac{1}{\ep} \;
\Bigl( \;\sum_{i \, \in C} \,H^{(2)}_i 
\Bigr. 
\Bigl. \left.
- H^{(2)}_{\wp}  \right. \Bigr)
\right\}
+ {\bf \Delta}_{m \,C}^{(2;\,2)}(\ep)  \;\;,
\eeeq
where
\beeq
\label{d22mc}
{\bf \Delta}_{m \,C}^{(2;\,2)}(\ep) &=& \left(\f{\as(\mu^2)}{2\pi}\right)^2  
\left( \,- \, \f{1}{2 \,\ep^2} \right) \;\pi \; f_{abc} \,
\sum_{i \in C} \;\sum_{\substack{j, k \,\in NC \\ j \,\neq\, k}} 
\,T_i^a \,T_j^b \, T_k^c \;\Theta(-z_i) \;
{\rm sign}(s_{ij}) \;\Theta(-s_{jk}) \nn \\
&\times& \ln\left(- \,\f{ s_{j\wp} \; s_{k\wp}}{s_{jk} \,\mu^2} -i0 \right) \;\;.
\eeeq
Since the flavour-dependent coefficients $H^{(2)}_i$ (the subscript $\wp$ in 
$H^{(2)}_{\wp}$ refers to the flavour of the parent collinear parton)
are $c$-numbers, the non-trivial colour-charge structure of $\imctwo$
is due to $\imc$ and ${\bf \Delta}_{m \,C}^{(2;\,2)}$. This structure produces
violation of strict factorization in the SL collinear regions. In the two-loop
splitting matrix $\sp^{(2, R)}$, the factorization breaking terms start to
contribute at ${\cal O}(1/\ep^3)$ (we recall that $\imc$ includes the
factorization breaking operator ${\bf \Delta}_{m \,C}^{(1)}(\ep)$ 
of Eq.~(\ref{del1})).

The colour-charge operator 
${\bf \Delta}_{m \,C}^{(2;\,2)}$ is responsible for some 
distinctive features of the multiparton collinear limit at two-loop
order.
This operator is closely analogous to the
corresponding operator 
${\widetilde {\bf \Delta}}_{C}^{(2)}$ (see Eq.~(\ref{d2til})) for the 
two-parton collinear limit. By direct inspection of Eq.~(\ref{d22mc}),
we see that the operator ${\bf \Delta}_{m \,C}^{(2;\,2)}$ produces IR
divergences at the level of double poles $1/\ep^2$, it is non-vanishing only 
in SL collinear regions, it is definitely non-abelian, and it leads to
violation of strict collinear factorization. These factorization breaking 
terms have the form of three-parton 
correlations: the colour charge of a
collinear parton is correlated to the colour charges of two non-collinear
partons (the partons $j$ and $k$) that have energies with {\rm opposite}
sign ($s_{jk} < 0$), and the intensity of the correlation is controlled by the
transverse-momentum scale ${\bom q}^2_{\perp \wp, j k}$ (see Eq.~(\ref{qtp})).
Since ${\bom q}^2_{\perp \wp, j k} > 0$, the logarithm on the right-hand side
of Eq.~(\ref{d22mc}) has a real and an imaginary part
(i.e. $\ln(-{\bom q}^2_{\perp \wp, j k}/\mu^2 -i0) = 
\ln({\bom q}^2_{\perp \wp, j k}/\mu^2) - i \pi$) and, correspondingly, the 
operator ${\bf \Delta}_{m \,C}^{(2;\,2)}$ has a hermitian and an 
antihermitian part.

Using Eq.~(\ref{d22mc}), the explicit expression of the antihermitian part of 
${\bf \Delta}_{m \,C}^{(2;\,2)}$ is
\beeq
\label{d22a}
{\bf \Delta}_{m \,C}^{(2;\,2)}(\ep) - 
\left[ \, {\bf \Delta}_{m \,C}^{(2;\,2)}(\ep) \right]^\dagger
\!\!\!&=&\!\!\! + \,i \;\f{\as^2(\mu^2)}{4 \,\ep^2}  \; f_{abc} \,
\sum_{i \,\in C} 
\;\sum_{\substack{j, k \,\in NC \\ j \,\neq\, k}}
\,T_i^a \,T_j^b \, T_k^c \;\Theta(-z_i) \;{\rm sign}(s_{ij})
\;\Theta(-s_{jk}) \nn \\
&& \\
\label{d22b}
\!\!\!&\simeq&\!\!\! - \,i \;\f{\as^2(\mu^2)}{4 \,\ep^2} \; f_{abc} \,
\sum_{\substack{i, \ell \,\in C \\ i \,\neq\, \ell}}
\;\sum_{j \,\in NC}
\,T_i^a \,T_{\ell}^b \, T_j^c \;\Theta(-z_i) \;\Theta(-s_{i \ell})
\;{\rm sign}(s_{j \wp}) \;.
\nn \\
&& 
\eeeq
The expressions (\ref{d22a}) and
(\ref{d22b}) are equivalent in the collinear limit (see Eq.~(\ref{3vs3}) and
the accompanying 
comments 
in the final part of this subsection). 
In particular, Eq.~(\ref{d22b}) shows that the 
antihermitian part of ${\bf \Delta}_{m \,C}^{(2;\,2)}$ can be expressed in terms
of {\em three-parton} correlations that involve a non-collinear parton and two
collinear partons (the partons $i$ and $\ell$) that have energies with 
opposite sign ($s_{i \ell} < 0$).

We add some brief comments on the two-loop contributions 
${\overline \sp}^{(2) \,{\rm div.}}$ (see Eq.~(\ref{sp2mcdiv}))
and $\sp^{(2) \,{\rm fin.}}$ (see Eq.~(\ref{sp2divfin})).
The term ${\overline \sp}^{(2) \,{\rm div.}}$, which is proportional to the
single pole $1/\ep$, 
is non-vanishing only 
in SL collinear regions and (analogously to 
${\bf \Delta}_{m \,C}^{(2;\,2)}$) it involves non-abelian factorization breaking
correlations between a collinear parton and 
two non-collinear partons (see Eq.~(\ref{sp2bardiv})).
The IR finite term $\sp^{(2) \,{\rm fin.}}$ is non-vanishing in both the TL and
SL collinear regions. In the SL collinear limit, 
$\sp^{(2) \,{\rm fin.}}$ receives factorization breaking contributions
from all types of 
colour correlations that appear in 
$\sp^{(2) \,{\rm div.}}$. In particular, $\sp^{(2) \,{\rm fin.}}$
also includes non-abelian correlations between a non-collinear parton and
two collinear partons (see, e.g., Eq.~(\ref{d22b})).

Factorization breaking terms that correlate the colour matrices of three partons
can be related by using the following identity:
\beq
\label{3vs3}
f_{abc} \,
\sum_{i \in C} \;\sum_{\substack{j, k \,\in NC \\ j \,\neq\, k}} 
\,T_i^a \,T_j^b \, T_k^c  \;
\,{\rm sign}(s_{ij}) \;\Theta(-s_{jk}) \;h_i
= f_{abc} \,
\sum_{\substack{i, \,\ell \,\in C \\ i \,\neq\, \ell}} \;\sum_{j \in NC}
\,T_i^a \,T_{\ell}^b \, T_j^c  \;
\,{\rm sign}(s_{ij}) \;h_i \;\;,
\eeq
where $h_i$ is an arbitrary $c$-number function that depends on the 
collinear parton $i$. The identity (\ref{3vs3}) relates terms that involve one
collinear parton and two non-collinear partons
to terms that involve two collinear partons and one non-collinear parton.
Note that this relation requires that the kinematical function $h_i$
is independent of the non-collinear partons (e.g., of the momenta of the
non-collinear partons).
If the function $h_i$ is simply $h_i= \Theta(-z_i)$,
we can implement the collinear limit $p_i \simeq z_i\wp$ and we can
rewrite the right-hand side of Eq.~(\ref{3vs3}) as follows:
\beq
\label{2c1ncvscol}
f_{abc} 
\sum_{\substack{i, \,\ell \,\in C \\ i \,\neq\, \ell}} \sum_{j \in NC}
T_i^a T_{\ell}^b T_j^c  
\,{\rm sign}(s_{ij}) \,\Theta(-z_i)
\simeq
- f_{abc} 
\sum_{\substack{i, \,\ell \,\in C \\ i \,\neq\, \ell}} \sum_{j \in NC}
T_i^a T_{\ell}^b T_j^c  
\,{\rm sign}(s_{j\wp}) \,\Theta(-z_i) \,\Theta(-s_{i\ell}) \,.
\eeq
The proof of Eq.~(\ref{3vs3}) (which follows from colour conservation) and the
derivation of the collinear relation (\ref{2c1ncvscol}) are presented in
Appendix~\ref{sec:appb} (see Eqs.~(\ref{3vs3a})--(\ref{2c1nccan})).
The expression
(\ref{d22b}) is obtained from (\ref{d22a}) by simply using Eqs.~(\ref{3vs3}) and
(\ref{2c1ncvscol}). The expression (\ref{d2tilantib}) is obtained from 
(\ref{d2tilanti}) by directly using Eq.~(\ref{3vs3}) (note that
${\rm sign}(s_{1j}) = -\,{\rm sign}(s_{2j})$ in the SL collinear region, since 
$s_{12} < 0$).


\setcounter{footnote}{2}
\section{
Squared amplitudes and cross sections}
\label{sec:square}

The perturbative QCD computation of cross sections (and related physical
observables)  requires the evaluation
of the square of the matrix element $\cm(p_1,p_2,\dots,p_n)$
and its integration over the phase space of the final-state partons.
In this section we consider the collinear limit of squared amplitudes.
In particular, we are interested in the implications of violation of strict
collinear factorization at the level of squared amplitudes and, possibly,
of cross sections.

\subsection{The collinear behaviour of squared amplitudes}
\label{sec:colsquare}

We consider the squared matrix element, $|\cm|^2$, summed
over the colours and spins of the external QCD partons
(see Eq.~(\ref{mel})):
\beq
\label{melsquared}
|\cm(p_1,p_2,\dots)|^2 \equiv 
\sum_{\{c_i\}} \, \sum_{\{s_i\}} \;
\Bigl[ \,\cm^{c_1,c_2,\dots;s_1,s_2,\dots}(p_1,p_2,\dots) \,
\Bigr]^\dagger
\;
\cm^{c_1,c_2,\dots;s_1,s_2,\dots}(p_1,p_2,\dots) \,
\;\;.
\eeq
Using the notation in colour+spin space (see Eq.~(\ref{cmvdef})),
$|\cm|^2$ can be written as
\beq
|\cm(p_1,p_2,\dots,p_n)|^2 =
\la \, \cm(p_1,p_2,\dots,p_n) \,
 | \, \cm(p_1,p_2,\dots,p_n)  \, \ra \;\;.
\eeq

The all-order singular behaviour of $|\cm|^2$, in a generic kinematical
configuration of $m$ collinear partons with momenta $\{p_1,\dots,p_m\}$,
is obtained by squaring the generalized factorization formula in 
Eq.~(\ref{factallL}). We have
\beq
\label{factallm2}
|\cm|^2 \;
\simeq \la \, \cmbar \,
 | \, \;{\bf P}(p_1,\dots,p_m;{\widetilde P};p_{m+1},\dots,p_n) 
\;\;\ket{\cmbar} \;\;,
\eeq
where the matrix ${\bf P}$ is the square of the all-order splitting 
matrix $\sp\,$:
\beq
\label{mpdef}
{\cal {\bf P}} \equiv
\left[\, \sp \,\right]^\dagger \,\sp \;\;.
\eeq
The loop expansion of the squared splitting matrix ${\bf P}$ is
\beq
\label{mploop}
{\bf P} = {\bf P}^{(0,R)} + {\bf P}^{(1,R)} + {\bf P}^{(2,R)} + \dots
\;\;,
\eeq
where ${\bf P}^{(k,R)}$ (with $k=0,1,2,\dots$) are the renormalized perturbative
contributions.
Inserting Eq.~(\ref{loopexspren}) in Eq.~(\ref{mpdef}),
we obtain 
the expression of ${\bf P}^{(k,R)}$ in terms of the 
perturbative contributions to $\sp\,$:
\beq
\label{mp0}
{\bf P}^{(0,R)} = \left(\, \sp^{(0,R)} \,\right)^\dagger \,\sp^{(0,R)} \;\;,
\eeq
\beq
\label{mp1}
{\bf P}^{(1,R)} = \left(\, \sp^{(0,R)} \,\right)^\dagger \,\sp^{(1,R)} \; 
+\; {\rm h.c.}
\;\;,
\eeq
\beq
\label{mp2}
{\bf P}^{(2,R)} = \left(\, \sp^{(1,R)} \,\right)^\dagger \,\sp^{(1,R)}
\,+ \left[ \left(\, \sp^{(0,R)} \,\right)^\dagger \,\sp^{(2,R)}
+\; {\rm h.c.} \right]
\;\;,
\eeq
where the abbreviation `h.c.' means hermitian conjugate.

The perturbative (loop) expansion of the all-order factorization formula 
(\ref{factallm2}) is obtained by using Eqs.~(\ref{loopexmren}) 
and (\ref{mploop}). Considering the expansion up to the two-loop level,
we have
\beq
\label{factm20}
|\,\cm^{(0,R)}\,|^2 \simeq 
\la \, \cmbar^{(0,R)} \,
 | \, \;{\bf P}^{(0,R)} 
\;\;\ket{\cmbar^{(0,R)}} \;\;,
\eeq
\beq
\label{factm21}
\la \, \cm^{(0,R)} \,\ket{\cm^{(1,R)}} +\; {\rm c.c.} \;\simeq
\left[ \la \, \cmbar^{(0,R)} \,
 | \, \;{\bf P}^{(0,R)} 
\;\ket{\cmbar^{(1,R)}} +\; {\rm c.c.} \;
\right] + \la \, \cmbar^{(0,R)} \,
| \, {\bf P}^{(1,R)} 
\;\ket{\cmbar^{(0,R)}} \;,
\eeq
\beeq
\label{factm22}
|\,\cm^{(1,R)}\,|^2 \!\!\!\!\!\!\!\!
&&+
\Bigl[ \;
\la \, \cm^{(0,R)} \,\ket{\cm^{(2,R)}} +\; {\rm c.c.} \;
\Bigr]
\simeq
\la \, \cmbar^{(1,R)} \,
 | \, \;{\bf P}^{(0,R)} 
\;\ket{\cmbar^{(1,R)}} \nn \\
&&+
\left[ \la \, \cmbar^{(0,R)} \,
 | \, \;{\bf P}^{(0,R)} 
\;\ket{\cmbar^{(2,R)}} +\; {\rm c.c.} \;
\right] 
+ 
\left[ \la \, \cmbar^{(0,R)} \,
 | \, \;{\bf P}^{(1,R)} 
\;\ket{\cmbar^{(1,R)}} +\; {\rm c.c.} \;
\right] \nn \\
&&
+ \;\la \, \cmbar^{(0,R)} \,
| \, {\bf P}^{(2,R)} 
\;\ket{\cmbar^{(0,R)}} \;\;,
\eeeq
where the abbreviation `c.c.' means complex conjugate.
The tree-level factorization formula (\ref{factm20})
depends on ${\bf P}^{(0,R)}$.
The one-loop factorization formula (\ref{factm21})
also depends on ${\bf P}^{(1,R)}$, whereas ${\bf P}^{(2,R)}$
enters the two-loop factorization formula (\ref{factm22}).
The expressions in Eqs.~(\ref{factm20}), (\ref{factm21}) and (\ref{factm22})
directly correspond to the terms that contribute to the order-by-order
perturbative calculation of cross sections.

The kernel ${\bf P}$ on the right-hand side of the generalized factorization
formula (\ref{factallm2}) is a matrix in colour+spin space.
The matrix acts on the vector space of the reduced matrix
element 
${\cmbar\,({\widetilde P},p_{m+1},\dots,p_n)}$.
The dependence
of ${\bf P}$ on the spin and colour indices can be denoted in the following
explicit form:
\beq
\label{pmat}
\bra{c^\prime,c^\prime_{m+1},\dots,c^\prime_{n}}
\left[\, {\bf P}(p_1,\dots,p_m;{\widetilde P};p_{m+1},\dots,p_n) 
\,\right]_{s^\prime s}
\ket{c,c_{m+1},\dots,c_{n}} \;\;,
\eeq
where $c_{m+1},\dots,c_{n}$ are the colour indices of the non-collinear partons
(with momenta $p_{m+1},\dots,p_{n}$) in $\ket{\cmbar}$,
whereas $c$ and $s$ are the colour and spin indices of the parent collinear
parton (with momentum $\widetilde P$) in $\ket{\cmbar}$
(the indices $c^\prime_{m+1},\dots,c^\prime_{n},c^\prime$ and $s^\prime$ refer
to the vector space of the complex conjugate matrix element $\bra{\cmbar}\,$).
The matrix structure of Eq.~(\ref{pmat}) follows from Eq.~(\ref{mpdef})
and from the matrix structure of the splitting matrix $\sp$.
We briefly comment on the dependence of Eq.~(\ref{pmat}) on the spin and colour
indices, in turn.

According to the generalized factorization formula (\ref{factallL}),
the splitting matrix $\sp$ is independent of the spin of the non-collinear
partons: indeed, $\sp$ only depends on the spin indices of the collinear partons
and of the parent collinear parton. Since the right-hand side of 
Eq.~(\ref{mpdef}) (or, equivalently, the left-hand side of Eq.~(\ref{factallm2}))
involves the (implicit) sum over the spins of the collinear partons,
the squared splitting matrix ${\bf P}$ can {\em only} depend on the spin 
indices,  $s$ and $s^\prime$, of the parent collinear parton
(as explicitly denoted in Eq.~(\ref{pmat})). The parent collinear parton
can be either a quark (antiquark) or a gluon, and we recall
(see, e.g., Ref.~\cite{Catani:1999ss}) that the spin structure
of $\left[\,{\bf P}\,\right]_{s^\prime s}$ is different in these two cases.
{\em Fermion} helicity is conserved by QCD radiation from {\em massless}
quarks (antiquarks). Therefore, if the parent collinear parton is a 
{\em fermion} (quark or antiquark), we can consider the helicity basis in spin
space and the squared splitting matrix ${\bf P}$ turns out to be {\em diagonal}
in this basis: 
actually, due to parity invariance,
we simply have 
$\left[\,{\bf P}\,\right]_{s^\prime s} \propto \delta_{s^\prime s}$.
An analogous reasoning cannot be applied if the parent collinear parton is a
gluon. In the {\em gluon} case, 
$\left[\,{\bf P}\,\right]_{s^\prime s}$ has a {\em non-trivial}
dependence on the gluon spin indices $s$ and ${s^\prime}$. This dependence 
has to carefully be taken into account 
(see, e.g., Refs.~\cite{csdip} and \cite{weinzierl,GehrmannDeRidder:2005cm})
to achieve the cancellation of IR, soft and collinear divergences in calculations
of cross sections.
Moreover (see Refs.~\cite{Catani:2010pd, Nadolsky:2007ba}), the dependence of 
$\left[\,{\bf P}\,\right]_{s^\prime s}$ on the gluon spin indices can lead to
physically-observable and {\em logarithmically-enhanced} effects in specific
kinematical configurations.

The colours of the collinear partons are (implicitly) summed
on the right-hand side of Eq.~(\ref{mpdef}) (or, equivalently, 
on the left-hand side of Eq.~(\ref{factallm2})) and thus, in general, the squared 
splitting matrix ${\bf P}$ depends on the colour indices of the parent collinear
parton and of the non-collinear partons 
(as explicitly denoted in Eq.~(\ref{pmat})). In the TL collinear region, strict
factorization is valid: since $\sp$ is independent of the colours of the 
non-collinear partons, the squared matrix ${\bf P}$ is also independent of them
and it can only depend on the colours, $c$ and $c^\prime$, of the parent
collinear parton. Owing to colour conservation, this residual colour dependence
is, however, trivial:
it is diagonal and simply proportional to $\delta_{c^\prime c}$. In
summary, considering the TL collinear region, ${\bf P}$ is proportional to the 
unit matrix in colour space, and the matrix structure in Eq.~(\ref{pmat})
simplifies as follows:
\beq
\label{pmattl}
\left[\, {\cal P} (p_1,\dots,p_m;{\widetilde P})\,\right]_{s^\prime s}
\;\delta_{c^\prime c} \;\delta_{c^\prime_{m+1} c_{m+1}} \;\dots 
\;\delta_{c^\prime_{n} c_{n}} \;\;, \;\;\quad ({\rm TL \,\;coll. \,\;lim.})
\;\;,
\eeq
where the matrix 
${\cal P}_{s^\prime s}$ only depends on the spin indices of the parent collinear
parton. The strictly factorized version of Eq.~(\ref{factallm2})
in the TL collinear region is
\beq
\label{factallm2tl}
|\cm|^2 \;
\simeq \sum_{s, s^\prime}
\; \left[\, {\cal P} (p_1,\dots,p_m;{\widetilde P})\,\right]_{s^\prime s}
\;\la \, \cmbar \,\ket{s^\prime} \;\la \, s \,\ket{\cmbar} \;\;,
\;\;\quad ({\rm TL \,\;coll. \,\;lim.}) \;.
\eeq
In the SL collinear region, the simplified structure of Eqs.~(\ref{pmattl})
and (\ref{factallm2tl}) is valid only at the tree level (i.e. it applies only 
to ${\bf P}^{(0,R)}$ in Eqs.~(\ref{mp0}) and (\ref{factm20})). Beyond the tree
level, the SL collinear limit violates strict factorization,
and
the perturbative contributions ${\bf P}^{(1,R)}$ and 
${\bf P}^{(2,R)}$ in Eqs.~(\ref{factm21}) and (\ref{factm22})
depend on the non-collinear partons (as generically denoted in
Eq.~(\ref{pmat})). In the following subsections, we present explicit expressions 
for the one-loop and two-loop collinear matrices, 
${\bf P}^{(1,R)}$ and ${\bf P}^{(2,R)}$, and we discuss the structure of the
terms that produce violation of strict collinear factorization
at the squared amplitude level in the SL collinear region.

\subsection{Two-parton collinear limit of squared amplitudes}
\label{sec:2square}

We consider the two-parton collinear splitting $A \to A_1A_2$ in a generic
(TL or SL) kinematical region. The corresponding tree-level collinear matrix
${\bf P}^{(0,R)}={\bf P}^{(0,R)}(p_1,p_2;\wp)$ 
is strictly factorized. It can be computed (see Eq.~(\ref{mp0}))
by squaring the splitting matrices in
Eqs.~(\ref{qqg0})--(\ref{ggg0}).
The four-dimensional expression for ${\bf P}^{(0,R)}$ is well known 
\cite{Altarelli:1977zs}, and the explicit $d$-dimensional expressions in various
variants of dimensional regularization are given in Ref.~\cite{Catani:1996pk}.

The one-loop collinear matrix ${\bf P}^{(1,R)}$ is computed by inserting
the splitting matrix $\sp^{(1,R)}$ of Eq.~(\ref{sp2oneren})
in Eq.~(\ref{mp1}). We straightforwardly obtain the result
\beq
\label{mp1two}
{\bf P}^{(1,R)}= \itp^{(1)}(\ep) \;{\bf P}^{(0,R)} +
\left[ \left(\, \sp^{(0,R)} \,\right)^\dagger \,\sp^{(1,R)}_H
+\; {\rm h.c.} \;\right] \;\;,
\eeq
where the IR divergent factor $\itp^{(1)}(\ep)$ is
\beq
\label{itp1}
\itp^{(1)}(\ep) = \itc^{(1)}(\ep) +\; {\rm h.c.} \;\;.
\eeq
Inserting the explicit expression of the colour operator $\itc^{(1)}$
(see Eq.~(\ref{it2cone})) in Eq.~(\ref{itp1}), the factor $\itp^{(1)}$
turns out to be a $c$-number (more precisely, $\itp^{(1)}$ is simply 
proportional
to the unit matrix in colour space). We explicitly obtain the following
expression:
\beeq
\label{itp1expl}
\itp^{(1)}(\ep)
&=& \f{\as(\mu^2)}{2\pi} \;\f{1}{2} \;\,{\widetilde c}_{\Gamma}
\; \left[ \left( \frac{-s_{12} -i0}{\mu^2} \right)^{-\ep} + \; {\rm c.c.} 
\;\right]
\nn \\
&\times&
\left\{  \; \frac{1}{\ep^2} \; \Bigl( C_{12} - C_1 - C_2 \Bigr)
+ \frac{1}{\ep} \; \Bigl( \gamma_{12} - \gamma_1 - \gamma_2 \Bigr)
\right. \nn \\
&-& \left. \; \frac{1}{\ep} \, \left[ \Bigl(C_{12} + C_1 - C_2 \Bigr) 
\;f_R(\ep;z_1)
 +  \Bigl(C_{12} + C_2 - C_1 \Bigr) \;f_R(\ep;z_2)\;
\right] \;\right\}   \;\;.
\eeeq

An important conclusion to be drawn from the result in Eq.~(\ref{mp1two})
is that ${\bf P}^{(1,R)}$ is {\em strictly factorized}, despite the fact that the
corresponding one-loop splitting matrix $\sp^{(1,R)}$ violates strict collinear
factorization in the SL collinear region.
The contribution $\sp_H^{(1,R)}$ in Eq.~(\ref{sp2oneren})
is strictly factorized, and the factorization breaking terms of 
$\sp^{(1,R)}$ are entirely embodied in $\itc^{(1)}$.
However, these terms are {\em antihermitian} and, therefore, they cancel
(see Eq.~(\ref{itp1})) in the computation of the squared splitting matrix 
${\bf P}^{(1,R)}$. The absence of factorization breaking terms in 
${\bf P}^{(1,R)}$ implies that, in the specific case of the {\em two-parton}
collinear
limit, the factorization structure of Eq.~(\ref{factallm2tl}) is valid 
up to the
one-loop level (this factorization structure can be implemented in the one-loop
formula (\ref{factm21})).

The result in Eqs.~(\ref{mp1two}) and (\ref{itp1expl}) also suggests a practical 
recipe to compute ${\bf P}^{(1,R)}$ (in both the TL and SL regions)
by-passing the violation of strict collinear factorization at 
the amplitude level. The recipe is: consider the expression of 
$\sp^{(1,R)}$ in the TL collinear region (this expression is strictly 
factorized, but it depends on the function $f(\ep;x)$ that is ill-defined in the
SL collinear region), replace $f(\ep;x)$ with its (well-defined) {\em real part}
$f_R(\ep;x)$ (see Eq.~(\ref{fepr})), and use the corresponding expression of  
$\sp^{(1,R)}$ to compute ${\bf P}^{(1,R)}$. 
This practical recipe, which gives the correct result in Eq.~(\ref{mp1two}),
coincides with the `effective prescription' proposed in Refs.~\cite{Bern:2004cz}
(see Sect.~7.4 therein) to perform the analytic continuation of 
${\bf P}^{(1,R)}$ from the TL into the SL collinear regions.
An extension of this practical recipe from one-loop to two-loop level is not
feasible. Indeed, as shown below, the two-loop matrix ${\bf P}^{(2,R)}$
is not strictly factorized in the SL collinear region.

The one-loop SL prescription (recipe) that we have just illustrated was
 used in practice in the actual NNLO computation of Ref.~\cite{deFlorian:2001zd}.
This prescription is also consistent with the one-loop calculation
of the initial-initial three-parton antenna functions
(see Eq.~(3.19) in Ref.~\cite{Gehrmann:2011wi}). Incidentally, we 
note that the initial-initial antenna functions derived in
Ref.~\cite{Gehrmann:2011wi} are based on explicit one-loop computations
of specific squared amplitudes with $n=3$ QCD partons (e.g., the squared amplitude
of the DY subprocess
$q {\bar q} \to \gamma^* g$). Therefore, the universality (process independence)
of those initial-initial three-parton antenna functions is eventually
a consequence of the strict factorization of the one-loop collinear matrix
${\bf P}^{(1,R)}$ for the SL collinear limit of $m=2$ partons.

The right-hand side of the two-loop collinear relation (\ref{factm22})
has four contributions. The first three contributions depend on
${\bf P}^{(0,R)}$ and ${\bf P}^{(1,R)}$ (which are strictly factorized),
and the last contribution depends on ${\bf P}^{(2,R)}$.
The two-loop collinear matrix ${\bf P}^{(2,R)}$ can be computed by inserting
Eqs.~(\ref{sp2oneren}), (\ref{sp2twoc}) and (\ref{i22part})
in Eq.~(\ref{mp2}). Performing some algebraic operations, we obtain
\beeq
\label{p22part}
{\bf P}^{(2,R)} &=& \itp^{(1)}(\ep) \;{\bf P}^{(1,R)}
+ \itp^{(2)}(\ep) \;{\bf P}^{(0,R)} 
+ \left( \sp_H^{(1,R)} \,\right)^\dagger \,\sp_H^{(1,R)}
\nn \\
&+& \left( \sp^{(0,R)} \,\right)^\dagger 
\;{\widetilde {\bf \Delta}}_{P}^{(2)}(\ep) \;\sp^{(0,R)}
+ \left[ \left( \sp^{(0,R)} \,\right)^\dagger
\,{\widetilde \sp}^{(2) \,{\rm fin.}} 
+\; {\rm h.c.} \;\right] \;\;,
\eeeq
where $\itp^{(1)}(\ep)$ is given in Eq.~(\ref{itp1expl}), and the two-loop
IR divergent factor $\itp^{(2)}(\ep)$ is
\beeq
\label{itp2expl}
\itp^{(2)}(\ep) &=& - \, \f{1}{2} \left[ \itp^{(1)}(\ep) \right]^2
+ \f{\as(\mu^2)}{2\pi} 
\left\{ \, \frac{1}{\ep} \,\cbet0
\left[ \, \itp^{(1)}(2\ep) - \itp^{(1)}(\ep) \right]
+ K \;\itp^{(1)}(2\ep) 
\right. \nn \\
&+& \, \f{\as(\mu^2)}{2\pi} 
\left[ \left(\f{-s_{12} -i0}{\mu^2} \right)^{\!-2\ep}
+\; {\rm c.c.} \;\right]
\left.
 \,\frac{1}{\ep} \;
\Bigl( 
\,H^{(2)}_1 + H^{(2)}_2 
\Bigr. 
\Bigl. \left.
- H^{(2)}_{12}  \right. \Bigr)
\right\} \;\;.
\eeeq
Since $\itp^{(2)}(\ep)$ is a $c$-number,  the first three terms on 
the right-hand
side of Eq.~(\ref{p22part}) are strictly factorized. 
The last two terms, instead, lead to violation of strict collinear
factorization. The last term, which depends on 
${\widetilde \sp}^{(2) \,{\rm fin.}}$, is IR finite.
The operator ${\widetilde {\bf \Delta}}_{P}^{(2)}(\ep)$
is IR divergent and it originates from the colour operator
${\widetilde {\bf \Delta}}_{C}^{(2)}$ on the right-hand side of 
Eq.~(\ref{i22part}); we have
\beq
\label{wdel2}
{\widetilde {\bf \Delta}}_{P}^{(2)}(\ep) = 
{\widetilde {\bf \Delta}}_{C}^{(2)}(\ep) +\; {\rm h.c.} \;\;.
\eeq
Inserting the explicit expression of ${\widetilde {\bf \Delta}}_{C}^{(2)}(\ep)$
(see Eq.~(\ref{d2til})) in Eq.~(\ref{wdel2}), we obtain
\beeq
\label{wd2til}
{\widetilde {\bf \Delta}}_{P}^{(2)}(\ep) 
&=& \left(\f{\as(\mu^2)}{2\pi}\right)^2  
\;\left( \f{- s_{12}}{\mu^2} \right)^{\!\!-2\ep}
2\;\pi \; f_{abc} \,
\sum_{i=1,2} 
\;\sum_{\substack{j, k = 3 \\ j \,\neq\, k}}^n
\,T_i^a \,T_j^b \, T_k^c \;\Theta(-z_i) \;
{\rm sign}(s_{ij}) \;\Theta(-s_{jk}) \nn \\
&\times& \ln\left( 
\,\f{ s_{j\wp} \; s_{k\wp} \,z_1 z_2}{s_{jk} \,s_{12}}  \right)
\;\left[ \,- \, \f{1}{2 \,\ep^{\,2}} \, +  \f{1}{\ep} 
\;\ln\left( \f{ - z_i }{1-z_i} \right)  \right]
\;\;. 
\eeeq
Note that the argument of the logarithm in Eq.~(\ref{wd2til}) can be replaced by
using the approximation
\beq
\f{ s_{j\wp} \; s_{k\wp} \,z_1 z_2}{s_{jk} \,s_{12}}
\simeq \f{ s_{j1} \; s_{k2} }{s_{jk} \,s_{12}} 
\simeq \f{ s_{j2} \; s_{k1} }{s_{jk} \,s_{12}} \;\;,
\eeq
which is valid in the collinear limit ($p_i \simeq z_i\wp,$ with $i=1,2$).

The expression (\ref{sp2twoc}) for $\sp^{(2,R)}$ includes several terms that
violate strict collinear factorization, and most of them cancel in the
computation of ${\bf P}^{(2,R)}$. Many factorization breaking terms of 
$\sp^{(2,R)}$ have a one-loop origin and are included in the operator
$\itc^{(1)}$. These one-loop terms do not appear in ${\bf P}^{(2,R)}$: their
cancellation is due to the iterative dependence of $\sp^{(2,R)}$
on $\sp^{(1,R)}$ and $\itc^{(1)}$, and to the fact that the factorization
breaking part of $\itc^{(1)}$ is antihermitian. The two-loop factorization
breaking operator ${\widetilde {\bf \Delta}}_{C}^{(2)}$ has instead a hermitian
component that leads to the operator ${\widetilde {\bf \Delta}}_{P}^{(2)}$
(see Eqs.~(\ref{wdel2}) and (\ref{wd2til})) and, thus, to ensuing factorization
breaking terms in ${\bf P}^{(2,R)}$.
Additional (though IR finite) 
factorization breaking contributions to ${\bf P}^{(2,R)}$
are produced by ${\widetilde \sp}^{(2) \,{\rm fin.}}$.

The IR divergent colour operator ${\widetilde {\bf \Delta}}_{P}^{(2)}(\ep)$
is non-abelian and it is non-vanishing only in the SL collinear region.
The main features of ${\widetilde {\bf \Delta}}_{P}^{(2)}$ are similar to those 
of the operator ${\widetilde {\bf \Delta}}_{C}^{(2)}$
(see the related discussion in Sect.~\ref{sec:2part2loop}).
In particular, ${\widetilde {\bf \Delta}}_{C}^{(2)}$ and, hence, 
${\widetilde {\bf \Delta}}_{P}^{(2)}$ vanish in the case of the SL collinear
limit of the amplitudes that are involved in lepton--hadron DIS.
Moreover, it is important to note (as discussed below) that 
${\widetilde {\bf \Delta}}_{P}^{(2)}$ gives a vanishing contribution to 
${\bf P}^{(2,R)}$ in the case of scattering amplitudes $\cm(p_1,\dots,p_n)$
with $n \leq 4$ QCD partons 
(and an arbitrary number of colourless external legs).
Therefore, to detect the effect of ${\widetilde {\bf \Delta}}_{P}^{(2)}$
at the squared amplitude level, we have to consider the SL collinear limit of
amplitudes with $n \geq 5$ QCD partons. 

The explicit expression of ${\widetilde {\bf \Delta}}_{C}^{(2)}$ for a generic
matrix element with $n = 4$ QCD partons 
(if $n = 3$, ${\widetilde {\bf \Delta}}_{C}^{(2)}$  vanishes trivially,
as noticed in Sect.~\ref{sec:2part2loop})
is given in Eq.~(\ref{d2til4prime}).
Independently of the specific kinematical configuration, colour conservation
implies that ${\widetilde {\bf \Delta}}_{C}^{(2)}$ and, hence,
${\widetilde {\bf \Delta}}_{P}^{(2)}$ are proportional to a single colour charge
operator; we have
\beq
\label{wd2til4}
{\widetilde {\bf \Delta}}_{P}^{(2)}(\ep) \propto \;
f_{abc}\; T_1^a \,T_2^b \, T_3^c \;\;, \quad \quad (n=4) \;\;.
\eeq
We note that, in the case of the two-parton collinear limit, the tree-level
splitting matrix $\sp^{(0,R)}$ fulfils the following colour charge relation:
\beq
\label{2partcorrcan}
\left( \sp^{(0,R)}(p_1,p_2;\wp) \right)^\dagger  \;
f_{abc} \,T_1^a \,T_2^b \,\; \sp^{(0,R)}(p_1,p_2;\wp) = 0
\eeq 
Therefore, Eqs.~(\ref{wd2til4}) and (\ref{2partcorrcan}) imply:
\beq
\left( \sp^{(0,R)} \right)^\dagger \;{\widetilde {\bf \Delta}}_{P}^{(2)} \;
\; \sp^{(0,R)} = 0 \;\;, \quad \quad (n=4) \;\;,
\eeq 
and, hence, ${\widetilde {\bf \Delta}}_{P}^{(2)}$ does not contribute to 
${\bf P}^{(2,R)}$ (see Eq.~(\ref{p22part})) if $n=4$.

The relation (\ref{2partcorrcan}) follows from colour conservation. The proof is
very simple. Using Eq.~(\ref{col2conp}) and the fact that the Casimir factor 
$C_{12}$ is a real $c$-number ($C_{12}^*=C_{12}$), we obtain 
\beq
\label{col2conpder}
\left( \sp^{(0,R)} \right)^\dagger \;\left[ \,T_1^c \,,
({\bom T}_1 + {\bom T}_2)^2 \,\right] \;\;\sp^{(0,R)}
= \left( \sp^{(0,R)} \right)^\dagger 
\;\left( \,T_1^c \,C_{12} - C_{12} \,T_1^c \,\right) \;\sp^{(0,R)} = 0 \;\;.
\eeq
Moreover, using elementary colour algebra, the commutator in 
Eq.~(\ref{col2conpder}) gives
\beq
\label{elemcomm}
i \,\left[ \,T_1^c \,, ({\bom T}_1 + {\bom T}_2)^2 \,\right] = 
2 \,f_{abc}\; T_1^a \,T_2^b \;\;.
\eeq
Inserting Eq.~(\ref{elemcomm}) in Eq.~(\ref{col2conpder}), we get the result
in Eq.~(\ref{2partcorrcan}).

\begin{figure}[htb]
 \begin{center}
 \begin{tabular}{c}
 \epsfxsize=12truecm
\epsffile{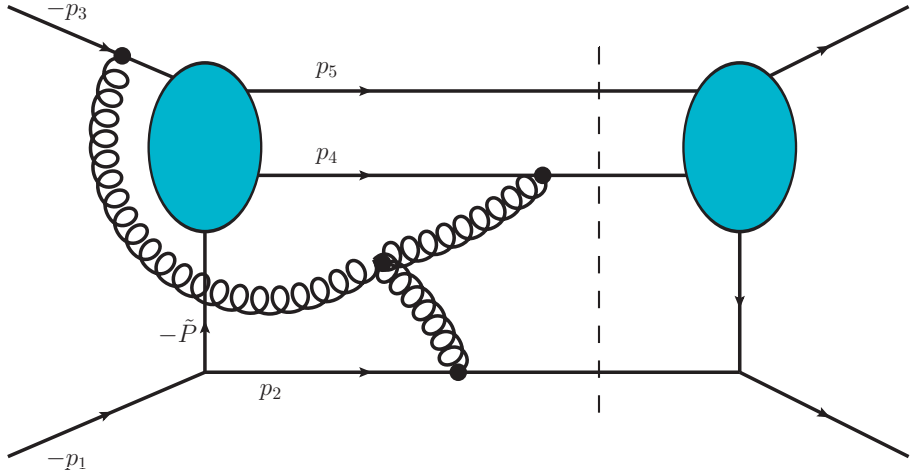} \\
 \end{tabular}
 \end{center}
 \caption{\label{fig:2loop}
 {\em Squared amplitude with $n=5$ QCD partons in parton--parton hard scattering
 (the dashed line cuts the final-state partons). Representative colour 
 structure of non-abelian factorization breaking correlations that accompany the
 two-parton SL collinear limit ($p_i \simeq z_i \wp, \;i=1,2$)
 at two-loop order.
 }}
 \end{figure}

The operator ${\widetilde {\bf \Delta}}_{P}^{(2)}$ contributes to the SL
collinear limit of the squared amplitudes (with $n \geq 5$ partons) that are
involved in hadron--hadron hard-scattering processes. We explicitly consider a
typical example with $n = 5$ partons: the process 
`parton + parton $\to$ 3 partons' (see Fig.~\ref{fig:2loop}) or, more generally, 
`parton + parton $\to X +$ 3 partons' ($X$ denotes non-QCD particles, e.g. a
vector boson), where one of the final-state partons is collinear to one of the
initial-state partons. We specify the parton momenta as follows:
the two initial-state partons have momenta $-p_1$ and $-p_3$ (i.e., 
the `energies' $p_1^0$ and $p_3^0$ are negative), and the three 
final-state partons have momenta $p_2, p_4$ and $p_5$ 
(i.e., $p_2^0, p_4^0$ and $p_5^0$ are positive).
Since $z_2 < 0 < z_1$ and $s_{45}>0$, there are only two different colour
operators that contribute to the expression (\ref{wd2til}) of
${\widetilde {\bf \Delta}}_{P}^{(2)}$: these operators are 
$f_{abc}\; T_2^a \,T_3^b \,T_4^c$ and
$f_{abc}\; T_2^a \,T_3^b \,T_5^c$. Using colour conservation
(i.e., ${\bom T}_5 = -( {\bom T}_1+{\bom T}_2+{\bom T}_3+{\bom T}_4 )$)
and the identity (\ref{tipj}), we obtain
\beq
f_{abc}\; T_2^a \,T_3^b \,T_5^c = - f_{abc}\; T_2^a \,T_3^b \,T_4^c
- f_{abc}\; T_2^a \,T_3^b \,T_1^c \;\;, \quad \quad (n=5) \;\;,
\eeq
and we can write ${\widetilde {\bf \Delta}}_{P}^{(2)}$ in terms of the colour
operators
$f_{abc}\; T_2^a \,T_3^b \,T_4^c$ and
$f_{abc}\; T_1^a \,T_2^b \,T_3^c$. From Eq.~(\ref{wd2til}), we thus obtain
\beeq
\label{wd2til5}
{\widetilde {\bf \Delta}}_{P}^{(2)}(\ep) 
&=& \f{\as^2(\mu^2)}{\pi}
\;\left( \f{- s_{12}}{\mu^2} \right)^{\!\!-2\ep}
 \, f_{abc} \,T_2^a \,T_3^b \, T_4^c \;
\ln\!\left( 
\,\f{ s_{34} \; s_{5\wp}}{s_{35} \,s_{4\wp}}  \right) 
\,\left[ \,- \, \f{1}{2 \,\ep^{\,2}} \, +  \f{1}{\ep} 
\;\ln\left( -\f{z_2 }{z_1} \right)  \right] \nn \\
&+& f_{abc} \,T_1^a \,T_2^b \, T_3^c \;\times 
\bigl( \cdots \bigr)
\;\;. 
\eeeq
On the right-hand side, we have not written the explicit expression for the 
coefficient of the colour operator $f_{abc} \,T_1^a \,T_2^b \, T_3^c$: indeed,
as a consequence of the relation (\ref{2partcorrcan}), this colour operator
does not contribute to the collinear matrix ${\bf P}^{(2,R)}$ 
(see Eq.~(\ref{p22part})).
The term proportional to $f_{abc} \,T_2^a \,T_3^b \, T_4^c$ in 
Eq.~(\ref{wd2til5}) (see Fig.~\ref{fig:2loop})
produces a non-vanishing factorization breaking contribution,
$( \sp^{(0,R)} )^\dagger \;{\widetilde {\bf \Delta}}_{P}^{(2)} \;
\; \sp^{(0,R)}$,
to ${\bf P}^{(2,R)}$.

\vspace{0.5cm}
We note that, in the case of a generic scattering amplitude with $n$ QCD
partons, the colour structure of the contribution of 
${\widetilde {\bf \Delta}}_{P}^{(2)}$ to ${\bf P}^{(2,R)}$ 
(see Eq.~(\ref{p22part})) can be simplified.
We have
\beq
\label{2partdeltabar}
\left( \sp^{(0,R)} \right)^\dagger \;{\widetilde {\bf \Delta}}_{P}^{(2)}(\ep) 
\; \; \sp^{(0,R)} = \; {\overline {\bf \Delta}}_{P}^{(2)}(\ep)
\;\;{\bf P}^{(0,R)} \;\;,
\eeq 
where the colour operator ${\overline {\bf \Delta}}_{P}^{(2)}$ acts on
the colour space of the $n-1$ external QCD partons of the reduced
matrix element $\cmbar$ (we recall that ${\bf P}^{(0,R)}$ is strictly factorized
and, hence, it is proportional to the unit matrix in colour space).
The simplified colour structure in Eq.~(\ref{2partdeltabar}) follows from the
fact that we can explicitly evaluate the action of the colour charge 
${\bom T}_i$ ($i=1,2$) of the collinear partons on the tree-level
two-parton splitting matrix $\sp^{(0,R)}$. Using basic colour algebra relations,
we straightforwardly find ($i=1,2$) 
\beq
\label{2partticharge}
\left( \sp^{(0,R)}(p_1,p_2;\wp) \right)^\dagger  \;
T_i^a \; \sp^{(0,R)}(p_1,p_2;\wp) = \;T^a_{(i)\,\wp} \;\;{\bf P}^{(0,R)}
\;\;,
\eeq 
where the colour operator ${\bom T}_{(i)\,\wp}$ is proportional to the colour
charge of the parent collinear parton $A$ with momentum $\wp$.
The explicit expression of ${\bom T}_{(i)\,\wp}$ depends on the flavour of the
collinear parton $i$ and of the parent collinear parton. Considering the flavour
structure
of the various splitting processes $A \to A_1 \,A_2$, we have
\beeq
{}\!\!\!\! f \to f\,g \;\;(f=q \;{\rm or} \;{\bar q}):&&
\quad \;\;
{\bom T}_{(f)\,f} = \left( 1 - \frac{C_A}{2 C_F} \right) \;{\bom T}_{f} \;\;,
\;\;\;\; {\bom T}_{(g)\,f} = \frac{C_A}{2 C_F}  \;{\bom T}_{f} \;\;, \\
{}\!\!\!\! g \to g\,g \;:&& 
\quad \;\;
{\bom T}_{(g)\,g} = \frac{1}{2}  \;{\bom T}_{g} \;\;, \\
{}\!\!\!\! g \to q\,{\bar q} \;:&&
\quad \;\;
{\bom T}_{(q)\,g} = \frac{1}{2}  
\;\left( \;{\bom T}_{g} + {\bom d}_{g} \right) \;\;,
\;\;\;\; {\bom T}_{({\bar q})\,g} = \frac{1}{2}  
\;\left( \;{\bom T}_{g} - {\bom d}_{g} \right) \;\;,
\eeeq
where ${\bom T}_{f}$ and ${\bom T}_{g}$ are the customary colour charges of a
fermion and a gluon, while ${\bom d}_{g}$ is the gluon colour matrix in a
{\em symmetric} `octet' state and its matrix elements are defined as
\beq
\left( d^{\,a} \right)_{bc} = \;d_{bac} \;\;, 
\quad d_{abc} \equiv 2 \;{\rm Tr}\left( t^a\{t^b,t^c\}\right) \;\;,
\eeq
and $d_{abc}$ is the fully-symmetrized trace of $t^a t^b t^c$.
The explicit expression of the colour operator
${\overline {\bf \Delta}}_{P}^{(2)}$ in Eq.~(\ref{2partdeltabar}) is
\beeq
\label{bard2}
{\overline {\bf \Delta}}_{P}^{(2)}(\ep) 
&=& \left(\f{\as(\mu^2)}{2\pi}\right)^2  
\;\left( \f{- s_{12}}{\mu^2} \right)^{\!\!-2\ep}
2\;\pi \; f_{abc} \,
\sum_{i=1,2} 
\;\sum_{\substack{j, k = 3 \\ j \,\neq\, k}}^n
\,T^a_{(i)\,\wp} \,T_j^b \, T_k^c \;\Theta(-z_i) \;
{\rm sign}(s_{ij}) \;\Theta(-s_{jk}) \nn \\
&\times& \ln\left( 
\,\f{ s_{j\wp} \; s_{k\wp} \,z_1 z_2}{s_{jk} \,s_{12}}  \right)
\;\left[ \,- \, \f{1}{2 \,\ep^{\,2}} \, +  \f{1}{\ep} 
\;\ln\left( \f{ - z_i }{1-z_i} \right)  \right]
\;\;. 
\eeeq
This expression is directly obtained by inserting Eqs.~(\ref{wd2til}) and
(\ref{2partticharge}) in the left-hand side of Eq.~(\ref{2partdeltabar}).


\subsection{Multiparton collinear limit of squared amplitudes}
\label{sec:msquare}

In this subsection we consider the multiparton splitting $A \to A_1 \dots A_m$ 
of $m$ ($m \geq 3$) collinear partons.
The corresponding tree-level collinear matrix ${\bf P}^{(0,R)}$
fulfils strict factorization. The explicit expressions of
${\bf P}^{(0,R)}$ for all the
flavour configurations of $m=3$ collinear partons were
computed in Refs.~\cite{Campbell:1997hg, Catani:1998nv, Catani:1999ss}.

The one-loop collinear matrix ${\bf P}^{(1,R)}$ is obtained by inserting 
Eqs.~(\ref{sp1df}) and (\ref{sp1div}) in Eq.~(\ref{mp1}). We have
\beq
\label{p1mpart}
{\bf P}^{(1,R)}= \left(\, \sp^{(0,R)} \,\right)^\dagger \;\imp(\ep) 
\;\sp^{(0,R)}
+ \left[ \left(\, \sp^{(0,R)} \,\right)^\dagger \,\sp^{(1) \,{\rm fin.}}
+\; {\rm h.c.} \;\right] \;\;,
\eeq
where $\imp$ is given in terms of the colour operator $\imc$
in Eqs.~(\ref{sp1div}) and (\ref{i1mcexd}):
\beq
\label{imp1m}
\imp(\ep) = \imc(\ep) + \; {\rm h.c.} \;\;.
\eeq
Using Eq.~(\ref{i1mcexd}), the explicit expression of $\imp$ is
\beeq
\label{i1mpexd}
\imp(\ep) 
&=&\f{\as(\mu^2)}{2\pi} \;
\left\{ \left( \,\frac{1}{\ep^2} \,C_{\wp} + \frac{1}{\ep} \,\gamma_{\wp} \right)
- \sum_{\substack{i \,\in \, C \\ {}}}
\left( \,\frac{1}{\ep^2} \,C_i + \frac{1}{\ep} \,\gamma_i
- \,\frac{2}{\ep} \,C_i \,\ln |z_i| \right)
\right. \nn \\
&-& 
\frac{1}{\ep} \left.\sum_{\substack{i,\ell \,\in C \\ i \,\neq\, \ell}} 
{\bom T}_i \cdot {\bom T}_\ell
\, \ln\left(\f{ s_{i \ell} }{z_i \, z_{\ell} \, \mu^2}\right)
 \right\}  \;\;.
\eeeq

The factorization breaking contribution $\dmc$ (see Eq.~(\ref{del1}))
to $\imc$ is antihermitian and, thus, it cancels in Eq.~(\ref{imp1m}).
The operator $\imp$ depends on the colour charges of the collinear partons, but
it is independent of the non-collinear partons. Therefore,
the first term on the right-hand side of Eq.~(\ref{p1mpart})
does not violate strict collinear factorization.
This term produces a strictly factorized and IR divergent contribution to
${\bf P}^{(1,R)}$.

The second term on the right-hand side of Eq.~(\ref{p1mpart}) is IR finite,
since it depends on the IR finite contribution $\sp^{(1) \,{\rm fin.}}$ 
to the one-loop splitting matrix. As mentioned in Sect.~\ref{sec:irone},
$\sp^{(1) \,{\rm fin.}}$ contains factorization breaking terms in the SL
collinear region and, therefore, their interference with  
$\sp^{(0,R)}$ can produce violation of strict collinear factorization 
at the squared amplitude level.
In this respect, the main difference between the
two-parton and multiparton collinear limits is that in the latter case several
different colour structures contribute to the splitting matrix 
$\sp$ already at the tree level (in the two-parton case, $\sp^{(0)}$ involves a
single colour structure: see Eq.~(\ref{spvssplit})).
The result in Eqs.~(\ref{p1mpart}) and (\ref{i1mpexd})
for the multiparton collinear limit shows that the IR divergent part of
${\bf P}^{(1,R)}$ is strictly factorized, while 
the IR finite part 
can contain terms that produce violation of strict
factorization in the squared amplitudes at the one-loop level
(see Eq.~(\ref{factm21})).

The two-loop collinear matrix ${\bf P}^{(2,R)}$ (see Eq.~(\ref{mp2}))
of the multiparton collinear limit can be computed by using
Eqs.~(\ref{sp1df}), (\ref{sp1div}),
(\ref{sp2divfin}), (\ref{sp2mcdiv}) 
and the explicit form of the colour operators
$\imc$ and $\imctwo$
(see Eqs.~(\ref{i1mcexd}) and (\ref{i2mexpl})). 
Performing some straightforward (though, slightly cumbersome)
algebraic operations, we can write the result in the following form:
\beq
\label{p22mpart}
{\bf P}^{(2,R)}= {\bf P}^{(2,R)}_{\rm f.} + {\bf P}^{(2,R)}_{\rm n.f.} \;\;,
\eeq
where the contribution ${\bf P}^{(2,R)}_{\rm f.}$ is strictly factorized,
while the term 
${\bf P}^{(2,R)}_{\rm n.f.}$ includes {\em all} the contributions 
that violate
strict collinear factorization (this term also includes additional 
contributions that are strictly factorized).

The strictly-factorized term ${\bf P}^{(2,R)}_{\rm f.}$ does not depend on the
non-collinear partons. Its explicit expression is
\beeq
\label{p2mf}
{\bf P}^{(2,R)}_{\rm f.} &=& \left(\, \sp^{(0,R)} \,\right)^\dagger
\left\{ \left[ \,\f{1}{2} \;\imp(\ep) \,{\bom I}^{(1)}_{m\,C,\,{\rm f.}}(\ep)
+\; {\rm h.c.} \;\right] \right. \nn \\
&+&\left. 
\f{\as(\mu^2)}{2\pi} 
\left( \, \frac{1}{\ep} \,\cbet0 
\left[ \imp(2\ep) - \imp(\ep) \right]
+ K \;\imp(2\ep) 
\right) \right. \nn \\
&+& \, \left( \f{\as(\mu^2)}{2\pi} \right)^2 
\left. \left.
 \,\frac{2}{\ep} \;
\Bigl( \;\sum_{i \, \in C} \,H^{(2)}_i 
\Bigr. 
\Bigl. 
- H^{(2)}_{\wp}  \right. \Bigr)
\right\}\;\sp^{(0,R)}
\eeeq
where $\imp$ is the one-loop operator in Eq.~(\ref{i1mpexd}), and the colour
operator ${\bom I}^{(1)}_{m\,C,\,{\rm f.}}$ is obtained from 
$\imc$ (see Eq.~(\ref{i1mcexd}))
by removing its
factorization breaking part $\dmc$:
\beq
{\bom I}^{(1)}_{m\,C,\,{\rm f.}}(\ep) = \imc(\ep) - \dmc(\ep) \;\;.
\eeq

The contribution ${\bf P}^{(2,R)}_{\rm n.f.}$ on the right-hand side of
Eq.~(\ref{p22mpart}) has the following expression:
\beeq
\label{p2mnf}
{\bf P}^{(2,R)}_{\rm n.f.} &=& 
\left(\, \sp^{(0,R)} \,\right)^\dagger \;{\bf \Delta}_{m \,P}^{(2;\,2)}(\ep)
\;\sp^{(0,R)} \nn \\
&+& \left[ \left(\, \sp^{(0,R)} \,\right)^\dagger \left(
\,\imp(\ep) \;\sp^{(1) \,{\rm fin.}} + {\overline \sp}^{(2) \,{\rm div.}}
+ \sp^{(2) \,{\rm fin.}} \,\right)
+\; {\rm h.c.} \;\right] \nn \\
&+&  \left(\, \sp^{(1) \,{\rm fin.}} \,\right)^\dagger  
\;\sp^{(1) \,{\rm fin.}} \;\;,
\eeeq
where the colour operator ${\bf \Delta}_{m \,P}^{(2;\,2)}$ is
\beq
\label{del2mcvsp}
{\bf \Delta}_{m \,P}^{(2;\,2)}(\ep) = 
\left( \,{\bf \Delta}_{m \,C}^{(2;\,2)}(\ep) +\; {\rm h.c.}\;
\right) + \frac{1}{2} \left[ \;\imp(\ep), \; \dmc(\ep) \;
\right] \;\;.
\eeq

The last term on the right-hand side of Eq.~(\ref{p2mnf}) is IR finite,
since it 
is given by the square of  
the IR finite contribution $\sp^{(1) \,{\rm fin.}}$
to the one-loop splitting matrix (see Eq.~(\ref{sp1df})).
As already recalled, $\sp^{(1) \,{\rm fin.}}$ contains factorization breaking
terms in the SL collinear region.

The term in the square bracket on the right-hand side of Eq.~(\ref{p2mnf})
is IR divergent. This term depends on $\sp^{(1) \,{\rm fin.}}$,
on the colour operator $\imp$ in Eq.~(\ref{i1mpexd}), and on the contributions
$\sp^{(2) \,{\rm fin.}}$ (see Eq.~(\ref{sp2divfin}))
and ${\overline \sp}^{(2) \,{\rm div.}}$ (see Eq.~(\ref{sp2mcdiv}))
to the two-loop splitting matrix $\sp^{(2,R)}$.
As stated in Sect.~\ref{sec:mpart2loop},
${\overline \sp}^{(2) \,{\rm div.}}$ contains factorization breaking
contributions that are IR divergent at the level of single poles $1/\ep$.
The operator $\imp$ contains also a double-pole term with a $c$-number
coefficient (see Eq.~(\ref{i1mpexd})) and, thus, it produces the following IR 
divergent contribution to the square-bracket term in Eq.~(\ref{p2mnf}):
\beeq
\label{dpfact}
\left( \sp^{(0,R)} \,\right)^\dagger \! 
\imp(\ep) \,\sp^{(1) \,{\rm fin.}} +\, {\rm h.c.} \!\!\!&=&\!\!\!
\f{\as(\mu^2)}{2\pi}  \Bigl( C_{\wp}   
- \sum_{\substack{i \,\in \, C \\ {}}} C_i  \Bigr)
\frac{1}{\ep^2} \left[ \left( \sp^{(0,R)} \,\right)^\dagger
\!\sp^{(1) \,{\rm fin.}} +\; {\rm h.c.} \right] \nn \\
\!\!\!&+&\!\!\! {\cal O}(1/\ep) \;\;.
\eeeq
We note that the double-pole term in Eq.~(\ref{dpfact}) is simply proportional
to the factorization breaking contribution
to the one-loop collinear matrix ${\bf P}^{(1,R)}$ (see Eq.~(\ref{p1mpart})).

The colour operator ${\bf \Delta}_{m \,P}^{(2;\,2)}$ in Eq.~(\ref{del2mcvsp})
depends on the two-loop factorization breaking term 
${\bf \Delta}_{m \,C}^{(2;\,2)}$
(see Eqs.~(\ref{i2mexpl}) and (\ref{d22mc})), on
the one-loop operator $\imp(\ep)$, and
on the one-loop factorization breaking term $\dmc$
(see Eqs.~(\ref{i1mcexd}) and (\ref{del1})).
The commutator on the right-hand side of Eq.~(\ref{del2mcvsp})
originates from a {\em non-abelian}  interference of two types of one-loop
contributions: the one-loop absorptive (antihermitian) 
contribution to the scattering amplitude (the absorptive interaction involves a
collinear and a non-collinear parton) and the one-loop radiative (hermitian)
contribution to the complex conjugate amplitude
(the radiative interaction involves two collinear partons).
Using Eqs.~(\ref{del1}) and 
(\ref{i1mpexd}), we evaluate the commutator and we obtain the following 
explicit expression of ${\bf \Delta}_{m \,P}^{(2;\,2)}$:
\beeq
\label{del2mp}
{\bf \Delta}_{m \,P}^{(2;\,2)}(\ep) &=&\!\! \left(\f{\as(\mu^2)}{2\pi}\right)^2  
\left( \,- \, \f{1}{\ep^2} \right) \;\pi \; f_{abc} \nn \\
&\times&\!\! \left[ \;\sum_{i \in C} 
\;\sum_{\substack{j, k \,\in NC \\ j \,\neq\, k}} 
\,T_i^a \,T_j^b \, T_k^c \;\Theta(-z_i) \;
{\rm sign}(s_{ij}) \;\Theta(-s_{jk}) \;
\ln\left( \,\f{ s_{j\wp} \; s_{k\wp}}{s_{jk} \,\mu_0^2} \right) \right. 
\nn \\
&-& \left.
\;\sum_{\substack{i, \ell \,\in C \\ i \,\neq\, \ell}}
\;\sum_{j \in NC} \;\,T_i^a \,T_{\ell}^b \, T_j^c \;\Theta(-z_i) \;
\Theta(-s_{i \ell}) \; {\rm sign}(s_{ij})\;
\,\ln\left( \,\f{ s_{i \ell}}{z_i z_{\ell} \,\mu_0^2} \right) \;
\right] .
\eeeq
Note that the argument of the logarithms in the square bracket depends on 
the scale $\mu_0$, which is arbitrary (we can possibly set $\mu_0=\mu$).
However, the expression (\ref{del2mp}) is actually independent of $\mu_0$;
the independence of $\mu_0$ directly follows from the relation
(\ref{3vs3}) (to apply Eq.~(\ref{3vs3}), we note that the explicit constraint
$s_{i \ell} < 0$ can be removed from the second term in the 
square bracket of Eq.~(\ref{del2mp}); indeed, the sum of the terms with 
$s_{i \ell} > 0$ gives a vanishing contribution to Eq.~(\ref{del2mp})).

The operator ${\bf \Delta}_{m \,P}^{(2;\,2)}$ is IR divergent and proportional
to the double pole $1/\ep^2$. The first term in the square bracket 
of Eq.~(\ref{del2mp}) involves non-abelian contributions that correlate the
colour charges of {\em two non-collinear} partons (and a collinear parton).
In Eq.~(\ref{p2mnf}), these contributions produce factorization breaking terms 
that cannot be cancelled by the IR divergent term of 
Eq.~(\ref{dpfact}) (indeed, as mentioned in 
Sect.~\ref{sec:irone}, $\sp^{(1) \,{\rm fin.}}$ embodies factorization 
breaking correlations with a {\em single non-collinear} parton).
Therefore, the explicit expression (\ref{del2mp}) of 
${\bf \Delta}_{m \,P}^{(2;\,2)}$ shows that the two-loop multiparton collinear
matrix ${\bf P}^{(2,R)}$
necessarily includes non-abelian contributions that lead to violation of strict
collinear factorization at the squared amplitude level.

A special exception to this conclusion about violation of strict collinear
factorization regards the case of the SL collinear limit in lepton--hadron DIS.
In the DIS case, the operator ${\bf \Delta}_{m \,P}^{(2;\,2)}$
effectively takes a form that is independent of the non-collinear partons.
To show this, we consider a DIS matrix element 
$\cm(p_1,\dots,p_m,p_{m+1},\dots,p_n)$: the initial-state parton has `outgoing'
momentum $-p_1$ ($p_1^0 < 0$), the momenta of the final-state collinear 
partons are $p_2,\dots,p_m$, and the momenta of the final-state non-collinear
partons are $p_{m+1},\dots,p_n$. If $j,k \in NC$, we have $s_{jk} > 0$:
thus, the correlation terms of Eq.~(\ref{del2mp}) that depend on two
non-collinear partons vanish. As for the remaining 
correlation terms of Eq.~(\ref{del2mp}), the sign of $s_{ij}$ is independent
of $j$ if $j \in NC$ (actually, $s_{ij} > 0$ if $i \in C$ and $z_i < 0$);
therefore, we can perform the sum over $j$ by using colour conservation
$(\sum_{j\in NC} {\bom T}_j= - \sum_{r\in C} {\bom T}_r)$. The final expression
of ${\bf \Delta}_{m \,P}^{(2;\,2)}$ in lepton--hadron DIS is
\beeq
\label{del2mpdis}
{\bf \Delta}_{m \,P}^{(2;\,2)}(\ep) &=&\!\! 
\left(\f{\as(\mu^2)}{2\pi}\right)^2  
\left( \,- \, \f{1}{\ep^2} \right) \;\pi \; f_{abc} \nn \\
&\times&\!\!     
\;\sum_{\substack{i, \ell \,\in C \\ i \,\neq\, \ell}}
\;\,T_i^a \,T_{\ell}^b \;\sum_{r \in C}\, T_r^c \;\Theta(-z_i) \;
\Theta(-s_{i \ell}) \; 
\,\ln\left( \,\f{ s_{i \ell}}{z_i z_{\ell} \,\mu_0^2} \right) \;, 
\quad ({\rm DIS}) \;.
\eeeq
This expression is independent of the non-collinear partons and, thus, it 
effectively has a strictly-factorized form. This form is a consequence of a
colour coherence mechanism (due to DIS kinematics and colour conservation).

In generic kinematical configurations (typically, those that occur in
hadron--hadron hard-scattering processes), the two-loop multiparton collinear
matrix ${\bf P}^{(2,R)}$ is not strictly factorized. We note that the two-loop
collinear formula (\ref{factm22}) 
for the squared amplitudes includes factorization breaking 
contributions
that are due to both ${\bf P}^{(1,R)}$ and ${\bf P}^{(2,R)}$. 
The factorization breaking contributions
that are due to ${\bf P}^{(1,R)}$ involve correlations with one non-collinear
parton, while the contributions due to ${\bf P}^{(2,R)}$ also involve
correlations with two non-collinear partons.

\setcounter{footnote}{2}

\subsection{Cross sections and violation of strict collinear factorization:
some remarks}
\label{sec:crosssec}

The applicability of perturbative QCD to the calculation of cross sections for
hard-scattering processes is based on the universal (process-independent) 
factorization theorem of mass singularities \cite{Collins:1989gx}.
According to this factorization picture, the {\em sole} uncancelled IR
divergences that are eventually encountered in the computation of inclusive
partonic cross sections are due to partonic states whose momenta are collinear 
to the momenta of the colliding partons or of triggered final-state partons;
these uncancelled divergences are factorizable in a process-independent
form and, therefore, they can be removed by a formal redefinition
(`renormalization') of the `bare' parton densities and parton fragmentation
functions. 

The violation of strict collinear factorization at the level of squared
amplitudes certainly challenges the validity of universal mass-singularity
factorization at the cross section level. 
We present some
comments and remarks on this issue.

We first present a general comment. The singular behaviour of the 
{\em two-parton} collinear limit at {\em one-loop} order is one of the key 
ingredients that are used to handle IR divergences and mass singularities
and to cancel the IR divergences in perturbative QCD computations of
hard-scattering processes at the NNLO
(see, e.g., Refs.~\cite{Anastasiou:2005qj, Aglietti:2008fe, Catani:2009sm, 
Weinzierl:2009yz, Czakon:2011ve, Gehrmann:2011wi, Boughezal:2011jf, glover} 
and references therein).
In this context, it is reassuring that the one-loop two-parton 
collinear matrix ${\bf P}^{(1,R)}$ is strictly factorized even in the SL
collinear region (see Eqs.~(\ref{mp1two})--(\ref{itp1expl})).
This result guarantees that the extension of NNLO methods from lepton--lepton
collisions to lepton--hadron and hadron--hadron collisions does not involve
additional conceptual difficulties related to the violation of strict collinear
factorization.
 
In the following, rather than considering the issue of mass-singularity 
factorization in completely general terms, we limit our
discussion to the simplest case in which our study has definitely uncovered
the presence of strict-factorization breaking effects at the squared
amplitude level. We thus consider the inclusive production of a high-$p_T$
hadron or jet (with at least one recoiling jet) in the collision between two
high-energy hadrons.

At the leading order
(LO) in QCD perturbation theory, this production process is controlled by the
square of the tree-level amplitude (with $n=4$ partons) of the corresponding
partonic subprocess
\beq
\label{prolo} 
{\rm ~parton + parton} \to 2~{\rm partons~} \;\;. 
\eeq 
Part of the higher-order QCD corrections to this `$2 \to 2$' partonic subprocess
are obtained by considering the squared amplitude of the process
\beq
\label{prohoc}
 {\rm ~parton + parton} \to 3~{\rm partons~  }  \;\;,
\quad \;\;({\rm one\;low-}p_T \;\,{\rm final\!-\!state \;parton}) \;,
\eeq  
in the kinematical region where one of the three final-state partons 
(the `low-$p_T$ parton') is collinear to one of the initial-state partons. 
The IR divergences produced by the phase space integration over the SL collinear
region of the low-$p_T$ final-state parton have to be factorized with respect to the
corresponding LO `$2 \to 2$' partonic subprocess, and the IR factor has to
be strictly factorized, namely, it has to be independent of the kinematical and
colour flow structures of the LO  partonic subprocess in Eq.~(\ref{prolo}).
As explicitly shown in Sect.~\ref{sec:2square}, the squared amplitude of the
process in Eq.~(\ref{prohoc}) violates strict factorization in the SL collinear
region. This violation of strict factorization starts at the two-loop level 
(see Eqs.~(\ref{p22part}) and (\ref{wd2til5}), and Fig.~\ref{fig:2loop})
and, therefore, it may invalidate the factorization theorem of mass singularities
starting at the 
N$^3$LO
in QCD perturbation theory.
The validity of the factorization theorem can be recovered only {\em if}
the N$^3$LO factorization breaking effects produced by the two-parton SL
collinear limit of the process in Eq.~(\ref{prohoc})
are {\em cancelled} by corresponding factorization breaking effects 
produced by other partonic subprocesses. An explicit quantitative proof of such a
cancellation would represent a highly non-trivial check of the validity of 
the factorization theorem beyond the 
NNLO.

We continue our discussion of the cancellation mechanism at the qualitative
level and, eventually, we shall identify a 
sole additional  
partonic configuration 
that can lead to factorization breaking correlations with 
the same structure as in Eq.~(\ref{wd2til5}) (see Fig.~\ref{fig:2loop}).
The contribution of this partonic configuration (after combination with all the
other IR divergent terms) can cancel the violation of strict collinear
factorization that is produced by the subprocess in Eq.~(\ref{prohoc}).

To reach this conclusion, we consider the various high-order subprocesses that
lead to the N$^3$LO IR divergent contributions to the cross section controlled by
the LO process in Eq.~(\ref{prolo}), and we group them in three classes:
$i)$ virtual and soft-parton subprocesses, $ii)$ collinear-parton subprocesses,
and $iii)$ mixed soft-collinear parton subprocesses.

The class $\,i)$ contains the virtual three-loop corrections to the 
`$2 \to 2$' process in Eq.~(\ref{prolo}) and the $(3-k)$-loop corrections to the
processes `parton + parton $\to$ 2 partons + $k$ soft partons' with $k=1,2,3$.
Owing to kinematics, the IR divergences produced by this class cannot cancel 
the factorization breaking effects due to the two-loop corrections of the process
in Eq.~(\ref{prohoc}). In fact, the virtual and soft-parton radiation that
accompanies the basic `$2 \to 2$' partonic process cannot match the kinematics 
of the process `parton + parton $\to$ 3 partons' in the region where the
final-state collinear parton has a large longitudinal-momentum fraction
(e.g., $p_2 \simeq z_2 \wp$ with $|z_2| \sim |z_1| \sim {\cal O}(1)$ in 
Eqs.~(\ref{p22part}) and (\ref{wd2til5}), and in Fig.~\ref{fig:2loop}).
The subprocesses of the class $\,i)\,$ can (partly) cancel the IR divergences
of the subprocess in Eq.~(\ref{prohoc}) only in the kinematical region
where the final-state collinear parton is also soft (e.g., at the kinematical
endpoint $z_1=1-z_2 \to 1$ in Eq.~(\ref{p22part})).

The class $\,ii)$ contains the subprocess in Eq.~(\ref{prohoc}) at the two-loop
level and two other contributions: the tree-level subprocess
`parton + parton $\to$ 5 partons' with three low-$p_T$ final-state partons,
and the one-loop subprocess `parton + parton $\to$ 4 partons' with two 
low-$p_T$ final-state partons. The subprocess `parton + parton $\to$ 5 partons'
produces IR divergent terms from the tree-level SL collinear limit of $m=4$
partons: this tree-level collinear limit is strictly factorized.
The  subprocess 
\beq
\label{prohomc}
{\rm ~parton + parton} \to 4~{\rm partons~} \;\;
\quad \;\;({\rm two\;low-}p_T \;\,{\rm final\!-\!state \;partons}) \;
\eeq  
produces IR divergent terms from the one-loop SL collinear limit of $m=3$
partons: this collinear limit violates strict factorization but, as discussed in
Sects.~\ref{sec:multi}, \ref{sec:irone} and \ref{sec:msquare}
(see Eq.~(\ref{p1mpart})), 
the corresponding factorization
breaking terms involve correlations with a single non-collinear parton.
Since the factorization breaking effects due to the two-loop
subprocess in Eq.~(\ref{prohoc}) also involve correlations with two non-collinear
partons (see Eq.~(\ref{wd2til5}) and Fig.~\ref{fig:2loop}), these effects cannot
be fully
cancelled by the contributions of the other subprocesses in the class~$\,ii)$.

\begin{figure}[htb]
 \begin{center}
 \begin{tabular}{c}
 \epsfxsize=12truecm
 \epsffile{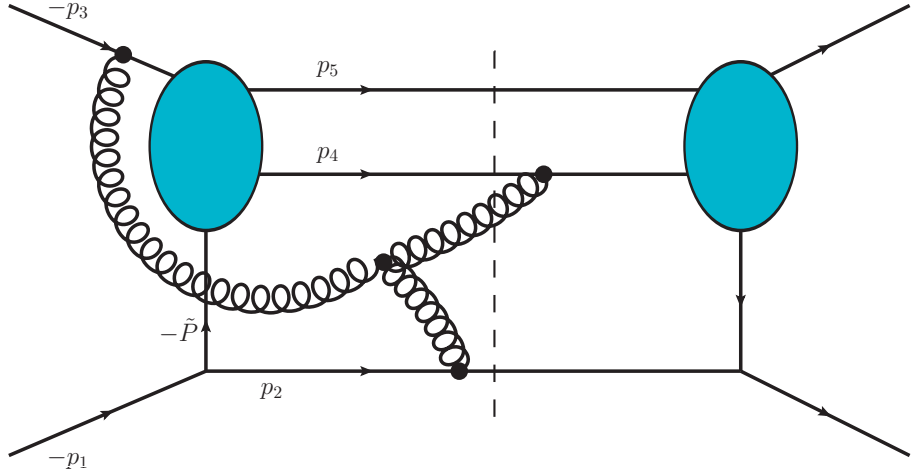} \\
 \end{tabular}
 \end{center}
 \caption{\label{fig:1loop}
 {\em Squared amplitude for parton--parton hard-scattering with three QCD partons
 and one soft gluon in the final state (the dashed line cuts the final-state
 partons). Representative colour structure of non-abelian
 factorization breaking correlations
 that accompany the
 mixed soft-collinear limit (with two collinear partons,
 $p_i \simeq z_i \wp, \;i=1,2$, in the SL region)
 at one-loop order.
  }}
 \end{figure}

The class $\,iii)$ contains two tree-level subprocesses:
the subprocess
`parton + parton $\to$ 4 partons + $1$ soft parton',
which produces IR divergent terms from the SL collinear limit of $m=3$ partons
(two of the non-soft final-state partons have low $p_T$), and the 
subprocess
`parton + parton $\to$ 3 partons + $2$ soft partons',
which contributes through the SL collinear limit of $m=2$ partons
(one of the non-soft final-state partons has low $p_T$).
Owing to colour coherence (see 
Sects.~3.4 and 3.5 in Ref.~\cite{Catani:1999ss}), the singular collinear factors
are strictly factorized in these mixed soft-collinear limits of {\em tree-level}
QCD amplitudes.
The remaining contribution to the class $\,iii)$ is due to the subprocess
\beq
\label{prohos}
{\rm ~parton + parton} \to 3~{\rm partons + 1~soft \;parton~} \;\;,
\quad \;\;({\rm one\;low-}p_T \;\,{\rm final\!-\!state \;parton}) \;,
\eeq  
in the kinematical region where one of the three non-soft final-state partons
is collinear to one of the initial-state partons (Fig.~\ref{fig:1loop}). 
The subprocess in 
Eq.~(\ref{prohos}) contributes to one-loop order. The structure of the
absorptive part of the one-loop interaction limits the coherent action of the two
collinear partons in the SL region (the mechanism is analogous to that discussed
in Sect.~\ref{sec:caus}).
As a consequence, the {\em one-loop} mixed soft-collinear limit of the process
in Eq.~(\ref{prohos}) leads to a singular SL collinear factor that violates
strict factorization. Moreover, this singular factor can also produce 
correlations
with two non-collinear partons (see Fig.~\ref{fig:1loop}), 
whose structure is analogous
to that in Eqs.~(\ref{p22part}) and (\ref{wd2til5}) (see Fig.~\ref{fig:2loop}).

\vspace{0.5cm}

In summary, the two-loop radiative corrections to the subprocess in 
Eq.~(\ref{prohoc}) and the one-loop radiative corrections to the subprocesses in 
Eqs.~(\ref{prohomc}) and (\ref{prohos}) produce violation of strict factorization
in the SL collinear region. The factorization breaking effects are due to the
singular partonic configurations with $m=2$ collinear partons, 
$m=3$ collinear partons, and $m=2$ collinear partons plus one radiated soft
parton. The factorization breaking terms involve correlations with {\em one}
and {\em two} (in the case of the subprocesses in 
Eqs.~(\ref{prohoc}) and (\ref{prohos})) hard non-collinear partons.

In the inclusive hadroproduction of a single high-$p_T$ jet (or hadron),
this violation of strict collinear factorization leads to N$^3$LO contributions
that are separately
IR divergent; the total contribution of these IR divergent terms should vanish 
to guarantee the validity of the factorization theorem of mass singularities.

Even if the IR cancellation occurs, the violation 
of strict collinear factorization leads to residual IR finite terms.
These terms can produce observable contributions that are logarithmically
enhanced (and, therefore, large) in particular kinematical configurations
where virtual (see Eq.~(\ref{prohoc})) and real 
(see Eqs.~(\ref{prohomc}) and (\ref{prohos})) radiative corrections are highly
unbalanced. For instance, in these kinematical configurations
	 the cancellation of the IR pole term of ${\cal O}(1/\ep^2)$
	 in Eq.~(\ref{wd2til5}) can lead to a residual
	 double-logarithmic contribution.
The logarithmically-enhanced terms due to the violation 
of strict collinear factorization have a distinctive signature:
the factorization breaking correlations with the non-collinear partons
(see, e.g., Eq.~(\ref{wd2til5}) and Figs.~\ref{fig:2loop} and \ref{fig:1loop})
produce `{\em entangled} logarithms', namely, logarithmic terms whose coefficients get
tangled up with the colour flow and kinematical structure of the 
lowest-order hard-scattering subprocess (e.g., Eq.~(\ref{prolo})). In the
following, we mention some specific examples of processes that can exhibit
entangled logarithms at various perturbative orders.

The partonic subprocess in Eq.~(\ref{prolo}) also controls the LO inclusive
hadroproduction of a pair of nearly back-to-back high-$p_T$ hadrons or jets.
In the kinematical region where the total transverse momentum ${\bf Q}_T$
of the pair is small (e.g., much smaller than the invariant mass of the pair),
the perturbative QCD calculation leads to large contributions that are
proportional to powers of $\ln Q_T$. Owing to the violation 
of strict collinear factorization in the subprocesses 
of Eqs.~(\ref{prohoc})--(\ref{prohos}),
part of these logarithmic contributions can be
due to entangled logarithms (the entangled logarithms are absent if the 
small-$Q_T$ triggered final-state system includes no QCD partons 
\cite{Collins:1984kg, Catani:2010pd}) that first appear at the N$^3$LO.
The presence of these small-$Q_T$ entangled logarithms 
is also related to issues that arise in the context of factorization of
transverse-momentum dependent distributions \cite{Bomhof:2004aw, 
Bacchetta:2005rm, Rogers:2010dm}.

The perturbative QCD calculation of the inclusive hadroproduction of three
high-$p_T$ jets is also affected by large logarithms, $\ln p_T^{\,\rm min}$,
in the kinematical region where the transverse momentum $p_T^{\,\rm min}$
of the lowest-$p_T$
jet is small (e.g., much smaller than the transverse momenta of the other two
jets). In this three-jet production process, the partonic subprocess in 
Eq.~(\ref{prohoc})
enters at the LO, and the partonic subprocesses in Eqs.~(\ref{prohomc}) and
(\ref{prohos}) first contribute at the 
NLO.
The two-loop (in the case of the subprocess in Eq.~(\ref{prohoc})) and
one-loop (in the case of the subprocesses in Eqs.~(\ref{prohomc}) and 
(\ref{prohos})) violation of strict collinear factorization can produce 
small-$p_T^{\,\rm min}$ entangled logarithms starting from the 
NNLO perturbative calculation.

Another example of entangled logarithms is represented by the super-leading
(`non-global') logarithmic terms discovered 
\cite{Forshaw:2006fk}
in the calculation
of the cross section for the hadroproduction of a pair of jets with a rapidity gap
between them. These super-leading logarithms occur at the N$^4$LO in QCD
perturbation theory.


\setcounter{footnote}{2}

\section{Summary}
\label{sec:fin}

We have studied the singular behaviour of QCD scattering amplitudes
in the kinematical configurations where the momenta of two or more external
partons become collinear. We have shown that, beyond the tree-level,
strict (process-independent) collinear factorization is violated in the SL
collinear region. We have introduced a generalized form of collinear
factorization (see Eq.~(\ref{fact12onegen}) in Sect.~\ref{sec:gencol}, and 
Eqs.~(\ref{factallL}) and (\ref{loopexsp}) in Sect.~\ref{sec:allorder})
of the all-order scattering amplitudes, which is valid in both the TL and SL
collinear regions. In the SL region, the singular collinear factor retains some
process dependence, since it depends on the momenta and colour charges of the 
non-collinear partons. Owing to this colour dependence, the formulation of
collinear factorization directly in colour space, through the use of collinear
splitting matrices, is particularly suitable. An equivalent formulation in terms
of colour subamplitudes and collinear splitting amplitudes is feasible (see
Appendix~\ref{sec:appa}).
In the TL collinear region, strict 
factorization is recovered because
of colour coherence. In the SL collinear region, colour coherence is limited
by the causality structure of long-range gauge interactions (roughly, the
distinction between initial-state and final-state interactions), and this
produces absorptive contributions 
that eventually originate in the strict-factorization breaking phenomenon
(see Sect.~\ref{sec:caus}).

In the case of the {\em two-parton} SL collinear limit, we have computed the
($d$-dimensional) one-loop splitting matrix 
(Eqs.~(\ref{spgen}) and  (\ref{ic12gensym}))
to all orders in the dimensional regularization parameter $\ep$. 
The SL result explicitly shows that it cannot be obtained from the previously
known TL result by using crossing symmetry. The strict-factorization breaking
terms are antihermitian (`imaginary') at one-loop level.
        The expressions in Eqs.~(\ref{sldy}) and (\ref{d12hadt}),
	 which refer to the SL collinear limit in hadron--hadron
	 collisions, are the simplest examples of terms that violate
        strict collinear factorization.
At two-loop level (Sect.~\ref{sec:2part2loop}),
we have explicitly computed all the IR divergent contributions (i.e., the $\ep$
poles) to the splitting matrix. The two-loop result shows the presence of both
hermitian (`real') and antihermitian
terms that violate strict collinear factorization.

We have derived the structure of the IR divergences of the {\em multiparton}
collinear limit to all orders in the loop expansion (Sect.~\ref{sec:IRall}).
We have explicitly computed one-loop (Sect.~\ref{sec:irone})
and two-loop (Sect.~\ref{sec:mpart2loop}) IR divergent contributions to the SL 
collinear limit of $m \geq 3$ partons, 
thus extending the results of the two-parton collinear limit.

The SL collinear limit of the scattering amplitudes that are involved in
lepton--hadron DIS (namely, parton radiation collinear to the sole initial-state
parton) is a special case, since all the non-collinear partons are produced in the
final state. The one-loop expression of the two-parton SL splitting matrix
`effectively' takes a strictly-factorized form, in which there is no explicit
dependence on the non-collinear partons (see Sect.~\ref{sec:SL4}).
This effective strict factorization of the SL collinear limit can be a general
feature of the DIS kinematics
(there are no initial-state interactions between collinear and non-collinear
partons). The one-loop and two-loop terms of the multiparton splitting matrix that
we have explicitly computed fulfill this effective strict factorization in the DIS
kinematical configuration.

In hadron--hadron collision configurations,
the two-loop SL splitting matrix has factorization breaking terms that are
definitely non-abelian (see Sects.~\ref{sec:2part2loop} and \ref{sec:mpart2loop}).
These terms involve correlations between the momenta and colour charges of three
partons and, in particular, between a collinear parton and two non-collinear
partons. These non-abelian factorization breaking effects appear in the SL
collinear limit of two-loop scattering amplitudes with at least five external legs
and, in particular, $n \geq 4$ external QCD partons.

Owing to their absorptive origin, strict-factorization breaking effects partly
cancel at the level of squared amplitudes. Nonetheless, 
strict factorization is certainly
violated in the SL collinear limit of two-loop squared amplitudes for
parton--parton hard scattering with at least three final-state
        partons (see Sect.~\ref{sec:2square} and, in particular,
        Eq.~(\ref{wd2til5})).
These factorization breaking effects have consequences in perturbative QCD
calculations of hard-scattering cross sections in hadron--hadron collisions
(Sect.~\ref{sec:crosssec}). 
The violation of strict factorization affects the non-abelian structure of
logarithmically-enhanced terms at NNLO and higher orders; it has implications on
various factorization issues and, in particular, it challenges the validity
of mass-singularity factorization in jet and hadron production, starting
from the N$^3$LO.


\appendix
\section{Appendix: 
Multigluon amplitudes and the structure of the collinear
limit}
\label{sec:appa}

In this Appendix we consider the specific case in which the matrix element
$\cm(p_1,p_2,\dots,p_n)$ is a multiparton scattering amplitude with $n$
external {\em gluons}. We recall the decomposition of the matrix element in
colour subamplitudes, and we illustrate the behaviour of the colour subamplitudes
in the collinear limit.

The colour indices of the $n$
external gluons of $\cm(p_1,\dots,p_n)$ are denoted by $a_1,\dots,a_n$.
At the tree level 
the pure multigluon amplitude $\cm^{(0)}$ (see Eq.~(\ref{loopexnog}))
can be expressed as 
follows~\cite{Mangano:1990by}
\beq
\label{coldectree}
\cm^{(0) \,a_1,\dots,a_n}(p_1,\dots,p_n)
= (\sqrt 2)^n \;{\rm Tr}\left(t^{a_1} \dots t^{a_n}\right) 
\;{\cal A}^{(0)}(1,\dots,n) + {\rm non-}{\rm cyclic \;perms.} \;\;,
\eeq
where $t^{a_i}$'s are colour matrices in the fundamental representation
(see Eq.~(\ref{tnor})),
and the sum of terms on the right-hand side extends over the $(n-1)!$
non-cyclic permutations of the set $\{1,2,\dots,n\}$ of the external legs.
This is the customary colour decomposition in colour subamplitudes. 
Each tree-level colour subamplitude, ${\cal A}^{(0)}(1,\dots,n)$, 
is independent of the 
colour indices, and it embodies the kinematical dependence on the momenta and
spin polarizations of the external gluons. Note, however, that the colour
subamplitudes are {\em colour ordered}, namely, the functional form
of ${\cal A}^{(0)}(1,\dots,i,\dots,j,\dots,n)$ depends on the
specific ordering $(1,\dots,i,\dots,j,\dots,n)$ of the external legs
in the argument of ${\cal A}^{(0)}$
(e.g., ${\cal A}^{(0)}(1,\dots,i,\dots,j,\dots,n) \neq {\cal
A}^{(0)}(1,\dots,j,\dots,i,\dots,n)$). The overall normalization of the
colour subamplitudes in Eq.~(\ref{coldectree}) is adjusted to that used in 
Refs.~\cite{Bern:1998sc, Kosower:1999rx}; the only difference between 
${\cal A}^{(0)}(1,\dots,n)$ and the corresponding subamplitude 
$A_n^{\rm tree}(1,\dots,n)$ in Refs.~\cite{Bern:1998sc, Kosower:1999rx}
is due to the overall factor $\g^{n-2}$, which we have included in the 
definition
of ${\cal A}^{(0)}(1,\dots,n)$.

At the one-loop level the colour structure of the multigluon amplitude
$\cm^{(1)}$ (see Eq.~(\ref{loopexnog}))
is \cite{Bern:1990ux, Bern:1998sc, Kosower:1999rx}
\beeq
\label{coldecloop}
\cm^{(1) \,a_1,\dots,a_n}(p_1,\dots,p_n)
&\!\!=\!\!&\left\{ (\sqrt 2)^n \;{\rm Tr}\left(t^{a_1} \dots t^{a_n}\right) 
\;{\cal A}^{(1)}(1,\dots,n) + {\rm non-}{\rm cyclic \;perms.} \right\} \nn \\
&\!\!+\!\!& {\rm double \; trace \; terms} \;\;.
\eeeq
The term in the curly bracket has the same colour structure of 
Eq.~(\ref{coldectree}). The subamplitude ${\cal A}^{(1)}(1,\dots,n)$, which is 
called
leading-colour subamplitude (or primitive amplitude),
is the one-loop analogue of the tree-level colour subamplitude
${\cal A}^{(0)}(1,\dots,n)$.
The remaining terms on the right-hand side of 
Eq.~(\ref{coldecloop}) are proportional to colour factors that involve the
product of two traces of $t^a$ matrices, namely, 
${\rm Tr}\left(t^{a_1} \dots t^{a_k}\right) \,
{\rm Tr}\left(t^{a_{k+1}} \dots t^{a_n}\right)$. 
Each double trace is multiplied by a corresponding kinematical factor  
called subleading-colour subamplitude (or subleading-colour 
`partial amplitude'). The subleading-colour partial amplitudes
are in fact not independent of the leading-colour subamplitudes
${\cal A}^{(1)}(1,\dots,n)$; rather, they can be expressed as sum over
permutations of the arguments of the latter. Owing to this linear dependence,
it suffices to examine the collinear limit of the 
leading-colour subamplitudes. In particular, we do not explicitly consider
subleading-colour partial amplitudes and their contribution to 
Eq.~(\ref{coldecloop}).

As in the case of the tree-level colour decomposition in Eq.~(\ref{coldectree}),
our normalization of the one-loop subamplitude ${\cal A}^{(1)}(1,\dots,n)$
is adjusted to that used in Refs.~\cite{Bern:1998sc, Kosower:1999rx}.
To be precise, at the one-loop level the authors of 
Refs.~\cite{Bern:1998sc, Kosower:1999rx}
decompose the QCD leading-colour subamplitude in two 
terms,
$A_n^{{\rm 1-loop} \,[1]}$
and
$A_n^{{\rm 1-loop} \,[1/2]}$: the subscripts $[1]$ 
and $[1/2]$ respectively refer to the contribution of a gluon and a quark
circulating in the loop. The relation between our ${\cal A}^{(1)}(1,\dots,n)$
and the subamplitudes of Ref.~\cite{Bern:1998sc} is
${\cal A}^{(1)}(1,\dots,n)= \g^n [ N_c \,A_n^{{\rm 1-loop} \,[1]}(1,\dots,n) 
+ N_f \,A_n^{{\rm 1-loop} \,[1/2]}(1,\dots,n) ]$.

To present the collinear behaviour of the colour subamplitudes, we first 
explicitly relate
the colour matrix ${\sp}_{g_1 g_2}$ of the splitting process $g \to g_1 g_2$
to its colour stripped component, the splitting amplitude
${\rm Split}$ (see Eq.~(\ref{spvssplit})). At the tree level we define:
\beq
\label{split0}
Sp_{g_1 g_2}^{(0)\;(a_1,a_2;\,a)}(p_1,p_2;{\widetilde P}) \equiv {\sqrt 2}
 \;\;i \;f_{a_1 a_2 \,a} \;\,{\rm Split}^{(0)}(p_1,p_2;\wp) \;\;,
\eeq
where the explicit expression of ${\rm Split}^{(0)}$ can be extracted by direct
inspection of Eq.~(\ref{ggg0}).
Analogously, at the one-loop level we consider the splitting matrix
$\sp^{(1)}_{H}$ (see Eqs.~(\ref{sp112gen}) and (\ref{spgen})),
and we define the splitting amplitude ${\rm Split}_H^{(1)}$ as follows:
\beq
\label{split1}
Sp_{H \;g_1 g_2}^{(1)\;(a_1,a_2;\,a)}(p_1,p_2;{\widetilde P}) \equiv {\sqrt 2}
 \;\;i \;f_{a_1 a_2 \,a} \;\,{\rm Split}_H^{(1)}(p_1,p_2;\wp) \;\;.
\eeq 
We note that, according to the notation used throughout this paper, the 
splitting
amplitudes on the right-hand side of Eqs.~(\ref{split0}) and (\ref{split1})
are still spin matrices: they act onto the spin (helicity) space of the  
two collinear gluons (with momenta $p_1$ and $p_2$)
and their parent gluon (with momentum $\wp$).

We recall that the colour subamplitudes 
${\cal A}^{(0)}(1,\dots,n)$ and ${\cal A}^{(1)}(1,\dots,n)$
in Eqs.~(\ref{coldectree}) and (\ref{coldecloop}) are colour ordered and, 
therefore, their kinematical structure depends on the specific
ordering $(1,\dots,n)$ of the gluon momenta of the external legs.
This features has consequences on the collinear behaviour of the 
colour subamplitudes. Considering the collinear limit of the
momenta $p_1$ and $p_2$, we can distinguish two types of configurations,
according to whether the two collinear gluons are adjacent or not adjacent
in the argument of the colour subamplitude.
The colour subamplitudes ${\cal A}(..,1,..,2,..)$, 
where the two collinear momenta
are not adjacent, are not singular in the collinear limit.
The colour subamplitudes ${\cal A}(..,1,2,..)$, 
where the two collinear momenta
are adjacent, are singular in the collinear limit.

At the tree level, the singular behaviour of the multigluon
colour subamplitudes 
with adjacent collinear legs is given by the following (helicity space)
factorization formula:
\beq
\label{facttreesub}
{\cal A}^{(0)}(\dots,k,1,2,j,\dots) \simeq \,{\rm Split}^{(0)}(p_1,p_2;\wp) 
\;\,{\cal A}^{(0)}(\dots,k,\wp,j,\dots) \;\;.
\eeq
The colour subamplitude on the right-hand side has $n-1$ external legs.
It is obtained from the colour subamplitude on the left-hand side
by replacing the two collinear gluons with a single gluon (with momentum
$\wp$). The relative ordering of the non-collinear legs in the argument of the
colour subamplitudes is left unchanged in going from the left-hand to the
right-hand sides. We have explicitly introduced the labels $k$ and $j$ of
two non-collinear legs to remark the unchanged ordering of the
non-collinear legs. Note, however, that the tree-level splitting amplitude
${\rm Split}^{(0)}$ is universal; it depends on the momenta (and helicities)
of the two collinear gluons and of the parent gluon, and it has no
dependence on the non-collinear legs. Moreover, the splitting amplitude 
${\rm Split}^{(0)}$ on the right-hand side of Eq.~(\ref{facttreesub})  
controls the singular behaviour of the collinear splitting subprocess 
$g \to g_1 g_2$ also in the case of colour subamplitudes with both gluons and 
quark--antiquark pairs in the external legs.

At the one-loop level, the singular behaviour of
the leading-colour subamplitudes
with adjacent collinear legs is given by the following {\em generalized}
factorization formula (in helicity space):
\beeq
\label{factonesub}
{\cal A}^{(1)}(\dots,k,1,2,j,\dots) &\simeq& 
\,{\rm Split}^{(0)}(p_1,p_2;\wp) \;\,
{\cal A}^{(1)}(\dots,k,\wp,j,\dots) \nn \\
&+& \, {\rm Split}^{(1)}(p_k,p_1,p_2,p_j;\wp) 
\;\, {\cal A}^{(0)}(\dots,k,\wp,j,\dots) \;\;,
\eeeq
and the one-loop splitting amplitude ${\rm Split}^{(1)}$ is
\beeq
\label{split1loop}
{\rm Split}^{(1)}(p_k,p_1,p_2,p_j;\wp) = {\rm Split}_H^{(1)}(p_1,p_2;\wp)
+ I_C(p_k,p_1,p_2,p_j;{\widetilde P}) \;{\rm Split}^{(0)}(p_1,p_2;\wp) \;\;,
\eeeq
where the `rational' part ${\rm Split}_H^{(1)}$ is defined in Eq.~(\ref{split1})
and the transcendental function that multiplies ${\rm Split}^{(0)}$ 
is
\beeq
\label{icsplit}
I_C(p_k,p_1,p_2,p_j;{\widetilde P}) &=& 
 \; \g^2 \; c_{\Gamma} \;
\left( \frac{-s_{12} -i0}{\mu^2} \right)^{-\ep} \;C_A \nn \\
&\times& \left\{  \; - \;\frac{1}{\ep^2}  
\right.  
- \left. \; \frac{1}{\ep} \, \Bigl[ \;f(\ep;z_1-i0s_{k1})
 +   \;f(\ep;z_2-i0s_{j2})\;
\Bigr] \right\} \;\;. 
\eeeq
The arguments $(\dots,k,1,2,j,\dots)$ and $(\dots,k,\wp,j,\dots)$ of the
colour subamplitudes in 
the factorization formula
(\ref{factonesub}) are the same as those in 
Eq.~(\ref{facttreesub}).

The tree-level factorization formula (\ref{facttreesub}) is a well-known result
\cite{Berends:1987me}.
At the one-loop level, considering the TL collinear limit (i.e. $s_{12} > 0$),
the momentum fractions $z_1$ and $z_2$ are positive, and  we can remove
the $i0$ prescriptions on the right-hand side of Eq.~(\ref{icsplit});
therefore, ${\rm Split}^{(1)}$ is universal (i.e., independent of $p_k$ and
$p_j$), and
Eqs.~(\ref{factonesub})--(\ref{icsplit}) 
give the one-loop factorized result derived in 
Refs.~\cite{Bern:1998sc, Kosower:1999rx}. The structure and the explicit form
of Eqs.~(\ref{factonesub})--(\ref{icsplit}) in the case of the SL collinear 
limit (i.e. $s_{12} < 0$) is a new result, which derives from the
application of the general results in Eqs.~(\ref{fact12onegen}), 
(\ref{spgen}) and (\ref{ic12gen}) (or (\ref{ic12gensym}))
to 
pure multigluon scattering amplitudes at one-loop order.

The main new feature of the SL collinear limit is that the one-loop splitting
amplitude 
${\rm Split}^{(1)}$ in Eq.~(\ref{factonesub}) is not `universal', since it
also depends on the non-collinear partons. More precisely, 
${\rm Split}^{(1)}$ depends on the two non-collinear gluon legs
$k$ and $j$ that are adjacent
(colour connected) to the two collinear gluons in the one-loop
leading-colour subamplitude
${\cal A}^{(1)}(\dots,k,1,2,j,\dots)$. The dependence 
(see Eqs.~(\ref{split1loop}) and (\ref{icsplit}))
is due the signs of 
$s_{k1}=2p_k\cdot p_1$ and $s_{j2}=2p_j\cdot p_2$, which control the imaginary
part of the analytic functions
$f(\ep;z_1-i0s_{k1})$ and 
$f(\ep;z_2-i0s_{j2})$ in Eq.~(\ref{icsplit}).

\section{Appendix: 
The IR divergences of the two-loop splitting matrix}
\label{sec:appb}

In this Appendix we illustrate the perturbative expansion of the IR
factorization formula (\ref{spfindef}). We recall that Eq.~(\ref{spfindef})
refers to the collinear limit of $m$ ($m \geq 2$) parton momenta in amplitudes
with $n$ ($n > m$) external QCD partons.

The splitting matrix operators ${\bom I}$ and ${\bom {\overline I}}$ in 
Eqs.~(\ref{vm1}) and (\ref{vbarm1}) have the following renormalized perturbative
expansions:
\beq
\label{iall}
{\bom I}(\ep) = \;{\bom I}^{(1)}(\ep) + {\bom I}^{(2)}(\ep) +
 \cdots \;\;,
\eeq
\beq
\label{ibarall}
{\bom {\overline I}}(\ep) = \;{\bom {\overline I}}^{(1)}(\ep) + 
{\bom {\overline I}}^{(2)}(\ep) +
 \cdots \;\;.
\eeq
The perturbative contributions ${\bom I}^{(k)}$ and 
${\bom {\overline I}}^{(k)}$ are obtained from the corresponding scattering
amplitude operators, ${\bom I}_M^{(k)}$ and ${\bom I}_{\Mbar}^{(k)}$,
according to Eqs.~(\ref{vdef}) and (\ref{vbardef}).
To be precise, we use the relations (\ref{ionedef}) and (\ref{ibaronedef})
at one-loop order ($k=1$), and analogous relations at two-loop order ($k=2$).
Performing the collinear limit (as specified in Eq.~(\ref{ionedef}))
of the expression (\ref{i1m}),
we obtain
\beeq
\label{ionefor}
{\bom I}^{(1)}(\ep) \!\!&=& \f{\as(\mu^2)}{2\pi} \;\frac{1}{2} \;
\left\{ - \sum_{i=1}^{n}
\left( \,\frac{1}{\ep^2} \,C_i + \frac{1}{\ep} \,\gamma_i\right)
+ \frac{2}{\ep} \sum_{\substack{j \,\in \, NC \\ {}}} \,C_j \,
\ln\left( \,\frac{-s_{j \wp} - i0}{\mu^2} \right)
\right. \nn \\
&-& 
\frac{1}{\ep} \sum_{\substack{i,\ell \,\in C \\ i \,\neq\, \ell}} 
{\bom T}_i \cdot {\bom T}_\ell
\, \ln\left(\f{-s_{i \ell} - i0}{\mu^2}\right)
\left.
- \;\frac{2}{\ep} \sum_{\substack{i \,\in C \\ j \,\in NC}} 
{\bom T}_i \cdot {\bom T}_j
 \, \ln\bigl( z_i - i0 s_{ij} \bigr) \right. \\
&-& \left. 
\frac{1}{\ep} \sum_{\substack{j,k \,\in NC \\ j \,\neq\, k}}
{\bom T}_j \cdot {\bom T}_k
\left[ \;\ln\left(\f{-s_{j k} - i0}{\mu^2}\right)
- \ln\left(\f{-s_{j \wp} - i0}{\mu^2}\right)
- \ln\left(\f{-s_{k \wp} - i0}{\mu^2}\right)
\right]
 \right\} \;, \nn
\eeeq
where we have also used colour conservation
($\,\sum_{i \,\in C} {\bom T}_i = - \sum_{j \,\in NC} {\bom T}_j\,$).
Considering Eq.~(\ref{i1m})
with the replacement $\cm \to \cmbar$, and applying the definition in
Eq.~(\ref{ibaronedef}), we obtain
\beeq
\label{ibaronefor}
{}\!\!\!\!{\bom {\overline I}}^{(1)}(\ep) 
\!\!&=&\!\!\! \f{\as(\mu^2)}{2\pi} \;\frac{1}{2} \;
\left\{ - 
\left( \,\frac{1}{\ep^2} \,C_{\wp} + \frac{1}{\ep} \,\gamma_{\wp}\right)
-  \sum_{\substack{j \,\in \, NC \\ {}}} 
\left[ \,\frac{1}{\ep^2} \,C_j + \frac{1}{\ep} \,\gamma_j 
- \frac{2}{\ep}\,C_j \,
\ln\left( \,\frac{-s_{j \wp} - i0}{\mu^2} \right) \right]
\right. \nn \\
&-&\!\!\! \left. 
\frac{1}{\ep} \sum_{\substack{j,k \,\in NC \\ j \,\neq\, k}}
\!\!{\bom T}_j \cdot {\bom T}_k
\left[ \;\ln\left(\f{-s_{j k} - i0}{\mu^2}\right)
- \ln\left(\f{-s_{j \wp} - i0}{\mu^2}\right)
- \ln\left(\f{-s_{k \wp} - i0}{\mu^2}\right)
\right]\!
 \right\} . 
\eeeq
Since the two-loop operator ${\bom I}_M^{(2)}$ (or ${\bom I}_{\Mbar}^{(2)}$)
is simply related to ${\bom I}_M^{(1)}$ (or ${\bom I}_{\Mbar}^{(1)}$)
by the expression (\ref{i2m}), the operators 
${\bom I}^{(2)}$ and ${\bom {\overline I}}^{(2)}$ are
\beeq
\label{i2ex}
{\bom I}^{(2)}(\ep) 
&=&  -\, \f{1}{2} \left[ {\bom I}^{(1)}(\ep) \right]^2
+ \f{\as(\mu^2)}{2\pi} 
\left\{ + \frac{1}{\ep} \,\cbet0 
\left[ {\bom I}^{(1)}(2\ep) - {\bom I}^{(1)}(\ep)\right]
+ K \;{\bom I}^{(1)}(2\ep) \right\} \nn \\
&+& 
\left(\f{\as(\mu^2)}{2\pi}\right)^2  \,\frac{1}{\ep} \;
\sum_{i=1}^{n} 
\; H^{(2)}_i \;\;,
\eeeq
\beeq
\label{ibar2ex}
{\bom {\overline I}}^{(2)}(\ep) 
&=&  -\, \f{1}{2} \left[ {\bom {\overline I}}^{(1)}(\ep) \right]^2
+ \f{\as(\mu^2)}{2\pi} 
\left\{ + \frac{1}{\ep} \,\cbet0 
\left[ {\bom {\overline I}}^{(1)}(2\ep) - {\bom {\overline I}}^{(1)}(\ep)\right]
+ K \;{\bom {\overline I}}^{(1)}(2\ep) \right\} \nn \\
&+& 
\left(\f{\as(\mu^2)}{2\pi}\right)^2  \,\frac{1}{\ep} \;
\left( H^{(2)}_{\wp} +
\sum_{j \,\in \, NC} 
\; H^{(2)}_j \right) \;\;.
\eeeq

The computation of the perturbative expansion of Eq.~(\ref{spfindef}) is
elementary and straightforward. We simply illustrate few intermediate steps.
To perform the expansion, we find it convenient to rewrite 
Eq.~(\ref{spfindef}) in the following equivalent form:
\beq
\label{spfindefeq}
\sp = \left[ \,1 -  {\bf {\overline V}}(\ep) \;{\bf V}^{-1}(\ep)
\,\right] \,\sp + {\bf {\overline V}}(\ep) \;\,\sp^{\,{\rm fin.}} 
\;\,{\bf {\overline V}}^{\,-1}(\ep)
\;\;.
\eeq
The two-loop perturbative expansion of the operator 
$1 -  {\bf {\overline V}} \,{\bf V}^{-1}$ is obtained by using 
Eqs.~(\ref{vm1}), (\ref{vbarm1}), (\ref{iall}) and (\ref{ibarall});
we have
\beeq
\label{1-vvbar}
1 -  {\bf {\overline V}}(\ep) \;{\bf V}^{-1}(\ep) &=&
{\bom I}^{(1)}(\ep) - {\bom {\overline I}}^{(1)}(\ep) \nn \\
&+& {\bom I}^{(2)}(\ep) - {\bom {\overline I}}^{(2)}(\ep)
+ {\bom {\overline I}}^{(1)}(\ep) 
\left( \,{\bom I}^{(1)}(\ep) - {\bom {\overline I}}^{(1)}(\ep) \,
\right)+ {\cal O}(\as^3) \;\;.
\eeeq
Using Eqs.~(\ref{spfinex}), (\ref{vbarm1}) and (\ref{ibarall}), the 
two-loop expansion of the second term on the right-hand side of 
Eq.~(\ref{spfindefeq}) is
\beq
\label{spfinbar}
{\bf {\overline V}}(\ep) \,\sp^{\,{\rm fin.}} 
\,{\bf {\overline V}}^{\,-1}(\ep) = \left\{ 
\sp^{(0, R)} + \sp^{(1) \,{\rm fin.}}
+ \sp^{(2) \,{\rm fin.}}
+  \left[ \,{\bom {\overline I}}^{(1)}(\ep) 
\,,\, \sp^{(1) \,{\rm fin.}} \,\right] \right\}
\left\{ 1+ {\cal O}(\as^3) \right\} \,.
\eeq
Note that in Eq.~(\ref{spfinbar}) we have used the fact that 
$\sp^{(0, R)}$ is strictly factorized and, hence, it commutes with 
${\bom {\overline I}}^{(k)}(\ep)$ (as already pointed out in 
Eqs.~(\ref{divonere}) and (\ref{divone2})).
Inserting the perturbative expansions (\ref{loopexspren}),
(\ref{1-vvbar}) and (\ref{spfinbar}) in Eq.~(\ref{spfindefeq}),
we reobtain the one-loop 
relations
(\ref{sp1df}) and (\ref{sp1div}), and we directly obtain the two-loop 
relations (\ref{sp2divfin}) and (\ref{sp2mcdiv}). The three
contributions on the right-hand side of Eq.~(\ref{sp2mcdiv}) depends on the
perturbative terms 
${\bom I}^{(k)}(\ep)$ and  ${\bom {\overline I}}^{(k)}(\ep)$ ($k=1,2$).
The one-loop operator $\imc(\ep)$ is given in Eq.~(\ref{i1mc}),
while $\imctwo(\ep)$ and 
${\overline \sp}^{(2) \,{\rm div.}}$ are given by the following expressions:
\beq
\label{i2mc}
\imctwo(\ep) = {\bom I}^{(2)}(\ep) - {\bom {\overline I}}^{(2)}(\ep)
+ {\bom {\overline I}}^{(1)}(\ep) 
\left( \,{\bom I}^{(1)}(\ep) - {\bom {\overline I}}^{(1)}(\ep) \,
\right) \;\;,
\eeq
\beq
\label{sp2bardiv}
{\overline \sp}^{(2) \,{\rm div.}} =  
\left[ \,{\bom {\overline I}}^{(1)}(\ep) \,,\, \sp^{(1) \,{\rm fin.}} \,\right]
\;\;.
\eeq
We discuss the structure of Eqs.~(\ref{i2mc}) and (\ref{sp2bardiv}), in turn.

Using elementary algebra, 
the expression (\ref{i2mc}) can be rewritten as follows:
\beeq
\label{i2mcsim}
\imctwo(\ep) &=& 
\left( \,{\bom I}^{(2)}(\ep) + \f{1}{2} \left[ {\bom I}^{(1)}(\ep) \right]^2
\right) - \left( \,{\bom {\overline I}}^{(2)}(\ep)
+ \f{1}{2} \left[ {\bom {\overline I}}^{(1)}(\ep) \right]^2 \right) \nn \\
&-& \f{1}{2} \left[ 
{\bom I}^{(1)}(\ep) - {\bom {\overline I}}^{(1)}(\ep)\right]^2
+ {\bf \Delta}_{m \,C}^{(2;\,2)}(\ep) \;\;,
\eeeq
where we have defined
\beq
\label{d22mcgen}
{\bf \Delta}_{m \,C}^{(2;\,2)}(\ep) \equiv \f{1}{2} 
\;\left[ \,{\bom {\overline I}}^{(1)}(\ep) \,,\, {\bom I}^{(1)}(\ep)
\,\right] \;\;.
\eeq
The form on the right-hand side of Eq.~(\ref{i2mcsim})
is useful to express the operator $\imctwo(\ep)$
in terms of the one-loop operator 
$\imc = {\bom I}^{(1)} - {\bom {\overline I}}^{(1)}$
(see Eq.~(\ref{i1mc})). Owing to the relations
(\ref{i2ex}) and (\ref{ibar2ex}), the linear combinations
$2\, {\bom I}^{(2)} + [ {\bom I}^{(1)} ]^2$ and
$2\, {\bom {\overline I}}^{(2)} + [ {\bom {\overline I}}^{(1)} ]^2$
are proportional to ${\bom I}^{(1)}$ and ${\bom {\overline I}}^{(1)}$,
respectively.
Therefore, by simple inspection of Eqs.~(\ref{i2ex}) and (\ref{ibar2ex}),
we see that Eq.~(\ref{i2mcsim}) corresponds to the result reported in
Eq.~(\ref{i2mexpl}).

The two-loop factorization breaking operator 
${\bf \Delta}_{m \,C}^{(2;\,2)}$ in Eq.~(\ref{i2mexpl}) originates from 
the colour-matrix commutator in Eq.~(\ref{d22mcgen}) (the commutator
trivially vanishes in QED and any abelian theories).
The commutator can be rewritten as follows:
\beeq
{\bf \Delta}_{m \,C}^{(2;\,2)}(\ep) &=& \f{1}{2} 
\,\left[ \,{\bom {\overline I}}^{(1)}(\ep) \,,\, {\bom I}^{(1)}(\ep)
- {\bom {\overline I}}^{(1)}(\ep)
\,\right] = \f{1}{2} 
\,\left[ \,{\bom {\overline I}}^{(1)}(\ep) \,,\, 
\imc(\ep)
\,\right] \nn \\
\label{d22mcnew}
&=& \f{1}{2} 
\,\left[ \,{\bom {\overline I}}^{(1)}(\ep) \,,\, 
\dmc(\ep)
\,\right]
\;\;.
\eeeq
Since the operator ${\bom {\overline I}}^{(1)}$ does not depend on the colour
matrices of the collinear partons, to obtain Eq.~(\ref{d22mcnew}) we have
exploited the fact that only $\dmc(\ep)$ (the factorization breaking part of 
$\imc(\ep)$ in Eq.~(\ref{i1mcexd})) contributes to the commutator.
Inserting the explicit expressions (\ref{del1}) and (\ref{ibaronefor})
in Eq.~(\ref{d22mcnew}) and computing the colour-charge commutator,
we straightforwardly obtain the result in Eq.~(\ref{d22mc}).

In the TL collinear region, the one-loop splitting matrix $\sp^{(1, R)}$
and its IR finite part,  $\sp^{(1) \,{\rm fin.}}$, in Eq.~(\ref{sp1df}) 
are strictly factorized, and they do not depend on the colour matrices of the
non-collinear partons.
Therefore, 
the commutator in 
Eq.~(\ref{sp2bardiv})
vanishes in 
the TL collinear limit.
In the SL collinear region, $\sp^{(1) \,{\rm fin.}}$ depends on the
colour matrices of the non-collinear partons 
(as mentioned in Sects.~\ref{sec:multi} and \ref{sec:irone},
$\sp^{(1) \,{\rm fin.}}$ has a linear dependence on these colour matrices)
and, therefore, the commutator term
${\overline \sp}^{(2) \,{\rm div.}}$ in Eq.~(\ref{sp2bardiv}) does not vanish.
Since $\sp^{(1) \,{\rm fin.}}$ is IR finite, the IR divergent part of 
${\overline \sp}^{(2) \,{\rm div.}}$ contains only single poles $1/\ep$
(the coefficient of the double pole in the expression (\ref{ibaronefor})
of ${\bom {\overline I}}^{(1)}(\ep)$ is a c-number and, hence, it gives a 
vanishing contribution to the commutator in Eq.~(\ref{sp2bardiv})).

In the specific case of the SL collinear limit of $m=2\;$ partons, 
we explicitly know $\sp^{(1, R)}$ to all orders in $\ep$ 
(see Sect.~\ref{sec:gencol}).
This information can be exploited to extract $\sp^{(1) \,{\rm fin.}}$
and, then, to explicitly compute ${\overline \sp}^{(2) \,{\rm div.}}$ 
in Eq.~(\ref{sp2bardiv}).
The multiparton collinear operator ${\bom I}^{(k)}_{m\,C}$ $(k=1,2)$
is denoted by ${\bom I}^{(k)}_{2\,C}$ 
in the case of 
$m=2\;$ collinear partons.
From the expressions in Eqs.~(\ref{i1mcexd}) and (\ref{del1}), we have
\beeq
\label{i12cexp}
\isc(\ep) 
&=&\f{\as(\mu^2)}{2\pi} \;\frac{1}{2} \;
\left\{ \frac{1}{\ep^2} \left( \,C_{12} - C_{1} -C_2 \right)
+ \frac{1}{\ep} \left( \,\gamma_{12} - \gamma_1 - \gamma_2 \right)
\right. \nn \\
&+& \frac{2}{\ep}\left( C_{1} \ln|z_1| + C_2 \ln|z_2|\right)
 - \frac{2}{\ep}  \;
{\bom T}_1 \cdot {\bom T}_2
\; \ln\left(\f{-s_{12} - i0}{|z_1| |z_2| \,\mu^2}\right)
\left. \right. \nn \\
&+&\left. i \,\f{2 \pi}{\ep} \;\sum_{j=3}^{n} \,\sum_{i=1,2}
\;{\bom T}_j \cdot {\bom T}_i \;\Theta(-z_i) \;{\rm sign}(s_{ij}) \right\} \;,
\eeeq
whereas ${\bom I}^{(2)}_{2\,C}$ is obtained from Eq.~(\ref{i2mexpl})
by the replacements ${\bom I}^{(k)}_{m\,C} \to {\bom I}^{(k)}_{2\,C} $
and ${\bf \Delta}_{m \,C}^{(2;\,2)} \to {\bf \Delta}_{2 \,C}^{(2;\,2)}$.
Using Eq.~(\ref{sp1df})
and, then, Eqs.~(\ref{sp2oneren}) and (\ref{sp1div}),
we have
\beq
\label{sp1fintwo}
\sp^{(1) \,{\rm fin.}} = \sp^{(1, R)} - \sp^{(1) \,{\rm div.}} =
\itc^{(1)}(\ep) \;
\sp^{(0,R)} + \sp^{(1,R)}_{H} - \itwoc(\ep) \;\sp^{(0, R)} \;\;.
\eeq
Inserting Eq.~(\ref{sp1fintwo}) in 
Eq.~(\ref{sp2bardiv}), we obtain
\beq
\label{sp2bardivtwo}
{\overline \sp}^{(2) \,{\rm div.}}
= \left[ \,{\bom {\overline I}}^{(1)}(\ep) \,,
\, \itc^{(1)}(\ep) - \itwoc(\ep) 
\,\right]\;\sp^{(0, R)} \equiv {\bf \Delta}_{2 \,C}^{(2; \,1)}(\ep) \;
\sp^{(0, R)} \;\;,
\eeq
where we have used the fact that ${\bom {\overline I}}^{(1)}(\ep)$ commutes with
both $\sp^{(1,R)}_{H}$ and $\sp^{(0,R)}$, and we have defined the colour
operator ${\bf \Delta}_{2 \,C}^{(2; \,1)}$. 
Using the explicit expressions
of ${\bom {\overline I}}^{(1)}$, $\itc^{(1)}$ and $\itwoc$
in Eqs.~(\ref{ibaronefor}), (\ref{it2cone}) and (\ref{i12cexp}),
we 
evaluate the commutator in Eq.~(\ref{sp2bardivtwo})
and we find
\beeq
\label{d2single}
{\bf \Delta}_{2 \,C}^{(2;\,1)}(\ep) &=&
\left[ \,{\bom {\overline I}}^{(1)}(\ep) \,,
\, \itc^{(1)}(\ep) - \itwoc(\ep) 
\,\right] \nn \\
 &=& \left(\f{\as(\mu^2)}{2\pi}\right)^2  
\; \f{1}{\ep} \;\pi \; f_{abc} \,
\sum_{i=1,2} \;\sum_{\substack{j, k \,\in NC \\ j \,\neq\, k}} 
\,T_i^a \,T_j^b \, T_k^c \;\Theta(-z_i) \;
{\rm sign}(s_{ij}) \;\Theta(-s_{jk}) \nn \\
&\times& \ln\left(- \,\f{ s_{j\wp} \; s_{k\wp}}{s_{jk} \,\mu^2} -i0 \right)
\;\ln\left( \f{ z_i \,s_{12}}{(1-z_i) \,\mu^2} \right) + {\cal O}(\ep^0)
\;\;.
\eeeq
Therefore, in the specific case of the two-parton collinear limit,
the computation of $\sp^{(2) \,{\rm div.}}$ 
(see Eqs.~(\ref{sp2divfin}) and (\ref{sp2mcdiv}))
is explicitly completed in the form
\beq
\label{sp2div2c}
\sp^{(2) \,{\rm div.}} =  \itwoc(\ep) \;\sp^{(1, R)} +
\left( \,\itwopart(\ep) + 
{\bf \Delta}_{2 \,C}^{(2;\,1)}(\ep)
\right)\;\sp^{(0, R)} \;\;, \quad (m=2) \;,
\eeq
where $\itwoc$, $\itwopart$ and ${\bf \Delta}_{2 \,C}^{(2;\,1)}$
are given in Eqs.~(\ref{i12cexp}), (\ref{i2mexpl}) and (\ref{d2single}).
In particular, the factorization breaking operators
${\bf \Delta}_{2 \,C}^{(2;\,2)}$ and ${\bf \Delta}_{2 \,C}^{(2;\,1)}$
in Eqs.~(\ref{d22mc}) and (\ref{d2single}) can be combined by defining
\beq
{\bf \Delta}_{2 \,C}^{(2)}(\ep) \equiv
{\bf \Delta}_{2 \,C}^{(2;\,2)}(\ep) + 
{\bf \Delta}_{2 \,C}^{(2;\,1)}(\ep) \;\;,
\eeq
and we obtain
\beeq
\label{d2tot}
{\bf \Delta}_{2 \,C}^{(2)}(\ep) &=& \left(\f{\as(\mu^2)}{2\pi}\right)^2  
\;\pi \; f_{abc} \,
\sum_{i=1,2} \;\sum_{\substack{j, k \,\in NC \\ j \,\neq\, k}} 
\,T_i^a \,T_j^b \, T_k^c \;\Theta(-z_i) \;
{\rm sign}(s_{ij}) \;\Theta(-s_{jk}) \nn \\
&\times& \ln\left(- \,\f{ s_{j\wp} \; s_{k\wp}}{s_{jk} \,\mu^2} -i0 \right)
\;\left[ \,- \, \f{1}{2 \,\ep^2} +  \f{1}{\ep}
\;\ln\left( \f{ z_i \,s_{12}}{(1-z_i) \,\mu^2} \right) + {\cal O}(\ep^0)
\right] \;\;.
\eeeq

We note that the expression of $\sp^{(2, R)}$ presented in 
Sect.~\ref{sec:2part2loop} (see Eq.~(\ref{sp2twoc}))
has a form that differs from the expression in Eq.~(\ref{sp2div2c}).
However, these two expressions are completely equivalent
since they lead to the same IR divergent contribution to 
$\sp^{(2, R)}$ (the differences in the IR finite part can be absorbed in the
definition of ${\widetilde \sp}^{(2) \,{\rm fin.}}$).
To be precise, comparing Eqs.~(\ref{sp2twoc}) and (\ref{sp2div2c}), we have
\beeq
\sp^{(2,R)} - \sp^{(2) \,{\rm div.}} &=& 
\left( \itc^{(1)}(\ep) - \itwoc(\ep)\right)
\;\,\sp^{(1,R)} \nn \\
\label{sp2twocdif}
&+&
\left( \itc^{(2)}(\ep) - \itwopart(\ep) - {\bf \Delta}_{2 \,C}^{(2;\,1)}(\ep)
\right)
\;\,\sp^{(0,R)}  
 + \,{\widetilde \sp}^{(2) \,{\rm fin.}}\;,
\eeeq
and it can be explicitly checked that the expression on the right-hand side is 
IR finite if $\ep \to 0$. This explicit check, which is left to the reader,
can be performed by using the expressions in Eqs.~(\ref{sp2oneren}), 
(\ref{it2cone}), (\ref{i22part}), 
(\ref{d2til}),
(\ref{i2mexpl}), 
(\ref{d22mc}),
(\ref{i12cexp}) and (\ref{d2single}).

We briefly illustrate the derivation of the relations in Eqs.~(\ref{3vs3})
and (\ref{2c1ncvscol}).

We first note that the explicit constraint $s_{jk} < 0$ can be removed from the
expression on the left-hand side of Eq.~(\ref{3vs3}). Indeed, the sum of the 
terms with $s_{jk} > 0$ gives a vanishing contribution to
that expression, as explicitly shown by the following relation:
\beq
\label{3vs3a}
f_{abc} \,
 \;\sum_{\substack{j,\, k \\ j \,\neq\, k}} 
 \,T_j^b \, T_k^c  \;
\,{\rm sign}(s_{ij}) \;\Theta(s_{jk}) 
= \f{1}{2} \;f_{abc} \,
 \;\sum_{\substack{j,\, k \\ j \,\neq\, k}} 
 \,T_j^b \, T_k^c  \;\left[
\;{\rm sign}(s_{ij}) - {\rm sign}(s_{ik}) \;\right] \;\Theta(s_{jk}) = 0 \;.
\eeq
Here, we have simply replaced the factor 
${\rm sign}(s_{ij}) \;\Theta(s_{jk})$ with its antisymmetric part with respect
to the exchange $j \leftrightarrow k$
(the symmetric part gives a vanishing contribution, since 
$f_{abc} \,T_j^b \, T_k^c$ is antisymmetric under the exchange 
$j \leftrightarrow k$) and, then, we have used the fact that $s_{ij}$
and $s_{ik}$ have the same sign if $s_{jk} > 0$.
Considering the left-hand side of Eq.~(\ref{3vs3}) and removing the constraint
$s_{jk} < 0$, we thus obtain
\beeq
f_{abc} \,
\sum_{i \in C} \;\sum_{\substack{j, k \,\in NC \\ j \,\neq\, k}} 
\!\!\!\!\!\!\!&& \!\!\!\!\!
\,T_i^a \,T_j^b \, T_k^c  \;
\,{\rm sign}(s_{ij}) \;\Theta(-s_{jk}) \;h_i
= f_{abc} \,
\sum_{i \in C} \;\sum_{\substack{j, k \,\in NC \\ j \,\neq\, k}} 
\,T_i^a \,T_j^b \, T_k^c  \;
\,{\rm sign}(s_{ij})  \;h_i \nn \\
\label{3vs3sim}
&=& - f_{abc} \,
\sum_{i \in C} \;\sum_{j \in NC} \,T_i^a \,T_j^b \,
\Bigl( \,\sum_{\substack{\ell \,\in C \\ \ell \,\neq\, i}} \,T_{\ell}^c
+ T_i^c + T_j^c \Bigr)
\,{\rm sign}(s_{ij})  \;h_i \;\;,
\eeeq
where we have performed the sum over $k$ by using
the colour conservation relation
\beq
\sum_{\substack{k \,\in NC \\ k \,\neq\, j}} \,T_k^c =
- \sum_{\ell \in C} \,T_{\ell}^c - T_j^c \;\;.
\eeq
Owing to the algebraic identity (\ref{tipj}), the final expression in
Eq.~(\ref{3vs3sim}) exactly corresponds to the expression on the right-hand 
side of Eq.~(\ref{3vs3}).

The left-hand side of Eq.~(\ref{2c1ncvscol}) can be written as
\beq
\label{2c1ncvsapp}
- \,f_{abc} 
\sum_{\substack{i, \,\ell \,\in C \\ i \,\neq\, \ell}} \;\sum_{j \in NC}
\,T_i^a \,T_{\ell}^b \,T_j^c  
\;{\rm sign}(s_{j\wp}) \;\Theta(-z_i)  \;\;,
\eeq
where we have simply used the collinear approximation 
$s_{ij} \simeq z_i\, s_{j\wp}$. The only difference between the right-hand side
of Eq.~(\ref{2c1ncvscol}) and the expression (\ref{2c1ncvsapp}) is due to the
presence of the explicit constraint $s_{i\ell} < 0$; however, this apparent
difference is
harmless, since the sum of the terms with $s_{i\ell} > 0$ gives a vanishing
contribution to the expression (\ref{2c1ncvsapp}).
The vanishing of this sum follows from the relation
\beq
\label{2c1nccan}
f_{abc} 
\sum_{\substack{i, \,\ell \,\in C \\ i \,\neq\, \ell}} 
\,T_i^a \,T_{\ell}^b   
 \;\Theta(-z_i)  \;\Theta( s_{i\ell}) =
\f{1}{2} \,f_{abc} 
\sum_{\substack{i, \,\ell \,\in C \\ i \,\neq\, \ell}} 
\,T_i^a \,T_{\ell}^b   
 \;\left[ \,\Theta(-z_i) - \Theta(-z_l) \,\right] \;\Theta( s_{i\ell}) = 0 \;\;.
\eeq
Here (analogously to the procedure used in Eq.~(\ref{3vs3a})),
we have simply replaced the factor 
$\Theta(-z_i) \;\Theta(s_{i\ell})$ with its antisymmetric part with respect
to the exchange $i \leftrightarrow \ell$
and, then, we have used the fact that $\Theta(-z_i)=\Theta(-z_l)$ if
$s_{i\ell} > 0$ (i.e., $z_i z_\ell > 0$).

\section{Appendix: TL collinear limit, strict factorization 
and requirement of IR consistency}
\label{sec:tlircon}

In this Appendix we consider the TL collinear region, and we
discuss how a strictly-factorized splitting matrix can be recovered from the 
all-order IR 
structure presented in Eq.~(\ref{spfindef}). 
Then, we present explicit expressions 
for the IR structure of the TL splitting matrix.

In the multiparton TL collinear limit, the all-order splitting matrix $\sp$ is
strictly factorized (see Eq.~(\ref{colsptlallL}))
and, thus, it does not depend on the colour matrices of the non-collinear 
partons.
On the contrary, owing to Eq.~(\ref{vbardef}), the matrix structure of 
${\bf {\overline V}}$ only depends on the colour matrices of the 
non-collinear partons. Therefore, $\sp$  commutes with ${\bf {\overline V}}$
and, inverting Eq.~(\ref{spfindef}), we obtain
\beq
\label{spfindefinv}
\sp^{\,{\rm fin.}} = {\bf V}^{\,-1}(\ep) \;\,\sp 
\;\,{\bf {\overline V}}(\ep) = {\bf V}^{\,-1}(\ep)\;\,{\bf {\overline V}}(\ep)
\;\,\sp \;\;.
\eeq
This relation can be inverted to give
\beq
\label{spinv}
\sp  = {\bf {\overline V}}^{\,-1}(\ep)
\;\, {\bf V}(\ep) \;\,\sp^{\,{\rm fin.}}  \;\;.
\eeq
Note that, in the derivation of Eqs.~(\ref{spfindef}) and (\ref{spinv}), we 
have
not used the property that $\sp^{\,{\rm fin.}}$ is strictly factorized
(in general, the assumption that $\sp^{\,{\rm fin.}}$ is strictly factorized
is not valid, even in the TL case). Nonetheless $\sp^{\,{\rm fin.}}$
is IR finite and, therefore, the IR divergent contributions to  $\sp$ 
are produced by the operator 
${\bf {\overline V}}^{\,-1}\, {\bf V}$ on the right-hand side of 
Eq.~(\ref{spinv}). Since $\sp$ and, hence, its IR divergent terms are strictly
factorized, Eq.~(\ref{spinv}) enforces a constraint on 
${\bf {\overline V}}^{\,-1}\, {\bf V}$: the IR divergent part of the operator 
${\bf {\overline V}}^{\,-1}\, {\bf V}$ has to be strictly factorized. 
More precisely, this operator must have the following form:
\beq
\label{vbarvtl}
 {\bf {\overline V}}^{\,-1}(\ep) \;\, {\bf V}(\ep) 
 = {\bf V}_{\rm TL}(\ep) \;\,{\bf V}_{\rm fin.}(\ep)  \;\;, 
\quad \quad \quad ({\rm TL \,\;coll. \,\;lim.}) \;\;,
\eeq
where the IR divergent operator ${\bf V}_{\rm TL}(\ep)$ is {\em strictly
factorized}, whereas the operator ${\bf V}_{\rm fin.}(\ep)$ is IR finite 
(but it is not necessarily strictly factorized) and its form is such that the
matrix ${\bf V}_{\rm fin.}(\ep) \,\sp^{\,{\rm fin.}}$ is {\em strictly
factorized}. This constrained structure of Eq.~(\ref{vbarvtl}) guarantees that
the right-hand side of Eq.~(\ref{spinv}) and, hence, the splitting matrix
$\sp$  are strictly factorized. Note that the IR divergent operators 
${\bf V}$ and ${\bf {\overline V}}$ separately depend on the momenta and colour
charges of the non-collinear partons. This separate dependence is certainly
constrained by Eq.~(\ref{vbarvtl}) (in particular, the dependence largely cancels
in Eq.~(\ref{vbarvtl})), since the IR divergent operator ${\bf V}_{\rm TL}$
is strictly factorized and, thus, it is completely independent of the 
non-collinear partons.

The constrained structure of Eq.~(\ref{vbarvtl}) has been derived by
using two properties of multiparton QCD scattering amplitudes: their strict
factorization in the multiple TL collinear limit of $m$ partons 
(Eqs.~(\ref{factallL}) and (\ref{colsptlallL}))
and their IR structure (according to Eq.~(\ref{mallir}) or, equivalently,  
Eq.~(\ref{mallfact})).
This constrained structure can be regarded as a requirement of consistency between
the TL collinear limit  and the IR properties of the QCD amplitudes. If at some
perturbative order Eq.~(\ref{vbarvtl}) is not valid, either strict TL collinear
factorization is violated or the IR structure in Eqs.~(\ref{mallir}) and 
(\ref{mallfact})
is not valid at the corresponding perturbative order.

In the case of $m=2$ collinear partons, the presence of a valuable
IR consistency constraint from the behaviour of the QCD amplitudes in the TL
collinear limit was pointed out in Ref.~\cite{Becher:2009qa}
and exploited also in Ref.~\cite{Dixon:2009ur}.
The discussion in this subsection on the TL collinear limit generalizes the
discussions in Sect.~5 of  Ref.~\cite{Becher:2009qa} and Sect.~4
of Ref.~\cite{Dixon:2009ur};
our generalization deals with the extension to $m$ $(m \geq 3)$ collinear
partons and to a generic all-order form of the IR operator
${\bf V}_M(\ep)$ (or ${\bf I}_M(\ep)$) of the scattering amplitudes.
Note that both the operators ${\bf V}$ and ${\bf {\overline V}}$
in Eq.~(\ref{vbarvtl}) originate from ${\bf V}_M(\ep)$ in Eq.~(\ref{mallfact})
(the form of ${\bf V}_{\Mbar}$ follows from ${\bf V}_{M}$ by simply reducing the
total number of the external parton legs in $\cm$)
through the collinear-limit procedure in Eqs.~(\ref{vdef}) and (\ref{vbardef}).
Therefore, the strict factorization requirement in Eq.~(\ref{vbarvtl})
eventually constrains the colour and kinematical structure of the IR operator
${\bf V}_M(\ep)$ at arbitrarily-high perturbative orders
\cite{Becher:2009qa, Dixon:2009ur}.

To sharpen our all-order discussion of the multiparton TL collinear limit,
we consider the case in which the IR operator ${\bf V}_M(\ep)$ has a `minimal
form', which includes only the terms proportional to the IR poles $1/\ep^k$
with $k \geq 1$, whereas additional terms of order $\ep^0, \ep, \ep^2$ and so
forth are absent (we do not specify the colour and kinematical dependence of 
${\bf V}_M$). The corresponding exponentiated operator in Eq.~(\ref{vexp})
is denoted by 
${\bom I}_{M,\,{\rm cor (min)}}$ and, analogously,  the 
exponentiated splitting matrix operators in Eqs.~(\ref{vm1}) and (\ref{vbarm1})
are denoted by 
${\bom I}_{{\rm cor (min)}}$ and 
${\bom {\overline I}}_{{\rm cor (min)}}$, respectively.
We also define the following operator:
\beq
\label{itlall}
{\bom I}_{\rm TL, \,cor}(\ep) = {\bom I}_{{\rm cor (min)}}(\ep)
- {\bom {\overline I}}_{{\rm cor (min)}}(\ep) \;\;,
\eeq
which, by definition, also has a minimal form. 
Having specified these definitions, we sharply state our main conclusion:
the strict factorization of the splitting matrix $\sp$ is {\em equivalent}
to the requirement of strict factorization of {\em both} the collinear matrix
$\sp^{\,{\rm fin.}}$ and the operator ${\bom I}_{\rm TL, \,cor}$.
The proof of this statement is given below.

If $\sp$ is strictly factorized, the strict factorization of $\sp^{\,{\rm fin.}}$
is a simple consequence of the minimal form of ${\bf V}_M$. Indeed, owing to
the relations (\ref{vdef}) and (\ref{vbardef}),
if ${\bf V}_M(\ep)$ has a minimal form, 
${\bf {V}}$ and ${\bf {\overline V}}$ also have a minimal form: 
this implies that
the IR finite operator ${\bf V}_{\rm fin.}$ in Eq.~(\ref{vbarvtl})
is trivial (i.e., ${\bf V}_{\rm fin.}(\ep)=1$) and, therefore, that the IR finite
matrix $\sp^{\,{\rm fin.}}$ in Eq.~(\ref{spinv})
is strictly factorized. The strict factorization of $\sp^{\,{\rm fin.}}$ then
implies that $\sp^{\,{\rm fin.}}$ commutes with 
${\bf {\overline V}}^{-1}$ and, using Eq.~(\ref{spfindef}),
we get
\beq
\label{spinvcom}
\sp  = {\bf V}(\ep) \;\, {\bf {\overline V}}^{\,-1}(\ep)
 \;\,\sp^{\,{\rm fin.}}  \;\;.
\eeq
Comparing Eqs.~(\ref{spinv}) and (\ref{spinvcom}),
we conclude that the operators ${\bf V}$ and ${\bf {\overline V}}^{\,-1}$
commute.
This commutation property is valid also for the
corresponding exponentiated operators 
${\bom I}_{{\rm cor (min)}}(\ep)$ and 
${\bom {\overline I}}_{{\rm cor (min)}}(\ep)$ in Eq.~(\ref{itlall})
and, therefore, the operator ${\bf V}_{\rm TL}$  in Eq.~(\ref{vbarvtl})
has the following minimal form:
\beeq
\label{vtlmin}
{\bf V}_{\rm TL}(\ep) &=& {\bf {\overline V}}^{\,-1}(\ep) \;\, {\bf V}(\ep) 
 =  \exp \Bigl\{ \, {\bom I}_{{\rm cor (min)}}(\ep)
- {\bom {\overline I}}_{{\rm cor (min)}}(\ep) \Bigr\} \nn \\
&=& \exp \Bigl\{ \,{\bom I}_{\rm TL, \,cor}(\ep) \Bigr\} \;\;, 
\quad \quad \quad \quad \quad \quad \quad \quad ({\rm TL \,\;coll. \,\;lim.}) 
\;\;.
\eeeq
Here, we have used the property $e^A e^B = e^{A+B}$, which is valid if $A$ and
$B$ are commuting matrices.
Since ${\bf V}_{\rm TL}$ is strictly factorized, Eq.~(\ref{vtlmin})
finally implies that the exponentiated operator
${\bom I}_{\rm TL, \,cor}(\ep)$ of Eq.~(\ref{itlall}) is strictly factorized.

If $\sp^{\,{\rm fin.}}$ and the 
operator  ${\bom I}_{\rm TL, \,cor}(\ep)$ in Eq.~(\ref{itlall})
are strictly factorized, we can easily proof 
that ${\bf V}_{\rm TL}$ and, hence, $\sp$ are strictly factorized. Indeed,
if ${\bom I}_{\rm TL, \,cor}$ in Eq.~(\ref{itlall}) is strictly factorized,
this operator commutes with ${\bom {\overline I}}_{{\rm cor (min)}}$. Therefore,
we have
\beeq
0 &=& \left[ \,{\bom I}_{\rm TL, \,cor}(\ep) \;, 
\;{\bom {\overline I}}_{{\rm cor (min)}}(\ep) \,\right]
= \left[ \,{\bom I}_{{\rm cor (min)}}(\ep)
- {\bom {\overline I}}_{{\rm cor (min)}}(\ep) \;, 
\;{\bom {\overline I}}_{{\rm cor (min)}}(\ep) \,\right] \nn \\
&=& \left[ \,{\bom I}_{{\rm cor (min)}}(\ep) \;, 
\;{\bom {\overline I}}_{{\rm cor (min)}}(\ep) \,\right] \;\;,
\eeeq
namely, 
${\bom I}_{{\rm cor (min)}}$ and 
${\bom {\overline I}}_{{\rm cor (min)}}$ are commuting operators.
This commutation property leads to the explicit expression 
(\ref{vtlmin}) of the operator ${\bf V}_{\rm TL}$, and this expression is
evidently strictly factorized.

In summary, the validity of strict factorization in the TL collinear limit
requires that the IR operator ${\bom I}_{\rm TL, \,cor}$ in Eq.~(\ref{itlall})
is independent of the non-collinear partons. Since 
${\bom I}_{{\rm cor (min)}}(\ep)$ and 
${\bom {\overline I}}_{{\rm cor (min)}}(\ep)$ derive from the collinear limit
(see Eqs.~(\ref{vdef}), (\ref{vbardef}), (\ref{vm1}) and (\ref{vbarm1}))
of the scattering amplitude operator ${\bf V}_M(\ep)$
(see Eq.~(\ref{mallfact})),
this requirement directly constrains the colour and kinematical structure
of the exponentiated operator ${\bom I}_{M,\,{\rm cor (min)}}$
(see Eq.~(\ref{vexp})) in its minimal form \cite{Becher:2009qa, Dixon:2009ur}.

Moreover, using Eqs.~(\ref{spinv}), (\ref{vbarvtl}) and (\ref{vtlmin}),
the IR structure of the splitting matrix $\sp$ for the multiparton
TL collinear limit can be presented in the all-order form
of Eqs.~(\ref{spallirtl}--\ref{itldef}).

In the derivation of Eqs.~(\ref{spallirtl})--(\ref{itldef}), we have used a
minimal form of the scattering amplitude operator 
${\bf V}_M(\ep)$ in Eqs.~(\ref{mallfact})--(\ref{vexp}).
This minimal form is explicitly presented in Eqs.~(\ref{i1m}) and (\ref{i2m})
at one-loop and two-loop orders, respectively
(we recall that ${\bom I}_{M,\,{\rm cor}}^{(1)}={\bom I}_{M}^{(1)}$
and $2{\bom I}_{M,\,{\rm cor}}^{(2)}= 2{\bom I}_{M}^{(2)} +
({\bom I}_{M}^{(1)})^2\,$).
Therefore, using Eq.~(\ref{itlall}),
we can explicitly compute the perturbative expansion of 
${\bom I}_{\rm TL, \,cor}$,
${\bf V}_{\rm TL}$ and ${\bf I}_{\rm TL}$ up to the two-loop order.
To be precise, we define the renormalized loop expansion of ${\bf I}_{\rm TL}$:
\beq
\label{itlperexp}
{\bf I}_{\rm TL}(\ep) = {\bf I}^{(1)}_{\rm TL}(\ep) +
{\bf I}^{(2)}_{\rm TL}(\ep) + \dots \;\;,
\eeq
and we straightforwardly find
\beeq
\label{i2tlexp}
{\bf I}^{(2)}_{\rm TL}(\ep) &=& - \, \f{1}{2} \left[ 
{\bf I}^{(1)}_{\rm TL}(\ep) \right]^2
+ \f{\as(\mu^2)}{2\pi} 
\left\{ + \frac{1}{\ep} \,\cbet0 
\left[ {\bf I}^{(1)}_{\rm TL}(2\ep) - {\bf I}^{(1)}_{\rm TL}(\ep) \right]
+ K \;{\bf I}^{(1)}_{\rm TL}(2\ep) 
\right. \nn \\
&+& \, \f{\as(\mu^2)}{2\pi} 
\left.
 \,\frac{1}{\ep} \;
\Bigl( \;\sum_{i \, \in C} \,H^{(2)}_i 
\Bigr. 
\Bigl. \left.
- H^{(2)}_{\wp}  \right. \Bigr)
\right\}
 \;\;,
\eeeq
where the explicit expression of the one-loop operator 
${\bf I}^{(1)}_{\rm TL}(\ep)$ exactly corresponds to the expression 
(\ref{i1mcexptl}) of the IR operator $\imc(\ep)$ in the TL collinear region.
The one-loop perturbative expansion of the TL factorization formula
(\ref{spallittl}) gives Eqs.~(\ref{sp1df}) and (\ref{sp1div})
with the obvious identification of ${\bf I}^{(1)}_{\rm TL}$ with $\imc$.
The two-loop perturbative expansion gives
\beq
\label{sp2rentl}
\sp^{(2, R)} = {\bf I}^{(1)}_{\rm TL}(\ep) \;\sp^{(1, R)}
+ {\bf I}^{(2)}_{\rm TL}(\ep) \;\sp^{(0, R)} + \sp^{(2) \,{\rm fin.}} \;\;,
\quad ({\rm TL \,\;coll. \,\;lim.}) \;\;,
\eeq
where ${\bf I}^{(2)}_{\rm TL}(\ep)$ is explicitly presented 
in Eq.~(\ref{i2tlexp}).
In the case of $m=2$ collinear partons, the results in 
Eqs.~(\ref{spallirtl})--(\ref{itldef}) and 
Eqs.~(\ref{itlperexp})--(\ref{sp2rentl})
exactly correponds to those presented in 
Refs.~\cite{Becher:2009qa, Dixon:2009ur}.

In this Appendix we have explicitly derived 
Eqs.~(\ref{spallirtl})--(\ref{itldef}) by considering IR divergent operators 
that have a minimal form.
However, Eq.~(\ref{spallirtl}) has the same IR factorization invariance as
Eq.~(\ref{mallfact}).
Specifically, the right-hand side of Eq.~(\ref{spallirtl})
is invariant under the joint transformations (redefinitions)
${\bf V}_{\rm TL}(\ep) \to {\bf V}_{\rm TL}(\ep) \,{\bf V}_{\rm fin.}(\ep)\,$ and
$\sp^{\,{\rm fin.}} \to {\bf V}_{\rm fin.}^{-1}(\ep) \,\sp^{\,{\rm fin.}}$,
where ${\bf V}_{\rm fin.}(\ep)$ is an
invertible IR finite operator that is {\em strictly factorized}.
Since this invariance can be used to redefine 
${\bf V}_{\rm TL}$ and $\sp^{\,{\rm fin.}}$, the structure of 
Eqs.~(\ref{spallirtl})--(\ref{itldef}) is actually valid independently 
of the minimal form of the IR divergent operator ${\bf V}_{\rm TL}$ (or, 
${\bf I}_{\rm TL}$ and ${\bom I}_{\rm TL, \,cor}$).

\noindent{\bf Acknowledgements}. This work was supported in part by
the Research Executive Agency (REA) of the European Union 
under the Grant Agreement number PITN-GA-2010-264564 (LHCPhenoNet) 
and by UBACYT, CONICET, ANPCyT, INFN,
INFN-MICINN agrement ACI2009-1061 and AIC10-D-000576,
MICINN (FPA2007-60323, FPA2011-23778 and CSD2007-00042)
and GV (PROMETEO/2008/069). D.dF. wishes to thank the Pauli Center for Theoretical Studies for support during his visit to the University of Z\"urich. 
We thank the Galileo Galilei Institute for Theoretical Physics for the
hospitality during the completion of this work.

\newpage
\noindent {\bf \large Note added}

After the completion of the present paper,
J.~Forshaw, M.~Seymour and A.~Siodmok made an important 
observation \cite{JMA}.
We would like to thank Jeff, Mike and Andrzej for discussions and the
communication of their result before its publication.
The observation regards the expectation value of the two-loop operator 
${\widetilde {\bf \Delta}}_{P}^{(2)}(\ep)$
(see Eq.~(\ref{wd2til})), which gives the (IR dominant) factorization
breaking contribution to the squared splitting matrix 
${\bf P}^{(2,R)}$ (see Eq.~(\ref{p22part})) for the 
two-parton collinear limit.
The expectation value onto the reduced matrix element $\cmbar$
is
\beeq
\label{vev}
&&  \!\!\!\!\!\! \!\!\!\!\!\!\!\!\! \la \, \cmbar \, |
\left( \sp^{(0,R)} \,\right)^\dagger
\;{\widetilde {\bf \Delta}}_{P}^{(2)}(\ep) \;\sp^{(0,R)} \;
\, \ket{\cmbar} =
 \la \, \cmbar^{(0,R)} \,
 |  
\left( \sp^{(0,R)} \,\right)^\dagger 
\;{\widetilde {\bf \Delta}}_{P}^{(2)}(\ep) \;\sp^{(0,R)} 
\;\ket{\cmbar^{(0,R)}} 
\nn \\
&&\;\; +\; \left[ \la \, \cmbar^{(1,R)} \,
 | \, 
\left( \sp^{(0,R)} \,\right)^\dagger 
\;{\widetilde {\bf \Delta}}_{P}^{(2)}(\ep) \;\sp^{(0,R)} 
\;\ket{\cmbar^{(0,R)}} + {\rm c.c.} \; \right] + {\rm higher \;orders}
\;\;,
\eeeq
where the right-hand side corresponds to the perturbative (loop) expansion
of $\cmbar$ ($ \cmbar = \cmbar^{(0,R)} +\cmbar^{(1,R)} + \dots$).

The key observation \cite{JMA} is that, at the lowest-order level (i.e.,
considering the expectation value onto $\cmbar^{(0,R)}$, as given
in the first term on the right-hand side of Eq.~(\ref{vev}))
the expectation value vanishes in {\em pure} QCD (i.e., if the lowest-order
reduced matrix element $\cmbar^{(0,R)}$ is produced by tree-level QCD
interactions). We refer to Ref.~\cite{JMA} for the explanation and discussion 
of this effect.

Note that ${\widetilde {\bf \Delta}}_{P}^{(2)}$ is not vanishing. The
lowest-order vanishing of the expectation value in Eq.~(\ref{vev})
can indeed be avoided by changing (and properly choosing) the lowest-order
matrix element $\cmbar^{(0,R)}$. For instance, we can consider tree-level 
quark--quark
scattering produced by electroweak interactions (with 
CP-violating electroweak couplings
and/or finite width of the $Z$ and $W^{\pm}$ bosons), 
or we can supplement tree-level QCD
scattering with one-loop (pure) QED radiative corrections.
Therefore, as a matter of principle, it remains true that the operator  
${\widetilde {\bf \Delta}}_{P}^{(2)}$ explicitly uncovers {\em two-loop}
QCD effects that lead to violation of strict collinear factorization at the
squared amplitude level.
In particular, the conceptual discussion presented in 
Sect.~\ref{sec:crosssec} (and briefly recalled at the end of
Sects.~\ref{sec:in} and \ref{sec:fin})
continues to be valid, although the lowest-order partonic subprocess 
$\,`{\rm ~parton + parton} \to 2~{\rm partons~}$' (see Eq.~(\ref{prolo}))
that is used as starting point of the discussion has to be interpreted in the
generalized sense mentioned at the beginning of this paragraph.
Of course, the fact that the lowest-order partonic subprocess
in Sect.~\ref{sec:crosssec} is not due to tree-level QCD interactions 
reduces the possible phenomenological consequences 
of the effects discussed in that section.
 
In the context of pure QCD, the expectation value in Eq.~(\ref{vev}) is 
{\em not}
vanishing at the next-to-lowest order (i.e., the second term on the right-hand
side of Eq.~(\ref{vev})), which is obtained by the one-loop QCD correction 
$\cmbar^{(1,R)}$ to
the reduced matrix element $\cmbar$. Therefore,
${\widetilde {\bf \Delta}}_{P}^{(2)}$ certainly contributes, through
the two-loop splitting matrix ${\bf P}^{(2,R)}$, to the SL collinear limit of
three-loop
QCD squared amplitudes. At the three-loop level, the violation of strict
collinear factorization produced by ${\bf P}^{(2,R)}$ (e.g., 
${\widetilde {\bf \Delta}}_{P}^{(2)}$) joins additional
factorization breaking effects that are produced by the 
three-loop collinear matrix ${\bf P}^{(3,R)}$.

This combined occurence of ${\bf P}^{(2,R)}$ and ${\bf P}^{(3,R)}$
in the collinear limit of QCD squared amplitudes at the three-loop level 
has prompted our preliminary investigation of three-loop effects.
The analysis and results of the present paper can be {\em readily}
and straightforwaldy  extended to higher-loop orders by exploiting the {\em
exponentiated} structure of the leading (i.e., ${\cal O}(\as^n/\ep^{n+1})$)
and next-to-leading (i.e., ${\cal O}(\as^n/\ep^{n})$) IR divergences
of multiparton QCD scattering amplitudes
(see Sect.~\ref{sec:IRall} and 
Refs.~\cite{leadingIR, Catani:1998bh, Sterman:2002qn}).
In the case of the SL collinear limit of two partons, we {\em know} the 
complete $\ep$ dependence of the one-loop splitting matrix $\sp^{(1,R)}$ and,
therefore, we have explicit control of three-loop factorization breaking 
effects starting from their dominant IR divergent terms
of ${\cal O}(\as^3/\ep^{4})$. Computing ${\bf P}^{(3,R)}_{\rm n.f.}$
(the part of  ${\bf P}^{(3,R)}$ that violates 
strict collinear factorization)
up to the accuracy of ${\cal O}(\as^3/\ep^{3})$, we obtain
\beeq
\label{p3nfleading}
{\bf P}^{(3,R)}_{\rm n.f.} &=& \left( \sp^{(0,R)} \,\right)^\dagger 
 \left\{
\left( 
\itp^{(1)}(\ep)  - \frac{\as(\mu^2)}{2 \pi} \,\frac{1}{\ep} \,\cbet0
\right)
\;{\widetilde {\bf \Delta}}_{P}^{(2)}(\ep) 
\right. \nn \\
&+& \left. \f{1}{6}
\;\left[ \, \left(\, {\bom {\overline I}}^{(1)}(\ep)
- \itc^{(1)}(\ep)
\, \right) - \;{\rm h.c.} \;\;, \;{\widetilde {\bf \Delta}}_{P}^{(2)}(\ep)
\,\right] \right\} \;\sp^{(0,R)} + {\cal O}\left(\frac{\as^3}{\ep^2}\right)
\;\;,
\eeeq
where $\itc^{(1)}, \itp^{(1)}$ and ${\bom {\overline I}}^{(1)}$
are the customary one-loop operators used throughout the paper
(see Eqs.~(\ref{it2cone}), (\ref{itp1}) and (\ref{ibaronefor}))
and $\cbet0$ is the first-order coefficient of the QCD $\beta$ function 
(see Eq.~(\ref{norbeta0})).
The three-loop
expression in Eq.~(\ref{p3nfleading}) is controlled by the iterated
action of ${\widetilde {\bf \Delta}}_{P}^{(2)}$ 
without (first term on the right-hand side of Eq.~(\ref{p3nfleading}))
and with (second term on the right-hand side of Eq.~(\ref{p3nfleading}))
an additional colour commutator with one-loop terms.

We note that the three-loop factorization breaking contribution from 
${\bf P}^{(2,R)}$ (the second term on the right-hand side of Eq.~(\ref{vev}))
cannot be cancelled by the
factorization breaking contribution from
${\bf P}^{(3,R)}_{\rm n.f.}$ (e.g., the lowest-order expectation value of the
second term on the right-hand side of Eq.~(\ref{p3nfleading})).
In particular, we remark that in pure QCD each of these two terms
produces an IR divergent three-loop
contribution of order $\as^3/\ep^3$.
In the case of scattering amplitudes with $n=5$ QCD partons, the colour
correlation structure of these two
non-cancelled factorization breaking contributions is analogous to the
commutator structures
that were found \cite{Forshaw:2006fk} 
in the N$^4$LO computation of super-leading logarithms.
This observation is consistent with the common physical mechanism
that originates the violation of strict collinear factorization
and the emergence of super-leading logarithms in `gaps--between--jets'
cross sections
(as mentioned at the end of Sect.~\ref{sec:crosssec} and remarked in
Ref.~\cite{{JMA}}).

\end{document}